\documentclass[onecolumn,showpacs,floatfix,prb]{revtex4-1}
\usepackage{float}
\usepackage{dcolumn}
\usepackage{hyperref} % For hyperlinks in the PDF
\hypersetup{bookmarks=true,unicode=true,colorlinks=true, urlcolor=blue,linkcolor=blue,citecolor=blue,filecolor=blue}
\usepackage{bussproofs}
\usepackage{graphicx,graphics}
\usepackage{fancyhdr}
\usepackage{amsmath}
\usepackage{subfigure}
\usepackage{epsfig}
\usepackage{amssymb}
\usepackage{bm}
\usepackage{units}
\usepackage{pifont}
\usepackage{textcomp}
\usepackage{gensymb}
\usepackage{enumerate}
\usepackage{color}
\usepackage{physics}

\begin{document}

\title{Unveiling signatures of topological phases in open Kitaev chains and ladders}

\author{A. Maiellaro, F. Romeo} 
\address{Physics Department "E.R. Caianiello", Universit\`{a} di Salerno
\\ Via Giovanni Paolo II, 132, I-84084 Fisciano (SA), Italy}
\author{C.A. Perroni, V. Cataudella} 
\address{CNR-SPIN and Physics Department "Ettore Pancini", Universit\`{a} degli Studi di Napoli Federico II
\\ Complesso Universitario Monte S. Angelo, Via Cintia, I-80126 Napoli, Italy}
\author{R. Citro}
\address{CNR-SPIN and Physics Department "E.R. Caianiello", 
Universita' degli Studi di Salerno,\\
Via Giovanni Paolo II, 132, I-84084 Fisciano (Sa), Italy }

%\corres{Correspondence: fromeo@sa.infn.it; Tel. +39 089-968218: (F.R.)}

\begin{abstract}
In this work the general problem of the characterization of the topological phase of an open quantum system is addressed. In particular, we study the topological properties of Kitaev chains and ladders under the perturbing effect of a current flux injected into the system using an external normal lead and derived from it via a superconducting electrode. After discussing the topological phase diagram of the isolated systems, using a scattering technique within the Bogoliubov-de Gennes formulation, we analyze the differential conductance properties of these topological devices as a function of all relevant model parameters. The relevant problem of implementing local spectroscopic measurements to characterize topological systems is also addressed by studying the system electrical response as a function of the position and the distance of the normal electrode (tip). The results show how the signatures of topological order affect the electrical response of the analyzed systems, a subset of such signatures being robust also against the effects of a moderate amount of disorder. The analysis of the internal modes of the nanodevices demonstrates that topological protection can be lost when quantum states of an initially isolated topological system are hybridized with those of the external reservoirs. The conclusions of this work could be useful in understanding the topological phases of nanowire-based mesoscopic devices.

\end{abstract}

\maketitle

\section{Introduction}
In the last decade, the properties of Majorana zero-energy modes (MZMs) hosted in topological superconductors have gathered considerable interest. Indeed, their non-abelian statistics has been proposed as the working principle of a fault-tolerant topological quantum computation \cite{refPachos}. However, a crucial challenge towards the topological quantum computer is to implement quantum operations of nearly degenerate quantum states by a dynamical process involving Majorana fermions; the main one is the braiding dynamics in superconducting nanowires. From the theoretical side, the simplest model for realizing MZMs is the one-dimensional spinless $p$-wave chain proposed by Kitaev \cite{refKitaev2}. A realistic implementation of the Kitaev model has been proposed lately in the pioneering work by Fu and Kane \cite{refFuKane} who have predicted the presence of MZMs as a result of the proximity effect between an s-wave superconductor and the surface states of a strong topological insulator.\\
Many experimental works have been performed to realize the theoretical predictions of the Kitaev model and, in particular, Mourik and coworkers \cite{refKouwenhoven} have shown evidences of MZMs in the tunnel conductance of an $InAs$ nanowire proximized by an s-wave superconductor. Two years later, by implementing a previous theoretical proposal \cite{YazdaniTh}, Nadj-Perge et al. \cite{refYazdani} exhibited an STM measurement of a long chain of iron atoms deposited on a lead substrate, in which local density of states (LDOS) highlighted the presence of MZMs localized at the system edges.\\
Despite the existence of a rich literature on the topological phases of closed systems, or on different probes of MZMs in nanostructures (see e.g. \cite{been,aguado,marra_2016}), it remains relevant to understand how the topological phases are modified by the measurement procedure which is realized, e.g., by coupling the system to the normal tip of a scanning tunneling microscope (STM). To this aim, we study the charge transport through a Kitaev chain (KC) and a Kitaev ladder (KL) coupled to a normal and a $p$-wave superconducting electrode acting, respectively, as source and drain of the topological nanodevice. Using a scattering matrix approach, we analyse in what extent the topological edge states are perturbed by the passage of a current, also varying the distance and the position of the tip (i.e., the normal electrode). To test the robustness of the topological phase, we also consider the effect of disorder on transport properties and show the persistence of a quantized zero-bias peak in the conductance. Beyond a robust zero-bias conductance peak, it is shown the existence of quasi-zero energy peaks that correspond to hybridized MZMs or quasi-Majorana modes, i.e. quantum states strongly coupled to the external leads which present a spatial distribution peaked at the edges and extended over the entire system. Our results can be useful for the interpretation of future experimental works involving STM characterization of topological superconductors. From the methodological viewpoint, the presented analysis can be relevant to characterize the transport properties of nanodevices based on innovative materials whose minimal model can be mapped into an effective multi-orbital Kitaev chain theory (e.g. the one-dimensional heterostructures exploiting the emergent properties of the LaAlO$_3$/SrTiO$_3$ interfaces \cite{grilli_2018}). Going beyond the context of condensed matter, our findings could be relevant for the emergent field of \textit{atomtronics} \cite{atomtronics} aiming at the high-precision control of atomic matter waves. Atomtronics experiments  can be implemented with bosonic or fermionic atoms under extremely controllable conditions and are the ideal playground to implement concepts borrowed by quantum electronics, both in equilibrium and in non-equilibrium conditions.\\
The paper is thus organized as follows. In Sec. \ref{sec:model} and Sec. \ref{sec:ladder} we summarize the main properties of the isolated Kitaev chain and ladder and discuss the topological phases. The tight-binding model for the normal/Kitaev-chain/superconductor (N-KC-SC) device is introduced and analyzed in Sec. \ref{sec:junction}, where we also report the results for the conductance and study the evolution of the zero-bias peak as a function of the parameters driving the topological phase transition. The charge transport through a normal/Kitaev-ladder/superconductor (N-KL-SC) device is analyzed in Sec. \ref{sec:junction-ladder}, also discussing the effect of disorder on the zero-bias peak (Sec. \ref{disorder}). Technical details on the analytical calculations and the tight binding Bogoliubov de Gennes (BdG) equations are given in the appendix \ref{appBdg}. Conductance lowering effects in branched quantum devices and the analogies with the Blonder-Tinkham-Klapwijk (BTK) theory of transport in N-SC junction are discussed in the appendix \ref{appWG}. Charge neutrality of hybridized Majorana modes is analyzed in appendix \ref{appCharge}. Finally, the conclusions are given in Sec. \ref{sec:conclusions}.

\section{Majorana Fermions in the Kitaev chain}
\label{sec:model}
\begin{figure}
\includegraphics[scale=0.8]{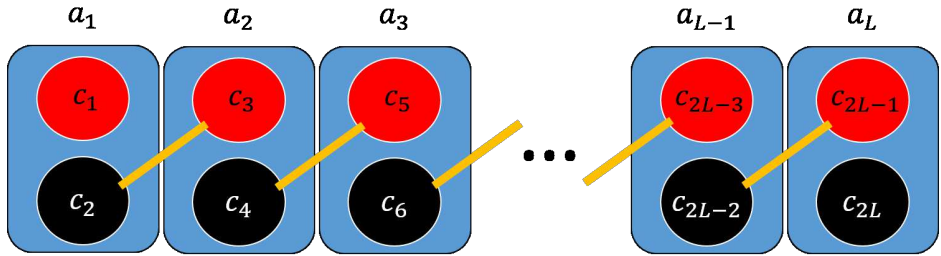}
\centering
\caption{A schematic representation of the ideal Kitaev chain model. Unpaired Majorana zero modes $c_1$ and $c_{2L}$ are localized at the system edges and do not enter the Hamiltonian $H$. The remaining Majorana modes, namely $c_{2j}$ and $c_{2j+1}$, recombine into ordinary fermionic excitations $a_j$.}
\label{Fig1}
\end{figure}

In this Section we briefly discuss the main topological properties of the Kitaev chain. This discussion provides the appropriate starting point before treating the open quantum systems and serves to fix the notation. Following the Kitaev seminal work \cite{refKitaev2}, we consider a system of spinless fermions confined to a one dimensional lattice and subject to a $p$-wave superconducting coupling described by the Hamiltonian:

\begin{equation}
\label{HKitaev}
H_K=\sum_{j=1}^L(-ta^\dagger_j a_{j+1}+\Delta a_ja_{j+1}+h.c.)-\mu \sum_{j=1}^{L} a^\dagger_j a_j,
\end{equation}

where $a_j$ ($a^\dagger_j$) is the fermionic annihilation (creation) operator for a site $j$ ($j=1,..., L$), $t>0$ is the amplitude of  nearest-neighbor hopping, $\Delta>0$ is the amplitude of the superconducting pairing, while $\mu$ represents the chemical potential. For the Kitaev-special case ($t=\Delta$, $\mu=0$), using the Majorana basis $c_{2j-1}=a^\dagger_j+a_j$ and $c_{2j}=-i(a_j-a^\dagger_j)$, the Hamiltonian in Eq.(\ref{HKitaev}) becomes:

\begin{equation}
H=it\sum_{j=1}^{L-1}c_{2j}c_{2j+1}\nonumber,
\end{equation}

where the hermitian Majorana operators fulfill the anticommutation relation $\{c_k,c_l\}=2\delta_{kl}$. In this case, the Hamiltonian has two zero-energy Majorana modes, namely $c_1$ and $c_{2L}$, which are located at the end of the wire and combine to form a zero-energy non-local fermion $f=(c_1+ic_{2L})/2$. The other $N-1$ fermionic modes (Bogoliubov modes) $\tilde{a}^\dagger_j=(c_{2j+1}+ic_{2j})$, $j\in [1, N-1]$,
are degenerate with energy $2t$ and connect the neighbouring sites in the bulk as depicted in Figure \ref{Fig1}.\\

For arbitrary values of $\Delta$, $t$, $\mu$, the Kitaev Hamiltonian $H_K$ can be diagonalized by the Bogoliubov transformation
$a_j=\sum_m(u_{j,m}\alpha_m+v^*_{j,m}\alpha^\dagger_m)$, where the quasiparticle fermionic annihilation and creation operators $\alpha_m$ and $\alpha^\dagger_m$ obey canonical anticommutation relations. The Bogoliubov amplitudes $u_{j,m}$ and $v_{j,m}$ satisfy the Bogoliubov-de Gennes (BdG) equations:

\begin{eqnarray}
\label{BdGKitaev}
-\mu u_{j, m}-t(u_{j+1, m}+u_{j-1, m})+\Delta(v_{j-1, m}-v_{j+1, m})=E_m u_{j,m}\nonumber\\
\mu v_{j, m}+t(v_{j+1, m}+v_{j-1, m})+\Delta(u_{j+1, m}-u_{j-1, m})=E_m v_{j, m},
\end{eqnarray}

with excitation energy $E_{m}\geq 0$. The diagonal form of $H_{K}$ reads:

\begin{equation}
H_{K}=E_0+\sum_m E_m \alpha^\dagger_m \alpha_m,
\end{equation}

where $E_0$ is the energy of ground state $|0\rangle$ of Bogoliubov modes so that $\alpha_{m}|0\rangle=0$ for all $\alpha_{m}$.\\
This model admits a topological phase for $|\mu|<2t$ with a robust zero-energy edge mode, which is here labeled by setting $m=M$. The energy $E_M$ of this mode is not exactly zero, as expected in thermodynamic limit, and presents an exponential decay with the system size $L$ (i.e., $E_M \backsim \exp(-L / \xi)$, where $\xi$ represents a characteristic decay length). The remaining non-topological modes of the spectrum are gapped and form a band.\\
The zero-energy Majorana modes are localised at the left/right edge of the wire and decay inside the bulk; the annihilation operators of such modes are given by:

\begin{eqnarray}
\gamma_L=\alpha_M+\alpha^\dagger_M=\sum_jf_{L,j}(a_j+a^\dagger_j)=\sum_j f_{L,j}c_{2j-1}\nonumber\\
\gamma_R=-i(\alpha_M-\alpha^\dagger_M)=-i\sum_jf_{R,j}(a_j-a^\dagger_j)=\sum_j f_{R,j}c_{2j}\nonumber
\end{eqnarray}

where the real valued eigenfunctions $f_{L,j}=(u_{j, M}+v^*_{j, M})$ and $f_{R,j}=(u_{j,M}-v^*_{j, M})$ are obtained by explicit solution of the BdG equations (\ref{BdGKitaev}). Following Ref. \cite{Baranov}, it is possible to show that $f_{L,j}=2|A|\rho^j \sin(j \theta)$ and $f_{R,j}=2|A|\rho^{L+1-j}\sin((L+1-j)\theta)$, where $\rho=\sqrt{\frac{t-\Delta}{t+\Delta}}<1$ and $\frac{-\mu\pm \sqrt{\mu^2-4(t^2-\Delta^2)}}{2(t+\Delta)}=\rho e^{\pm i \theta}$.\\
In the thermodynamic limit ($L\rightarrow\infty$), the presence of such edge modes results in an exact degeneration of the ground state that corresponds to the presence or absence of the non-local fermion $\alpha_M=(\gamma_L+i\gamma_R)/2$. In fact, in this limit, the two ground states (zero-energy states) $|+\rangle$,$|-\rangle$ have different parity and thus $\alpha_M|-\rangle=0$ and $\alpha^\dagger_M|-\rangle=|+\rangle$.

\begin{figure}
\centering
\includegraphics[scale=0.55]{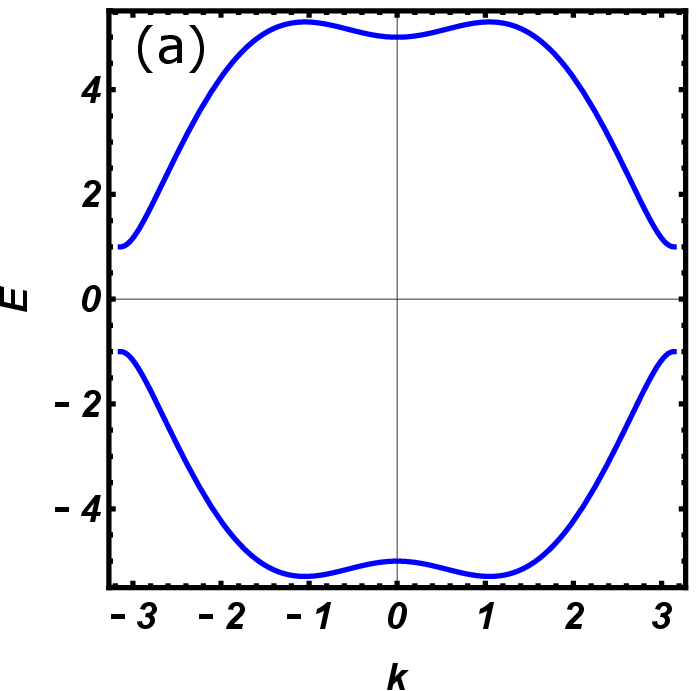}
\includegraphics[scale=0.55]{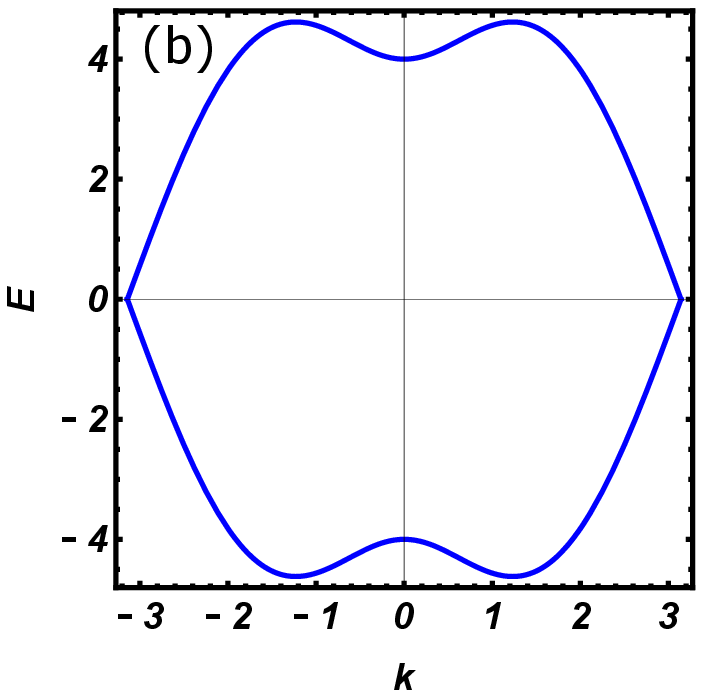}\\
\includegraphics[scale=0.55]{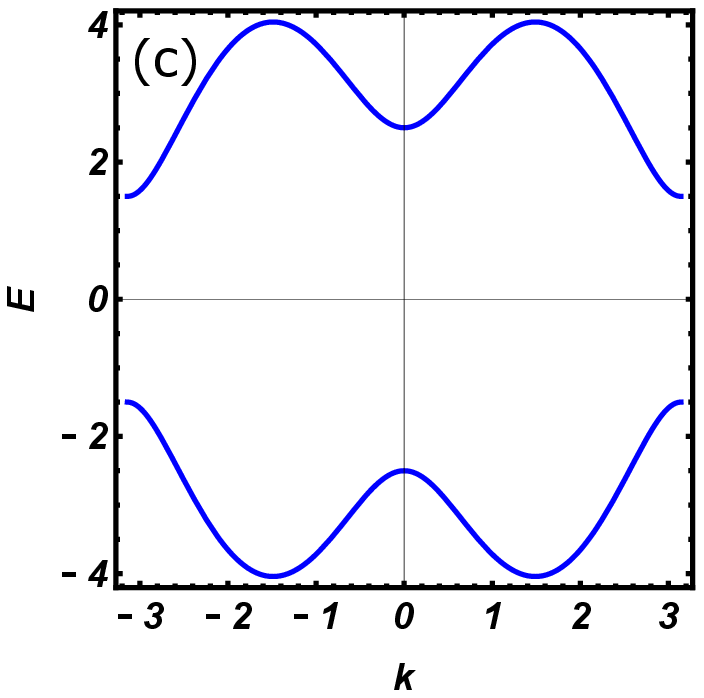}
\includegraphics[scale=0.55]{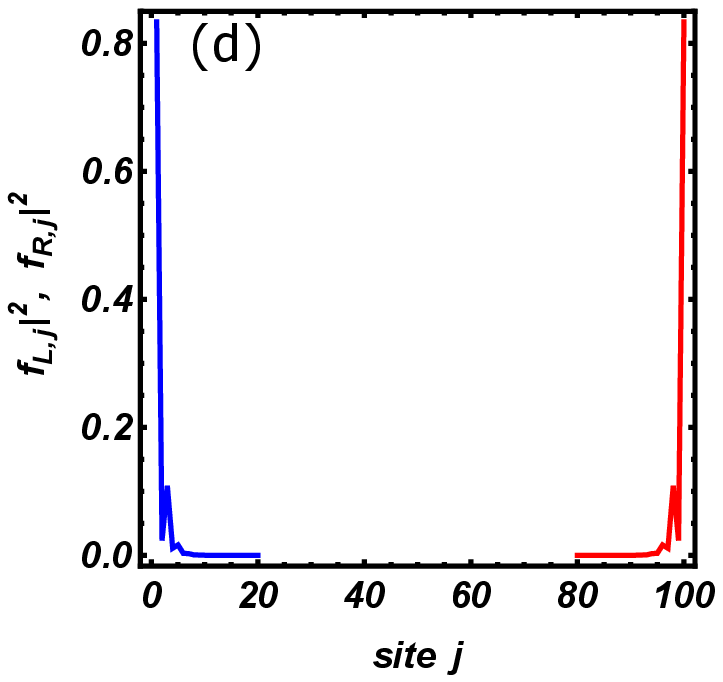}
\caption{Energy bands of the Kitaev chain in the non-topological phase $\mu=3$ (panel (a)), at the phase transition point $\mu=2$ (panel (b)), and inside the topological region $\mu=0.5$ (panel (c)). Panel (d) represents the modulus squared of Majorana zero-modes wave functions in the topological phase (c). Energy is expressed in units of the hopping amplitude $t$, while the remaining model parameters have been fixed as $L=100$, $\mu=0.5$, $\Delta=2$.}
\label{Fig2}
\end{figure}

\section{Ladder of two Kitaev chains}
\label{sec:ladder}
Various generalization of the single Kitaev chain have recently appeared \cite{potter_2010,benakker_2010,zhou_2011,Wakatsuki,refZhou,loss_2017}, the simplest one being obtained by coupling two single Kitaev chains with transversal hopping and pairing terms to form a Kitaev ladder (KL) \cite{Maiellaro}.
The Hamiltonian of the system is given by:

\begin{equation}
\label{Hamiltonian}
H=H_{K_1}+H_{K_2}+H_{K_1,K_2}
\end{equation}

where $H_{K_1}$ and $H_{K_2}$ are the Hamiltonians of the isolated Kitaev chains given in Eq.(\ref{HKitaev}), while $H_{K_1,K_2}$ describes the interchain coupling whose expression is given by:

\begin{equation}
\label{HCoupling}
H_{K_1,K_2}=\sum_{j=1}^L [-t_1 a^\dagger_{j,1}a_{j,2}+\Delta_1 a_{j,1}a_{j,2}+h.c.].
\end{equation}

The labels $1, 2$ in Eq. (\ref{HCoupling}) denote the two chains, $j$ is the site index, $t_1$ represents the transversal hopping amplitude and $\Delta_1$ is the transversal pairing term. All the model parameters, namely $t$, $t_1$, $\Delta$ and $\Delta_1$, are taken as real numbers. In the momentum representation the Hamiltonian (\ref{Hamiltonian}) can be written as

\begin{equation}
H=\frac{1}{2}\sum_{k}\Psi^\dagger(k)H(k)\Psi(k),
\end{equation}

where

\begin{equation}
\label{SpinorKspace}
\Psi(k)=(a_{k,1},a^\dagger_{-k,1},a_{k,2},a^\dagger_{-k,2})^t
\end{equation}

represents the Nambu spinor. The Hamiltonian is given by

\begin{eqnarray}
\label{BdGKspace}
H(k)=\left(
                               \begin{array}{cccc}
                                 \epsilon_k&i\Delta_k&t_1&-\Delta_1\\
                                 -i\Delta_k&-\epsilon_k&\Delta_1&-t_1\\
                                 t_1&\Delta_1&\epsilon_k&i\Delta_k\\
                                 -\Delta_1&-t_1&-i\Delta_k&-\epsilon_k\\
                               \end{array}
                             \right),
\end{eqnarray}

with $\epsilon_k=-2t\cos k-\mu$ and $\Delta_k=2\Delta \sin k$. By construction, the Hamiltonian (\ref{BdGKspace}) satisfies the particle-hole symmetry ($\Xi$)

\begin{equation}
\Xi H(k) \Xi^\dagger=-H(-k)\nonumber
\end{equation}

implemented by the operator

\begin{equation}
\Xi=\mathbb{I}\otimes\sigma_x \mathcal{K}\nonumber,
\end{equation}

which is defined in terms of the complex conjugation operator $\mathcal{K}$ and of the Pauli matrix $\sigma_x$, while $\mathbb{I}$ represents the identity operator. The Hamiltonian satisfies also the time-reversal symmetry ($\tau=\mathcal{K}$)

\begin{eqnarray}
\tau H(k) \tau^\dagger=H(-k) \nonumber
\end{eqnarray}

and the chiral symmetry ($\Pi=\Xi\tau=\mathbb{I}\otimes\sigma_x$)

\begin{eqnarray}
\Pi H(k)\Pi^\dagger=-H(k).\nonumber
\end{eqnarray}

According to the mentioned symmetry properties enjoyed by $H(k)$, the hamiltonian model belongs to the BDI symmetry class with Z topological index \cite{Atland}. The spectrum of the Hamiltonian can be easily obtained and is given by

\begin{equation}
E_k=\pm \sqrt{t^2_1+\Delta^2_1-\Delta^2_k+\epsilon^2_k\pm 2\sqrt{-\Delta^2_1 \Delta^2_k+t^2_1(\Delta^2_1+\epsilon^2_k)}}.\nonumber
\end{equation}

The topological phase diagram can be obtained by the calculation of the winding number \cite{ShortCourseTopologicalInsulator, refZhou}. Using the chiral basis, the Hamiltonian (\ref{BdGKspace}) assumes the following off-diagonal form \cite{ShortCourseTopologicalInsulator}

\begin{eqnarray}
\tilde{H}(k)=UH(k)U^{\dag}=
                               \begin{pmatrix}
                                 0&A_k\\
                                A^{\dag}_k&0\\
                               \end{pmatrix},
\end{eqnarray}

where $U$ is the basis change matrix and $A_k$ is the $2\times 2$ matrix

\begin{eqnarray}
A_k=
                               \begin{pmatrix}
                                 -\epsilon_k-i\Delta_k&-t_1-\Delta_1\\
                               -t_1+\Delta_1&-\epsilon_k-i\Delta_k\\
                               \end{pmatrix}.
\end{eqnarray}

\begin{figure}[!h]
\centering
\includegraphics[scale=0.55]{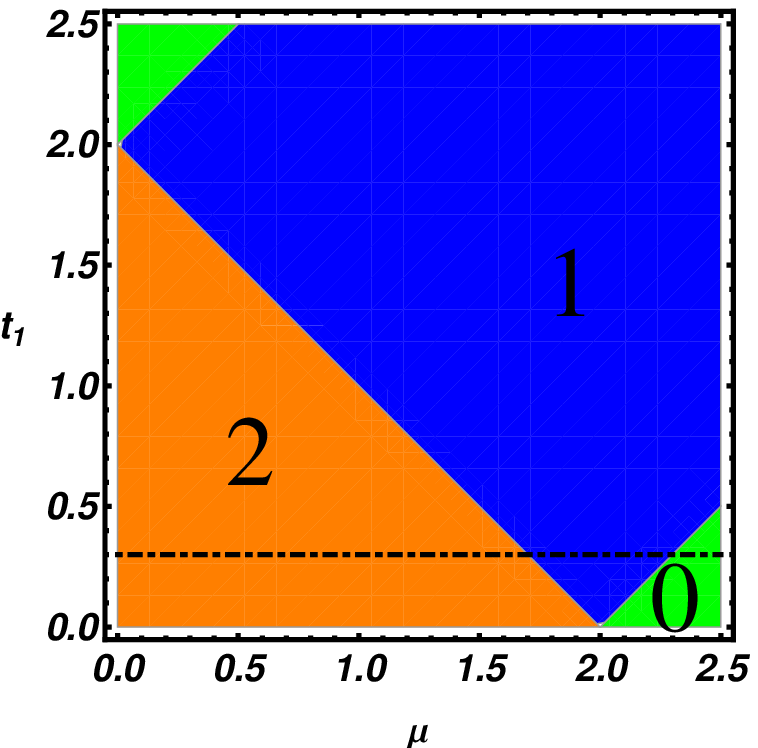}
\includegraphics[scale=0.55]{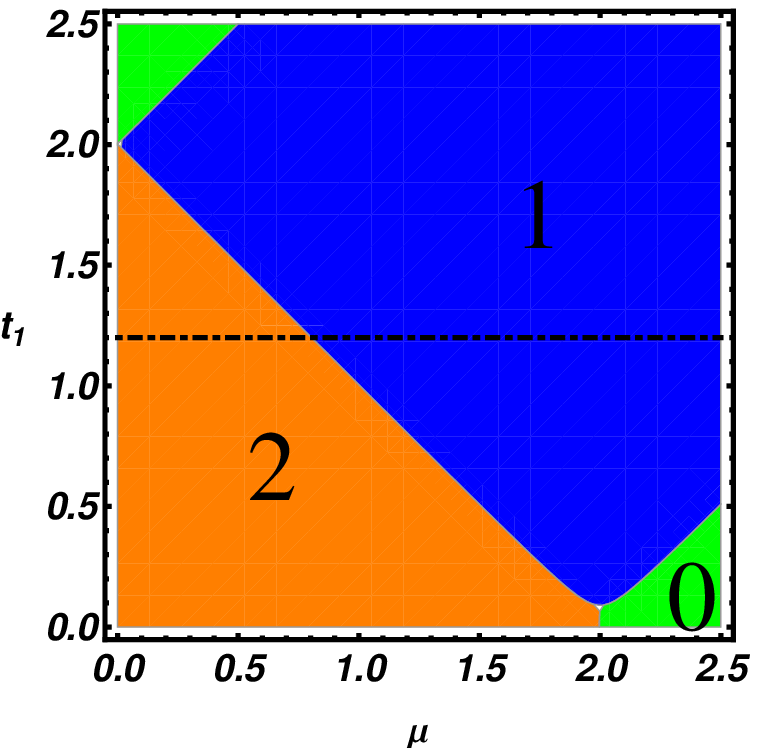}\\
\includegraphics[scale=0.55]{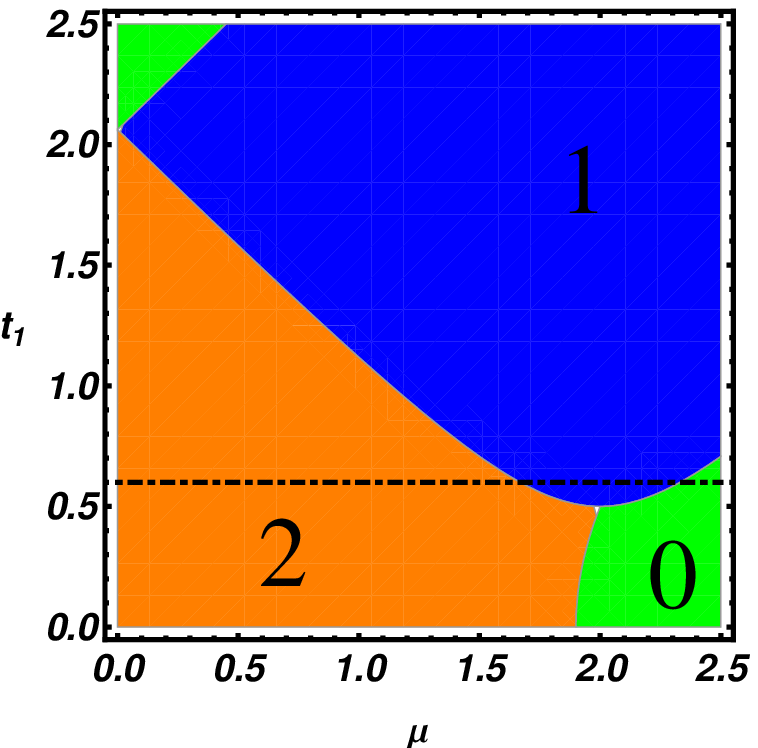}
\includegraphics[scale=0.55]{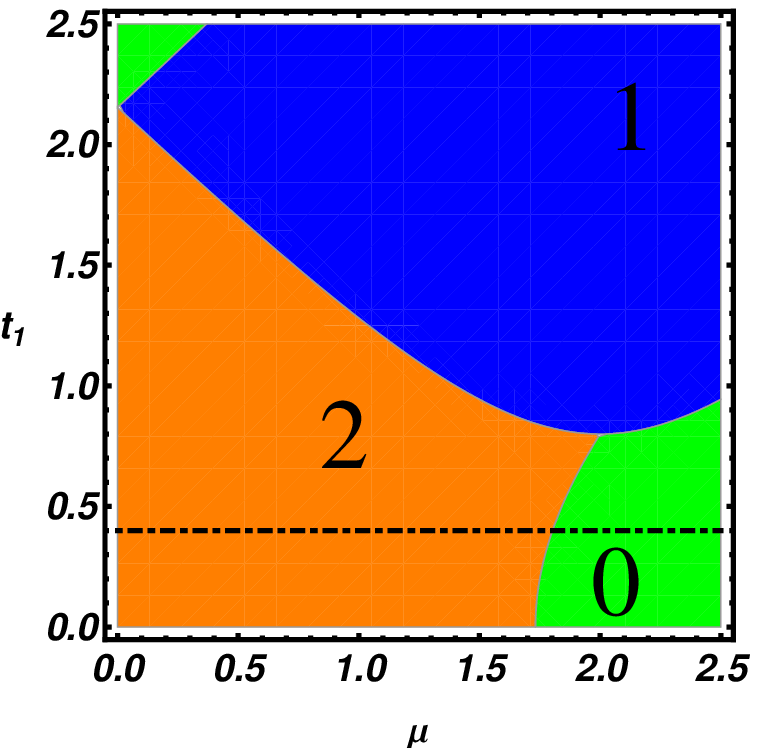}
\caption{Topological phase diagram of the ladder in the $(t_1, \mu)$ plane, given by the winding number for different values of $\Delta_1$, ($\Delta_1=0, 0.09, 0.5, 0.8$ from top left to bottom right) and for $\Delta=0.8$, $t=1$. The orange, blue and green regions are respectively the regions with $2$, $1$ and $0$ MZMs per edge. The black line represents the cut on which we take the spectra in Figure \ref{Fig5}.}
\label{Figure3}
\end{figure}

Following Zhou et al. \cite{refZhou}, we calculate the winding number as

\begin{equation}
W=Tr\int^\pi_{-\pi} \frac{dk}{2\pi i}\ \ A_k^{-1} \partial_k A_k=-\int^\pi_{-\pi}\frac{dk}{2\pi i}\ \ \partial_k\ln Det A_k
\end{equation}

and compute the topological phase diagram, which is systematically analyzed in the following discussion. To evidence the richness of the topological phases of the model, in the following we present our results using an extended parameters range, which in principle could be completely explored in cold atoms experiments rather than in condensed matter systems.
In Figure \ref{Figure3} we present the topological phase diagram in the ($t_1$, $\mu$) plane by fixing $\Delta=0.8$ and setting different values of $\Delta_1$ ($\Delta_1=0$, $0.09, $0.5, $0.8$, respectively), while taking $t$ as energy unit.

The integers in the figures correspond to the winding number and are related to the number of MZMs at every edge of the ladder. By direct inspection of Figure \ref{Figure3}, we notice that an increase of the pairing term $\Delta_1$ produces an enlargement of the trivial region, labeled by 0. The expansion of the trivial region changes the phase boundaries and induces a modification of the linear phase boundaries, which are peculiar to the $\Delta_1=0$ case. A further consequence of the enlargement of trivial region is the possibility, changing the chemical potential and setting appropriate $t_1$ values, to have a direct transition from the 2 MZMs region to the trivial phase without crossing the 1 MZMs phase. Interestingly, an additional consequence of the interchain coupling is that the critical value of the chemical potential, $\mu_c$, defining 2-0 phase boundary is progressively lowered as $\Delta_1$ increases. As a consequence $\mu_c$ becomes smaller than the one established by Kitaev for the 1-0 phase boundary of a single chain, i.e. $\mu_c<2t$.

\begin{figure}[!h]
\centering
\includegraphics[scale=0.55]{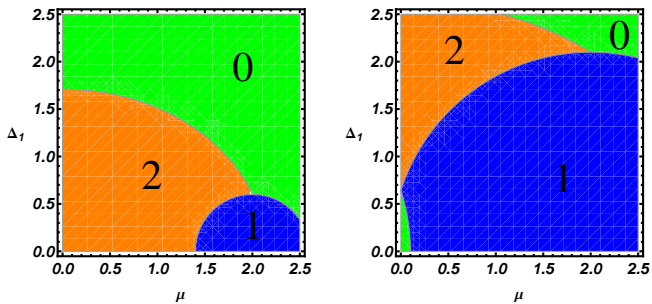}
\caption{Topological phase diagram of the ladder in the $(\Delta_1,\mu)$ plane given by the winding number. The two plots are realized for $\Delta=0.8$, $t=1$ and $t_1=0.6$ (left panel) and $t_1=2.1$ (right panel).}
\label{Fig4}
\end{figure}

In Figure \ref{Fig4} we present the topological phase diagrams in the ($\Delta_1$, $\mu$) plane. The different panels are realized by setting two values of $t_1$, namely $t_1=0.6$ (left panel) and $t_1=2.1$ (right panel), while retaining the remaining parameters as in Figure \ref{Figure3}. Direct observation of Figure \ref{Fig4} shows that for weak interchain hopping parameter ($t_1=0.6$) a phase with 2 MZMs is favored compared to the 1 MZM phase. On the other hand, when the transverse hopping is increased (Figure \ref{Fig4}, right panel), the phase with one Majorana mode per edge is favored and the system behaves like an effective two-orbital single Kitaev chain.

\begin{figure}[!h]
\centering
\includegraphics[scale=0.55]{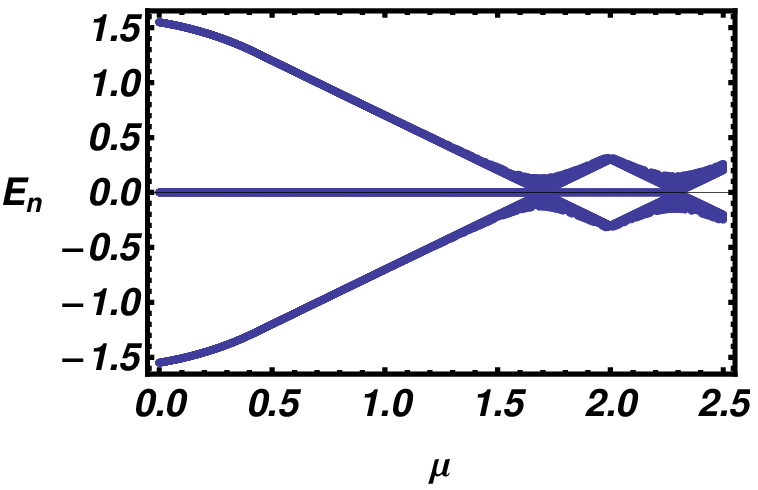}
\includegraphics[scale=0.55]{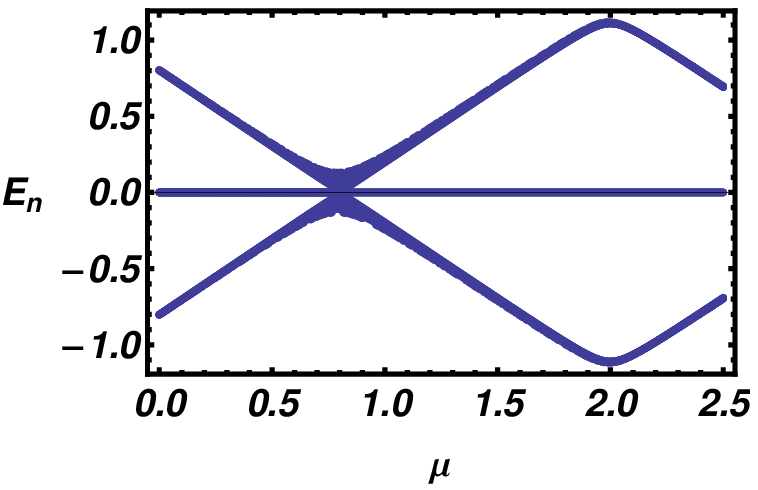}\\
\includegraphics[scale=0.55]{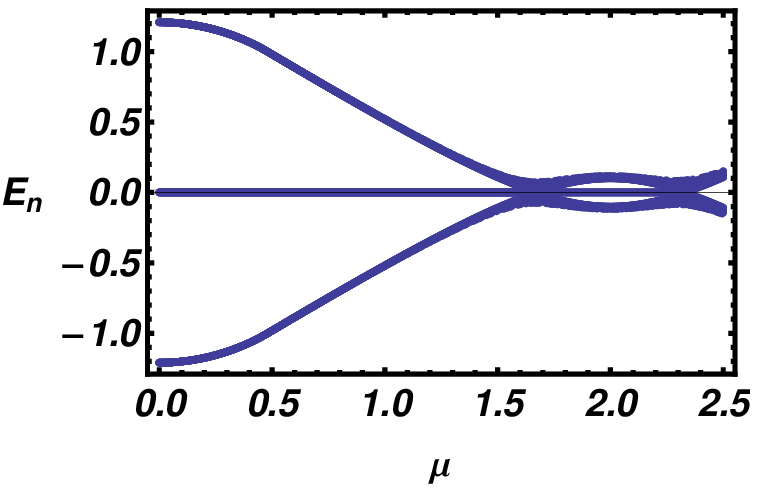}
\includegraphics[scale=0.55]{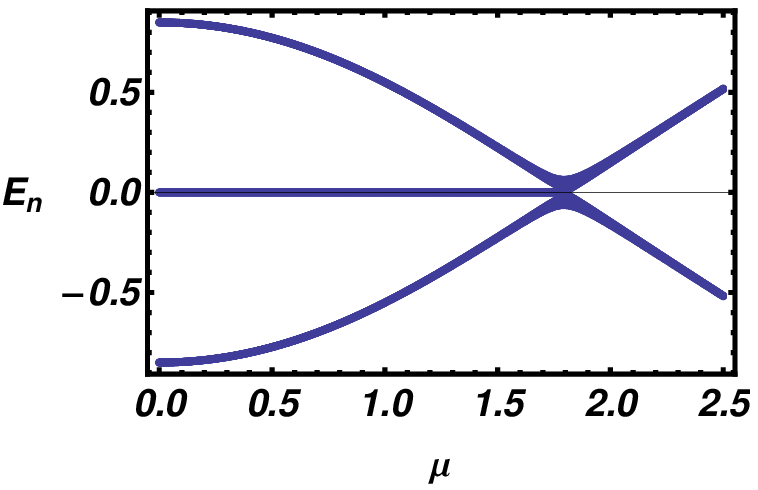}
\caption{Low energy part of the ladder energy spectra as a function of $\mu$, evaluated for different values of $\Delta_1$ and $t_1$. From the left to the right and from the top to the bottom the parameters are: $(\Delta_1, t_1)=(0, 0.3), (0.09, 1.2), (0.5, 0.6), (0.8, 0.4)$.}
\label{Fig5}
\end{figure}

\begin{figure}
\centering
\includegraphics[scale=0.7]{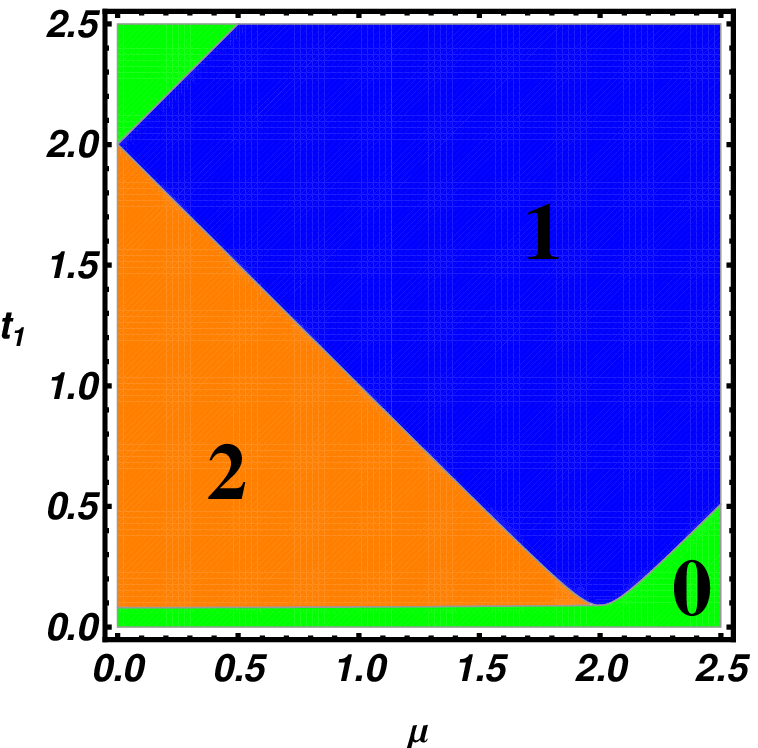}
\includegraphics[scale=0.7]{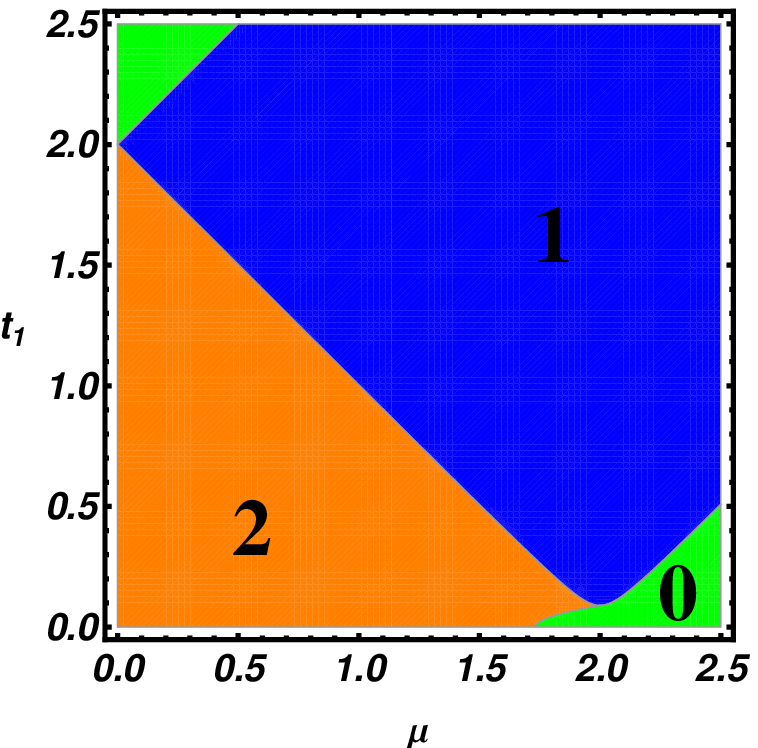}
\caption{Topological phase diagram of the ladder in the ($t_1$, $\mu$) plane. The orange, blue and green regions are respectively the regions with 2, 1 and 0 MZMs at one end. The fixed parameters for both the panels are $t=1$, $t_1=0.6$, while we set $\Delta=0.02$ and $\Delta_1=0.09$ for the left panel and  $\Delta=\Delta_1=0.09$ for the right panel.}
\label{Fig6}
\end{figure}

In order to verify the bulk-edge correspondence, in Figure \ref{Fig5} we present the low-energy part of the energy spectra of a ladder of $N=250$ lattice sites as a function of the chemical potential $\mu$. The analysis of the finite-size system confirms the presence and the multiplicity of zero-energy modes as prescribed by the winding number analysis.\\
Up to now we have explored an extended topological phase diagram in order to emphasize the richness of the topological phase plane of a ladder model. Hereafter, in Figure \ref{Fig6}, we specialize to the case of a condensed matter system and present phase diagrams showing a parameters region containing the device working point considered in the transport properties simulations (Figure \ref{Fig11}). In particular, in Figure \ref{Fig6} (left panel) the asymmetric pairing case (i.e. $\Delta=0.02$ and $\Delta_1=0.09$) is analyzed, while the equal pairing case ($\Delta=\Delta_1=0.09$) is presented in the right panel. In the asymmetric situation ($\Delta \neq \Delta_1$), a trivial region is found for $t_1 \leq 0.1$ and arbitrary $\mu$ values, the latter region being reduced in the symmetric case (right panel). Let us finally note that the situation $\Delta_1 > \Delta$, considered here, is realizable in multiorbital materials where the magnitude of different order parameters may well be different.

\section{Quantum transport through a Normal/Kitaev Chain/Superconductor device}
\label{sec:junction}
\begin{figure}[h!]
\includegraphics[scale=0.50]{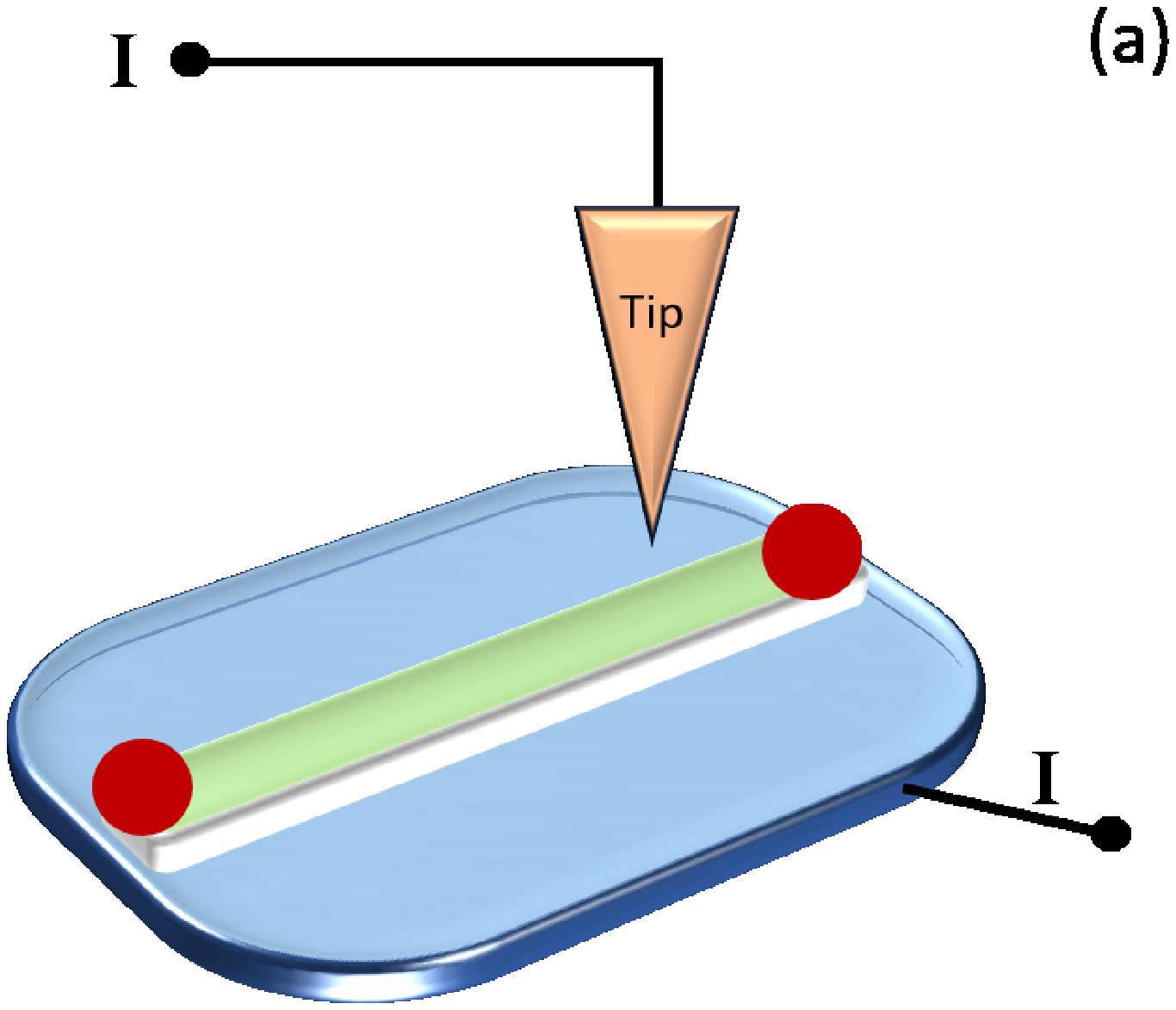}
\includegraphics[scale=0.50]{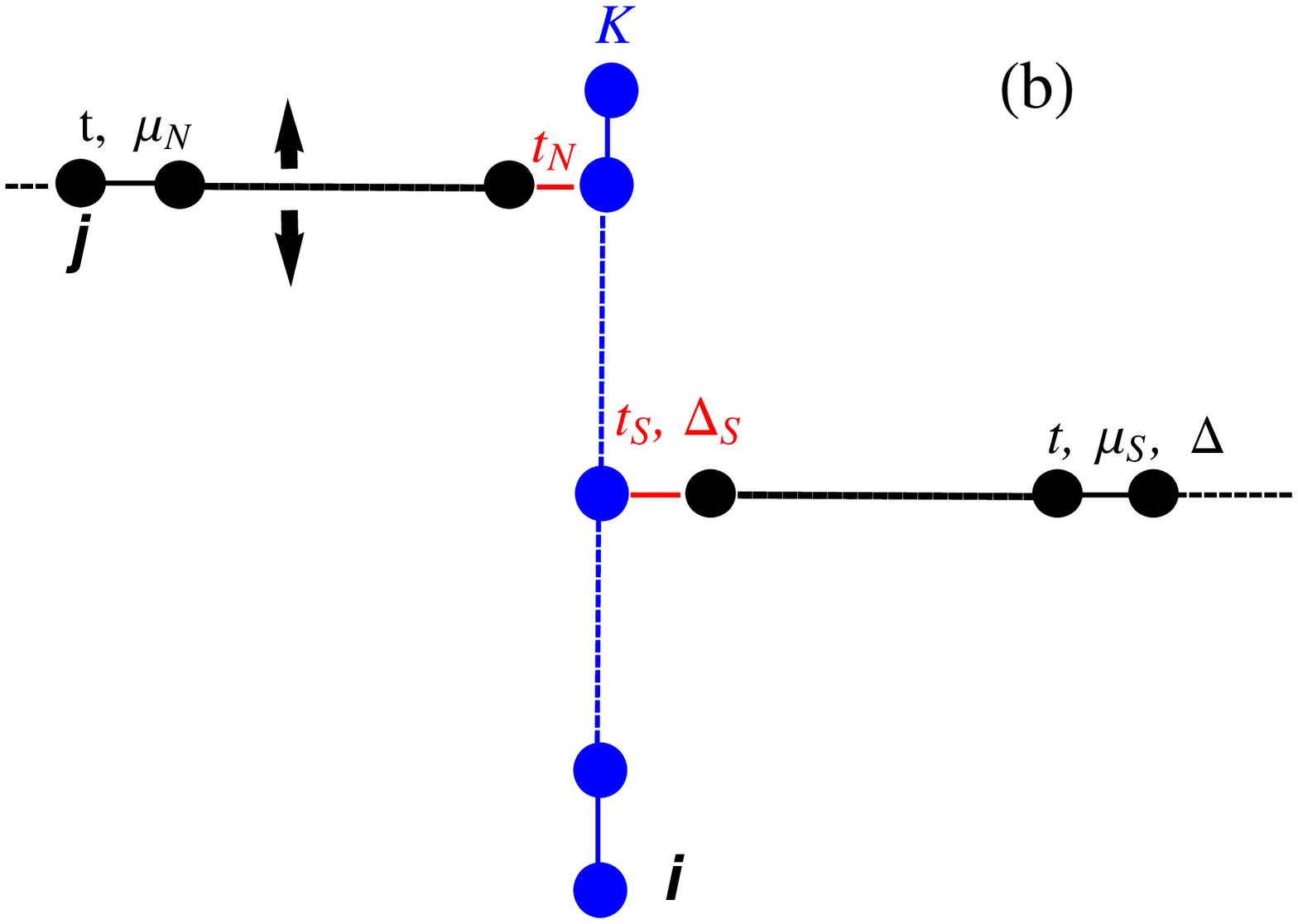}
\caption{Panel (a): Schematic of a tunnel conductance measurement setup where the normal tip position can be changed along the nanowire. Panel (b): tight binding model of the N-KC-SC device. The black chains represent the normal and superconducting $p$-wave leads with hopping amplitude $t$, chemical potential $\mu_N$, $\mu_S$ and pairing $\Delta$, respectively. The vertical finite line represents the Kitaev chain with parameters: $\mu, t, \Delta$. The couplings parameters between the leads and the Kitaev nanowire are given by $t_N$, $t_S$, $\Delta_S$.}
\label{Fig7}
\end{figure}

Using the scattering matrix approach, we study  the transport properties of a Kitaev chain coupled to a normal and a superconducting $p$-wave lead (Figure \ref{Fig7}). The Kitaev chain and the superconducting lead define a T-stub configuration (i.e. a waveguide with a closed sidearm \cite{Tstub}) with the junction point located in the middle of the Kitaev chain, while the normal electrode position can be changed to simulate an STM measurement configuration. The $t_N$ parameter defines the hopping between the normal and the KC, while $t_S$ represents the hopping between the KC and the superconducting lead, while $\Delta_S$ defines the pairing amplitude. In the following we study the topological signatures on the transport properties of the system considering a finite Kitaev chain of $N=121$ sites.

The scattering state of the normal electrode is given by (see appendix \ref{appbdgN} for details)

\begin{equation}
\label{scattering state}
\Psi_N(n)=\left(
\begin{array}{c}
1\\
0\\
\end{array}\right)e^{ik_e n}+r_{ee}\left(\begin{array}{c}
1\\
0\\
\end{array}\right)e^{-ik_e n}+r_{eh}\left(\begin{array}{c}
0\\
1\\
\end{array}\right)e^{ik_h n},
\end{equation}

which represents an electron coming from the normal bulk having an ordinary reflection probability $|r_{ee}|^2$ and an Andreev reflection probability given by $|r_{eh}|^2$. The wave vectors $k_{e/h}$ are obtained by the solution of the BdG equations:

\begin{equation}
\label{wavevectorsNormal}
k_{e/h} =\arccos\biggl({\frac{\mu_N \pm E}{-2t}}\biggr),
\end{equation}

where $E$ is the energy of the scattering process, $\mu_N$ the chemical potential and $t$ the hopping amplitude of the normal lead.
Inside the superconducting electrode the propagating state is given by (see appendix \ref{appbdgS} for details):

\begin{equation}
\label{propagatingstate}
\Psi_S(n)=t_{ee}\left(
\begin{array}{c}
u_e\\
iv_e\\
\end{array}\right) e^{iq_e n}+t_{eh}\left(
\begin{array}{c}
v_h\\
-iu_h\\
\end{array}\right) e^{-iq_h n},
\end{equation}

where $t_{ee}$ and $t_{eh}$ are transmission coefficients of an $e$-like and $h$-like quasiparticle. The BCS coherence factors $u_{e,h}$, $v_{e,h}$ for a $p$-wave superconductor are given by:

\begin{eqnarray}
u_{e/ h}=\sqrt{\frac{E+\sqrt{E^2-\Delta_{q_{e/ h}}^2}}{2E}}\nonumber \\
v_{e/h}=\sqrt{\frac{E- \sqrt{E^2-\Delta_{q_{e/h}}^2}}{2E}},
\end{eqnarray}

where $\Delta_{q_{e/h}}=2\Delta\sin q_{e/h}$ is the momentum-dependent superconducting gap.\\
Resorting to a numerical solution of the scattering problem, we obtain the transmission and reflection coefficients and the wave functions of the resonant modes along the Kitaev chain (we call these modes Kitaev modes by analogy with the Majorana zero-energy modes introduced by Kitaev \cite{refKitaev2}). The wave functions of modes belonging to the Kitaev chain, being a byproduct of the solution of the scattering problem, are not normalized with respect to the position since scattering states are normalized with respect to the particle flux.\\
The conservation of probability of all scattering processes can be written in terms of the scattering coefficients implying the following relation:

\begin{eqnarray}
|r_{ee}|^2+|r_{eh}|^2+(|u_e|^2-|v_e|^2)|t_{ee}|^2+(|u_h|^2-|v_h|^2)|t_{eh}|^2=1,
\label{conservationofprobability}
\end{eqnarray}

written under Andreev approximation $k_{e/h} \approx q_{e/h} \approx k$ which has been adopted in numerical evaluations. Once the scattering coefficients are known, the conduction properties of the system can be computed adopting a BTK-like approach. Accordingly, the zero-temperature differential conductance can be written as:

\begin{equation}
\label{conduttanzaformula}
G=\frac{dI}{dV}=\frac{2e^2}{h}(1-\mathcal{B}+\mathcal{A}),
\end{equation}

where $\mathcal{A}=|r_{eh}|^2$ and $\mathcal{B}=|r_{ee}|^2$ represent the Andreev reflection and the normal reflection probability, respectively \cite{Blonder}.

\begin{figure}[h]
\centering
\includegraphics[scale=0.25]{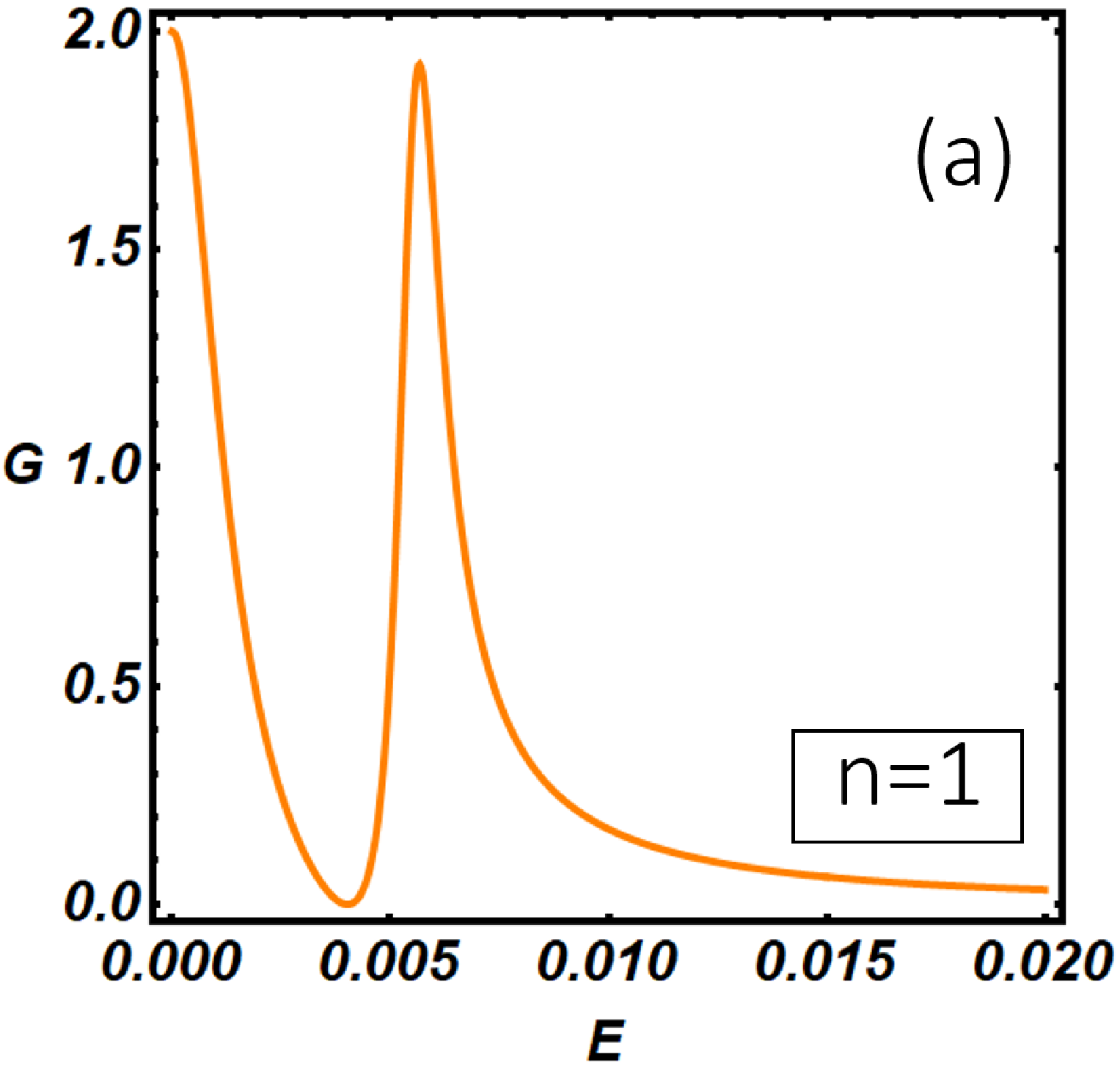}
\includegraphics[scale=0.25]{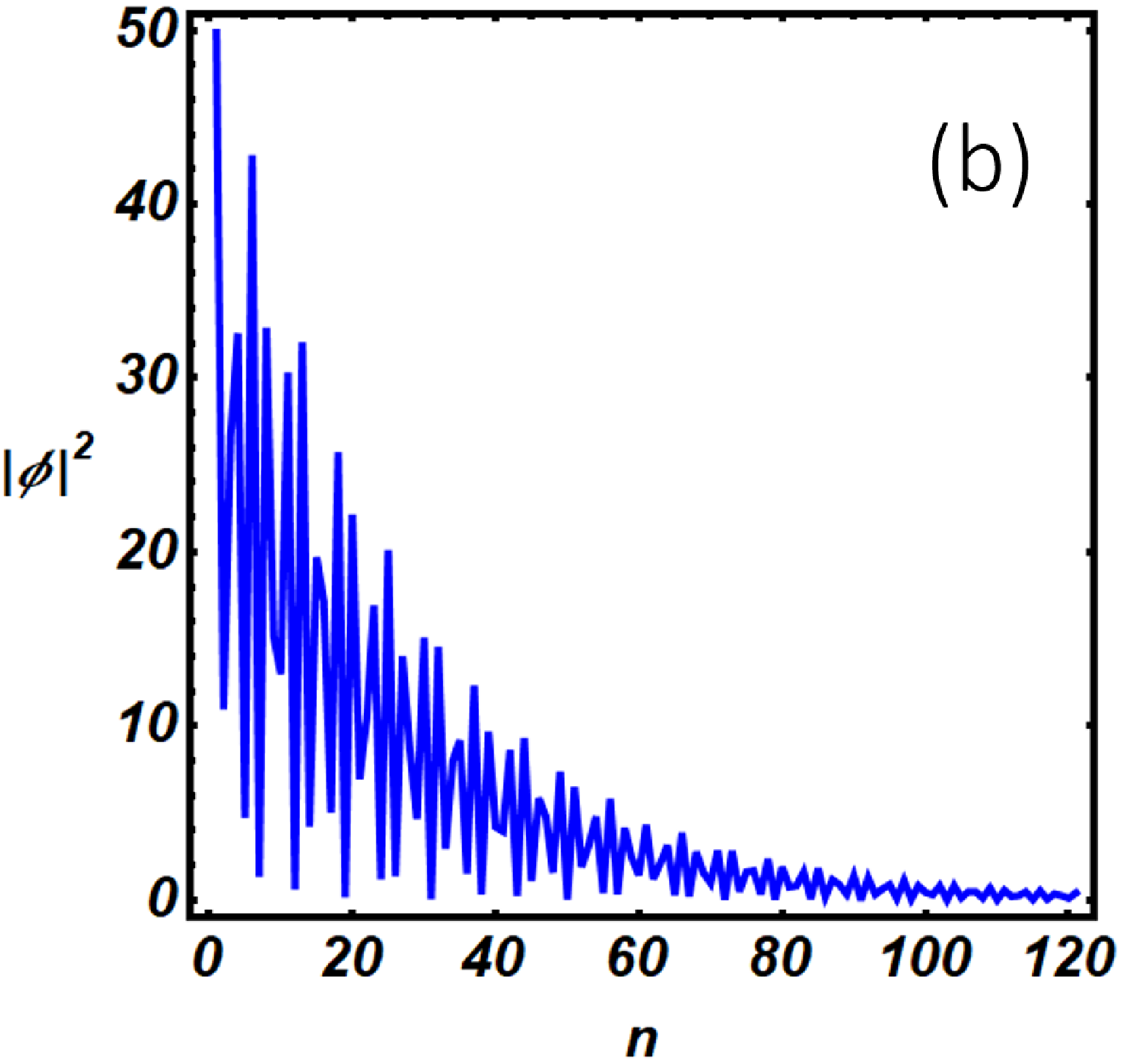}
\includegraphics[scale=0.25]{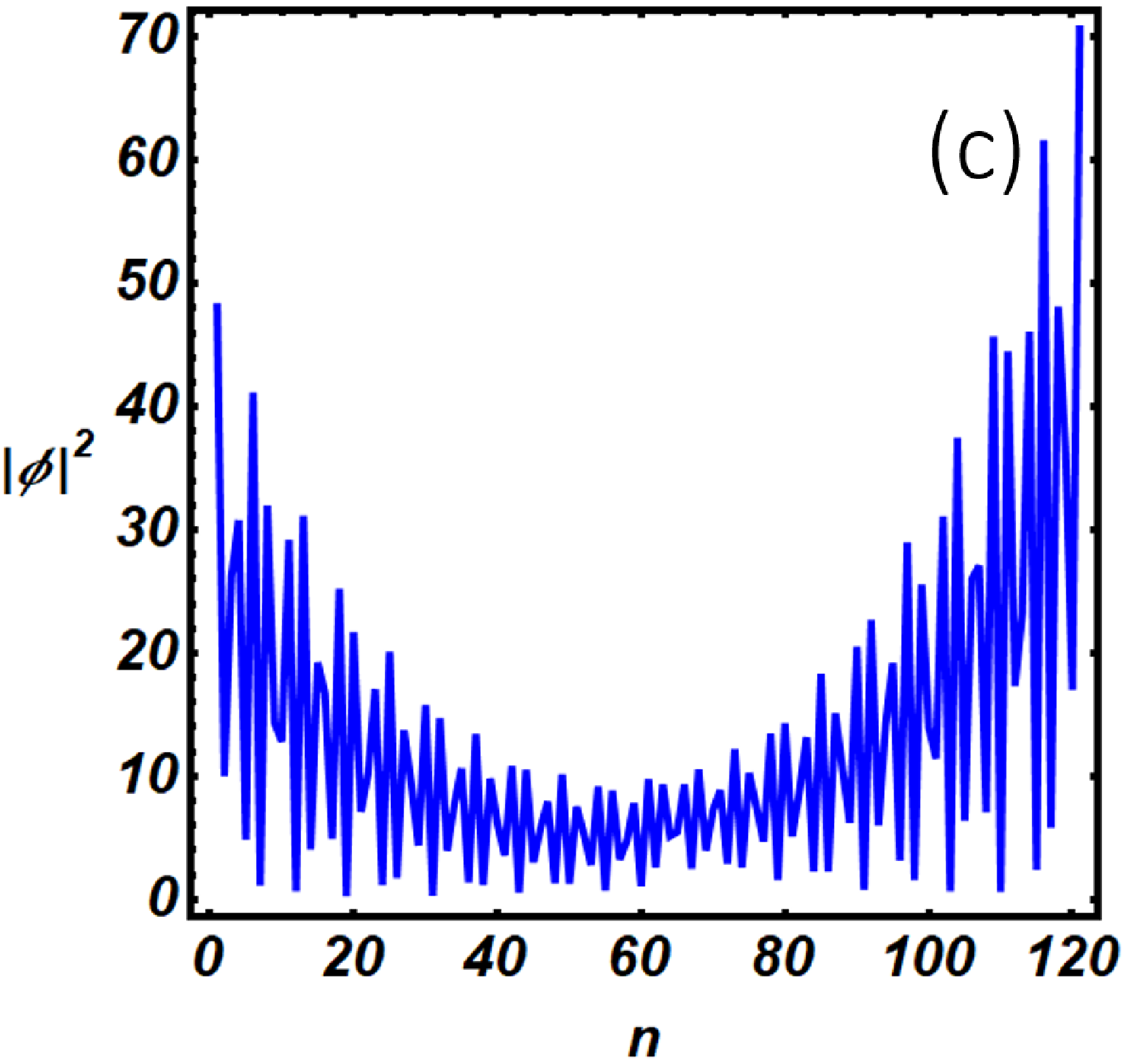}\\
\includegraphics[scale=0.25]{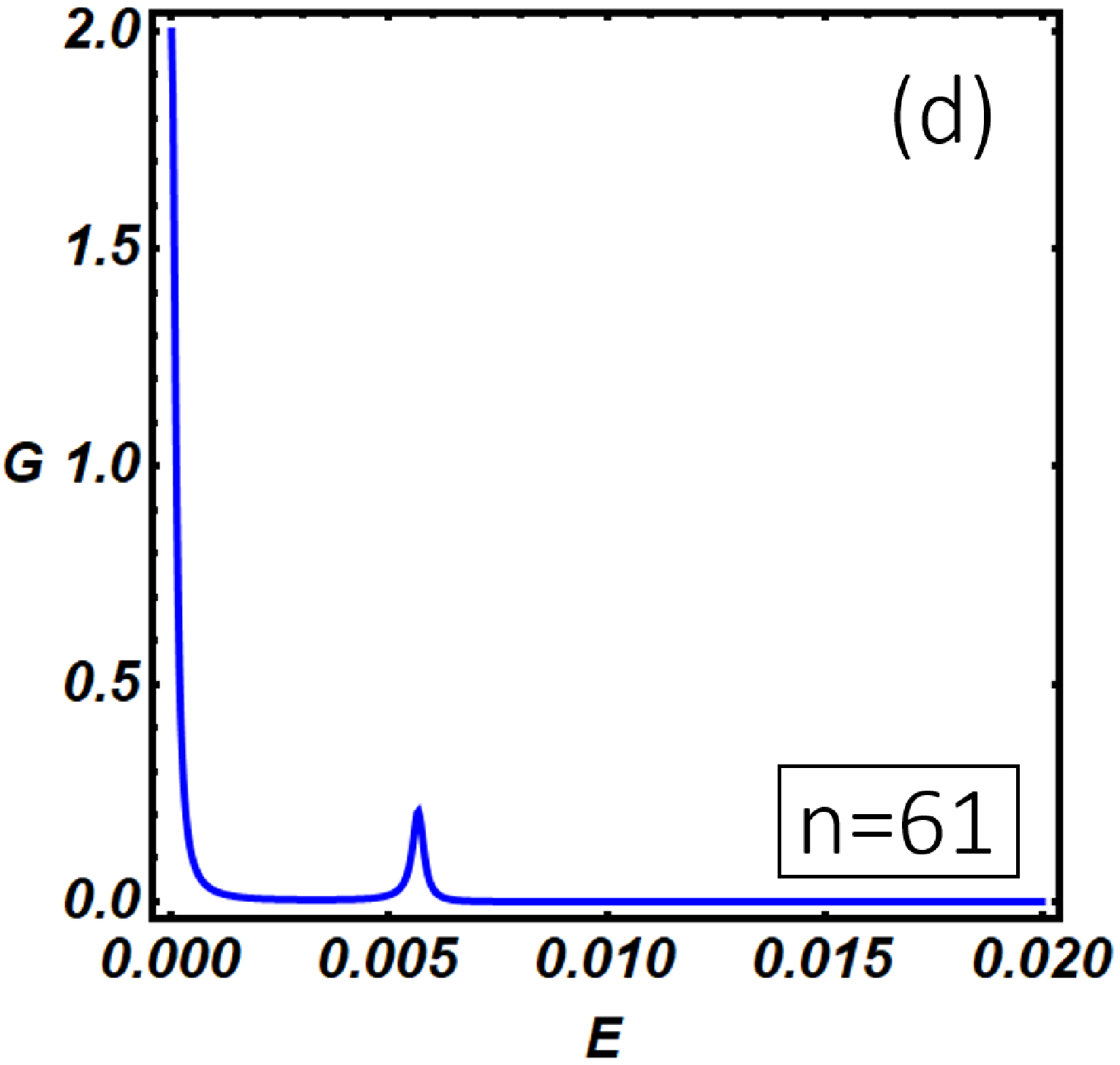}
\includegraphics[scale=0.25]{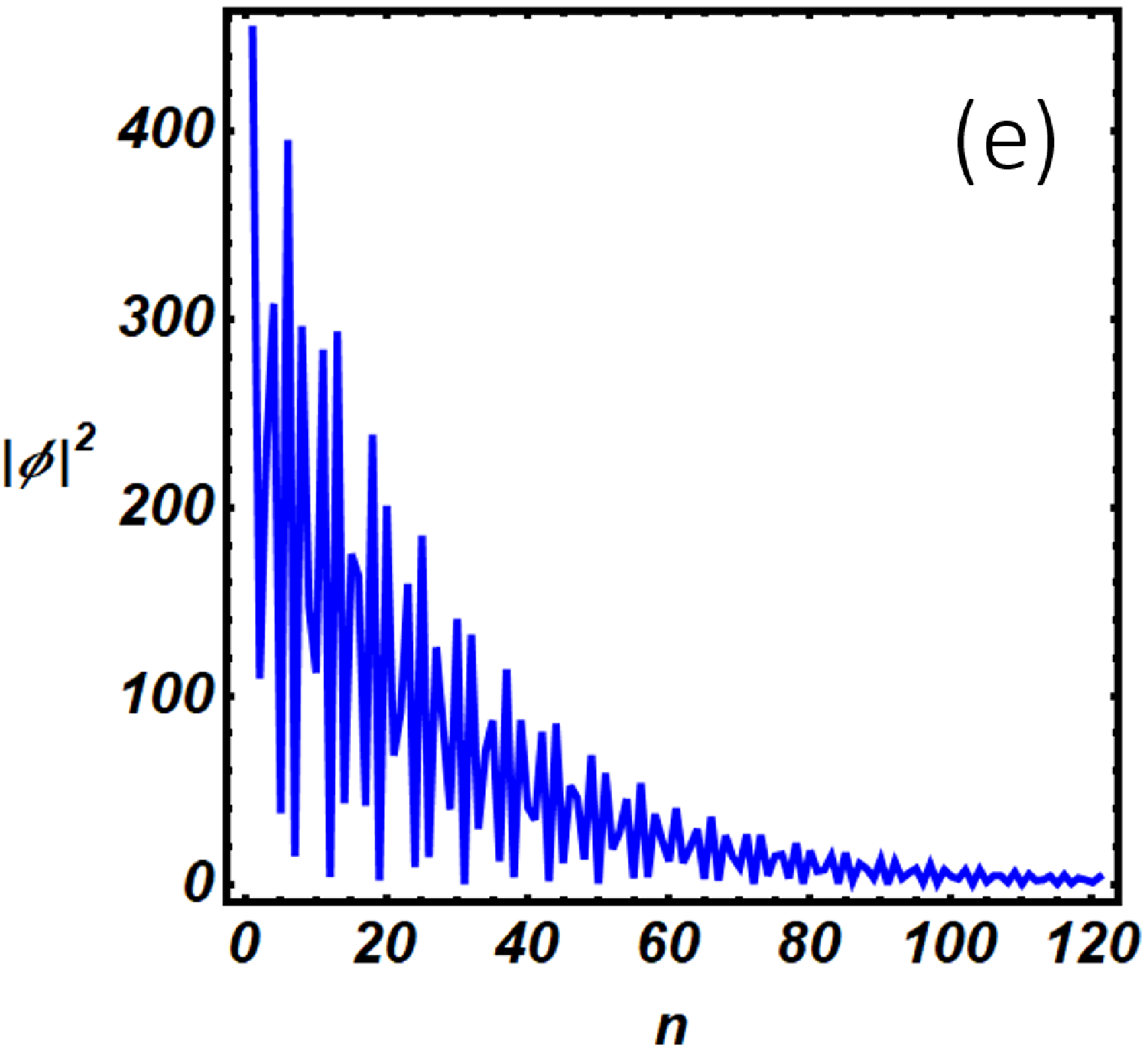}
\includegraphics[scale=0.25]{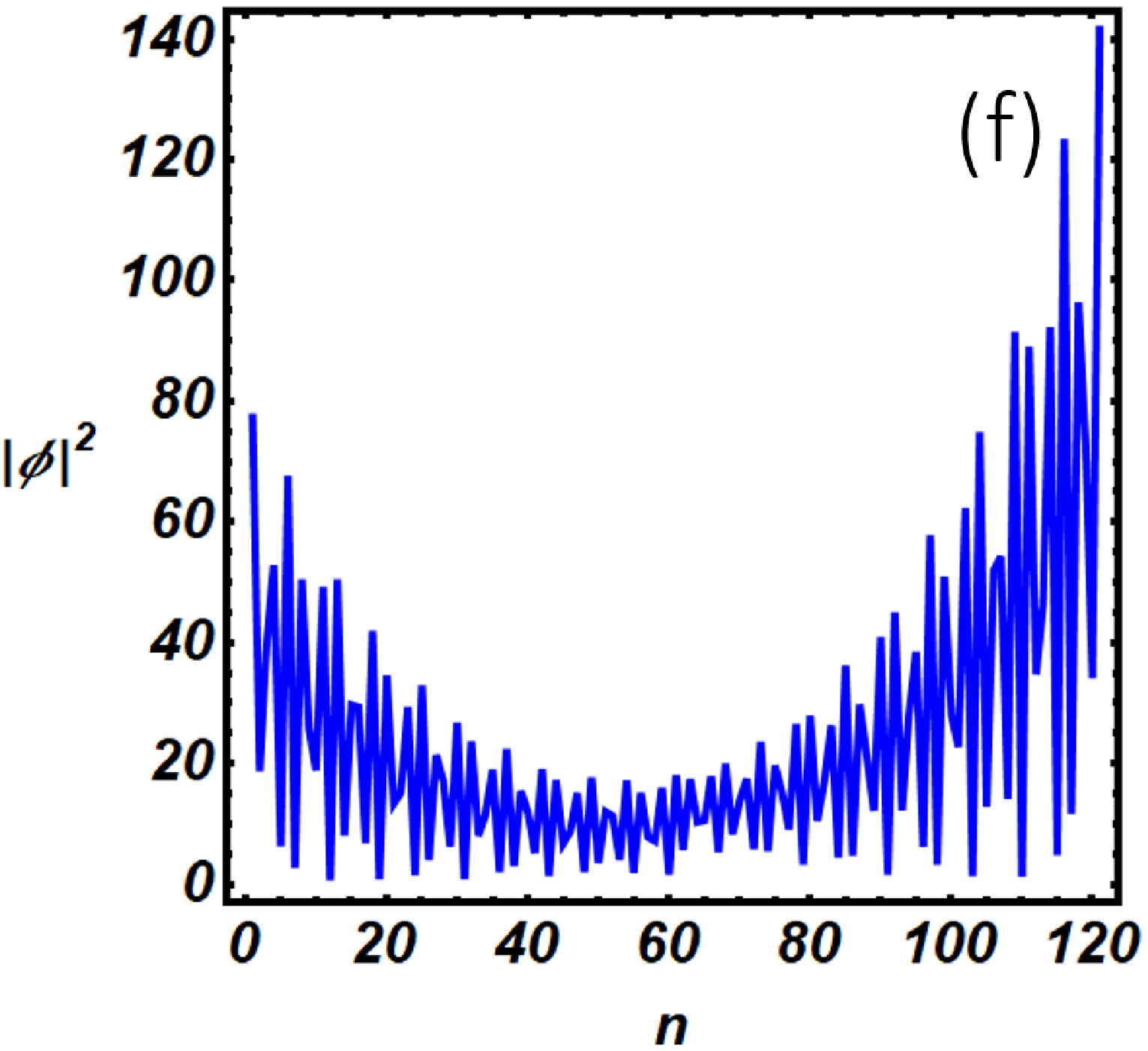}\\
\includegraphics[scale=0.25]{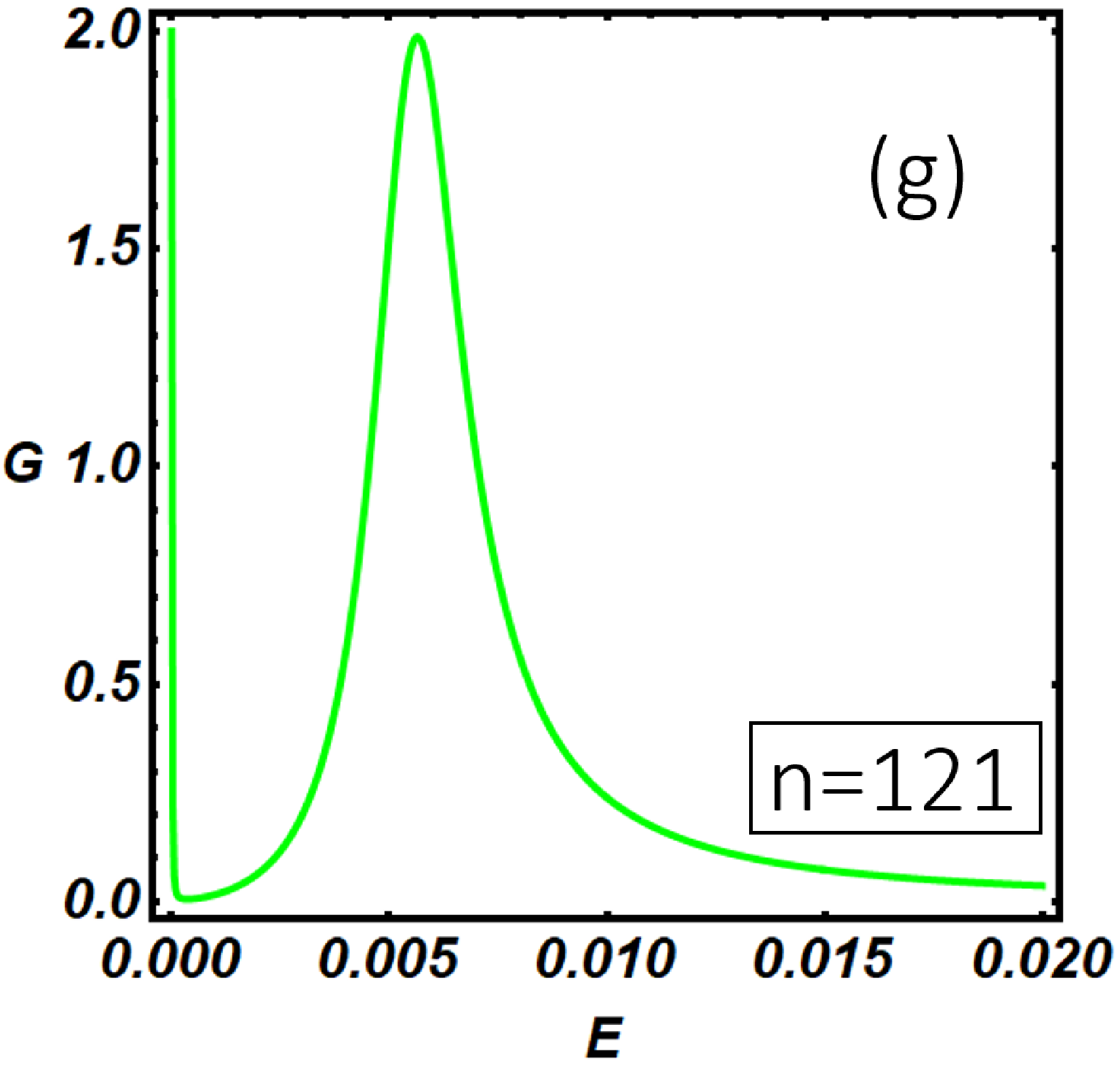}
\includegraphics[scale=0.25]{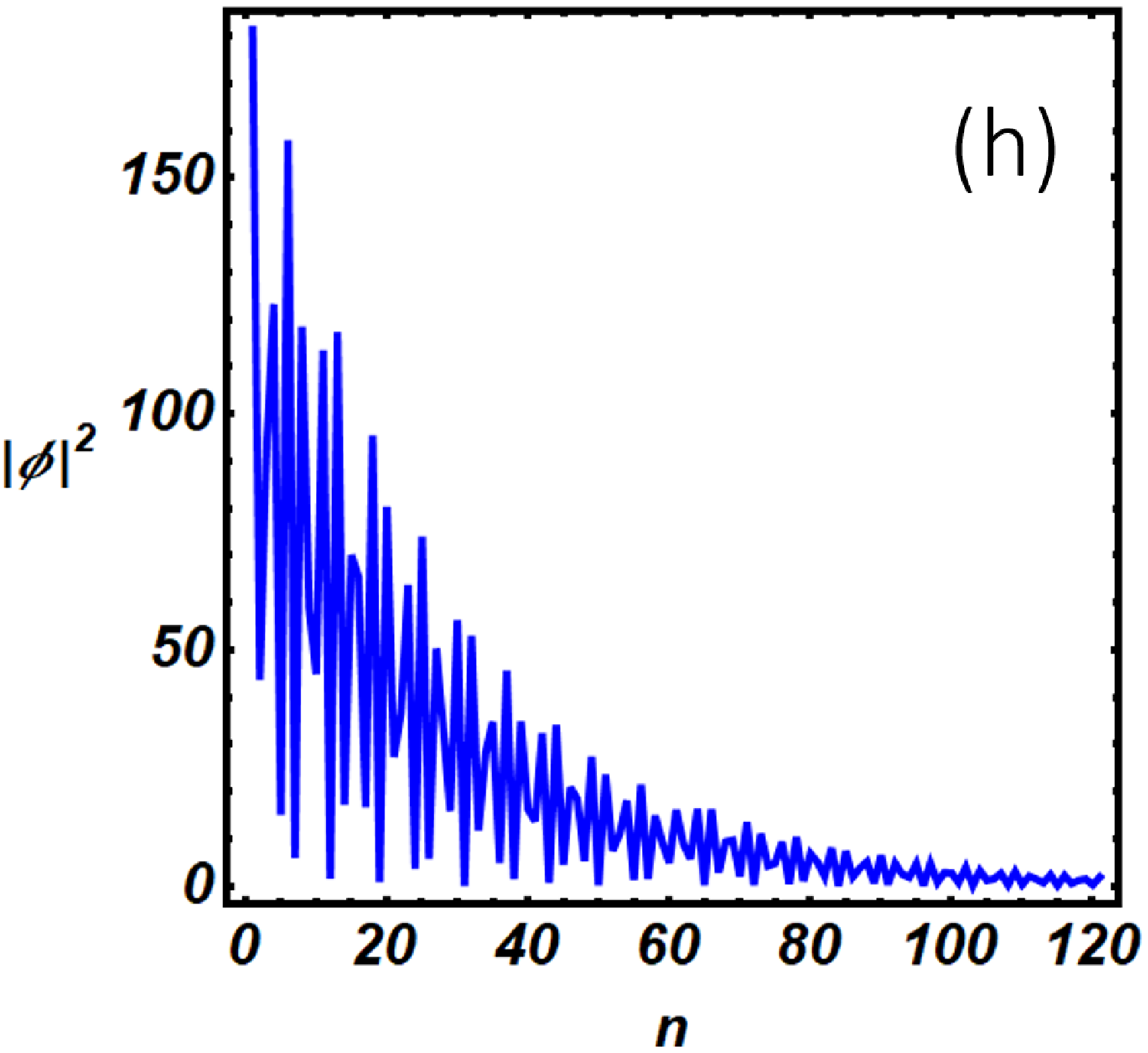}
\includegraphics[scale=0.25]{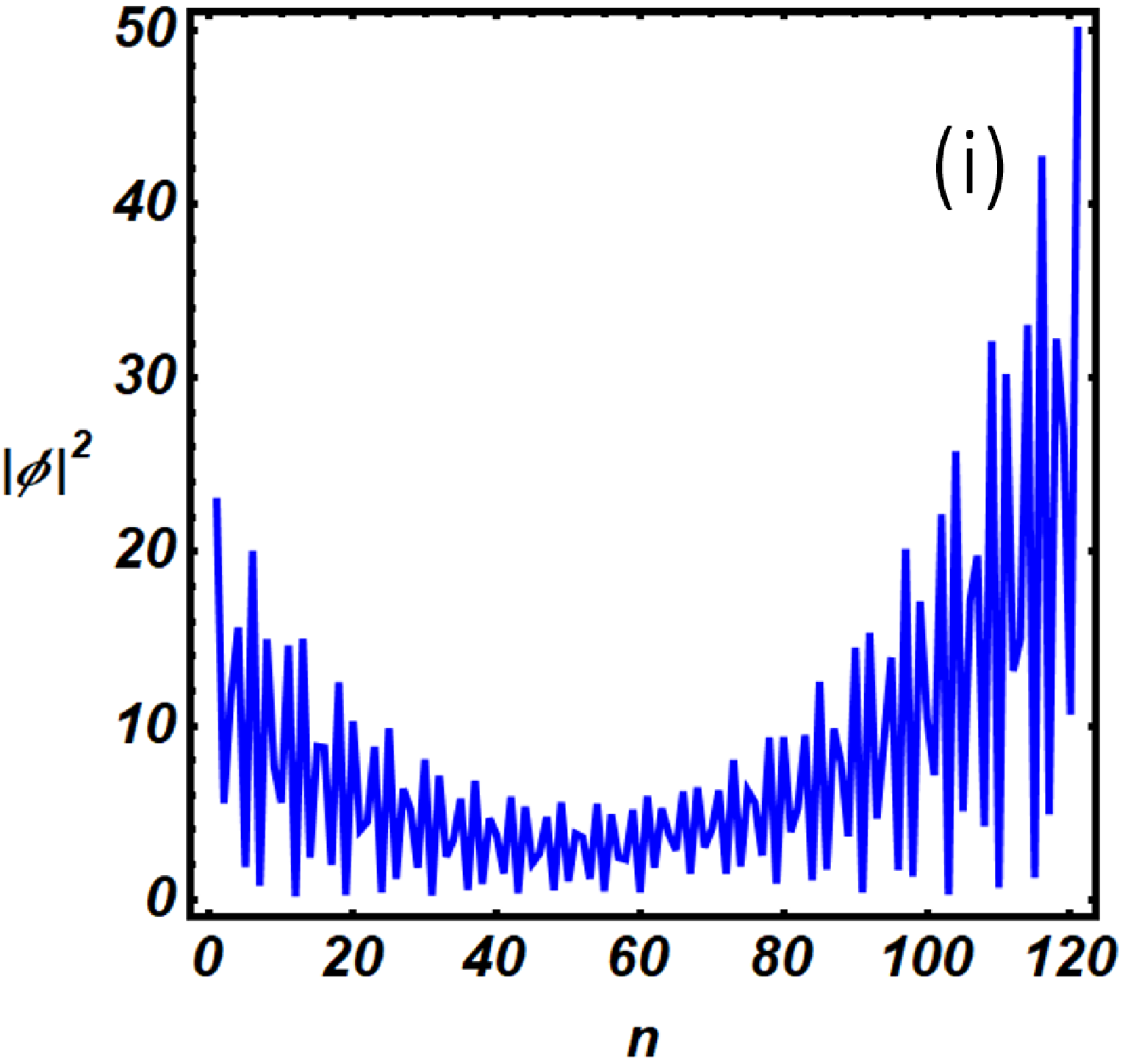}
\caption{N-KC-SC device: Zero-temperature differential conductance (in the unit of $\frac{2 e^2}{h}$) as a function of the energy in the sub-gap regime. Panels (a), (d), (g) are obtained by setting different linking positions, specified by $n$, between the normal lead and the Kitaev wire. The panels (b), (c), (e), (f), (h), (i) represent the modulus squared $|\Phi|^2\equiv |f_n|^2+|g_n|^2$ of the resonant modes along the Kitaev chain evaluated at energy values corresponding to the resonant sub-gap peaks in the conductance. From the left to the right and from the top to the bottom the parameters used are: (b): $n=1$, $E=5\cdot 10^{-5}$. (c): $n=1$, $E=6\cdot 10^{-3}$. (e): $n=61$, $E=5\cdot 10^{-5}$. (f): $n=61$, $E=6\cdot 10^{-3}$. (h): $n=121$, $E=5\cdot 10^{-5}$. (i): $n=121$, $E=6\cdot 10^{-3}$.  The remaining model parameters have been fixed as: $\Delta=0.02$, $t_N=t_S=0.2$, $t=1$, $\mu=0.5$, $\mu_S=\mu_N=0$.}
\label{Fig8}
\end{figure}

Using the quantum transport theory presented above, we have studied the differential conductance and the internal resonant modes of the N-KC-SC topological device.
In particular, in Figure \ref{Fig8}, we study the differential conductance of the N-KC-SC device as a function of the N-KC link position and setting the model parameters inside the topological non-trivial region, where Majorana modes are expected to be relevant. Panels (a), (d), (g) show the conductance curves at selected N-KC link positions (specified within the figures inset) corresponding to the left end ($n=1$), the middle ($n=61$) and the right end ($n=121$) of the nanowire. In Figure \ref{Fig8} (a), two peaks are clearly visible: a quantized zero-bias peak ($E=5 \cdot 10^{-5}$) and its satellite at higher energy ($E=6 \cdot 10^{-3}$). The analysis of the internal modes puts in evidence that the quantized zero-bias peak, which is expected when the system undergoes a topological phase transition (the so-called zero-bias anomaly (ZBA)), corresponds to a MZM peaked at the left end of the wire (panel (b)). On the other hand, the peak at higher energy ($E=6 \cdot 10^{-3}$) corresponds to hybridized Majorana modes which originate an U-shaped wave function peaked at the two ends of the wire (panel (c)) and extended along the whole length of the Kitaev chain. The width of the resonant conductance peaks depends on the amplitude of the internal mode wave function at the N-KC linking point, the latter quantity being relevant in defining the overlap between the scattering states of the normal electrode and the internal modes of the Kitaev chain. The overlap increases with the wave function amplitude and depends on the normal electrode position and on the conductance peak considered. As a consequence, a narrow (broaden) peak is observed when the normal lead is weakly (strongly) coupled with the Kitaev chain internal modes, the latter situation corresponding to poor (strong) overlap between the normal electrode and the nanowire quantum states. These overlap effects are clearly visible in Figure \ref{Fig8} (a), (b) and (c) in which the two broadened conductance peaks originates from the non-vanishing values of the wave function amplitudes at the $n=1$ position, i.e. the N-KC linking point. It is worth mentioning here that changing the linking point does not produce a qualitative change of the internal modes amplitude along the system. As a consequence, in panel (d) and (g) we observe the progressive and monotonic width reduction of the zero-bias peak which follows the lowering of the wave function amplitude as the normal electrode is moved towards the right end of the Kitaev chain (panels (e) and (h)). The broadening of the satellite of the zero-bias peak is not monotonic as a function of the N-KC link position and follows the U-shaped behavior characteristic of the wave function amplitude (see panels (c), (f), (i)) of the internal Kitaev modes. From the experimental side, these observations suggest the possibility to reconstruct the wave function amplitude profile of the internal modes of the Kitaev chain by measuring the conductance peak broadening as the normal tip of an STM is moved along the system, being the latter method an interesting tool in characterizing the survival of topological properties under non-equilibrium conditions. It is here worth mentioning that pure Majorana modes are neutral excitations and thus they are not able to sustain a charge current. For this reason a certain degree of hybridization of these topological modes is a necessary condition to have a finite charge current flowing through the system. The degree of hybridization of the Majorana modes inside the Kitaev chain can be quantified by studying the site-dependent charge density $\rho_n=|f_n|^2-|g_n|^2$ associated to the internal modes profile presented in Figures \ref{Fig8} (b), (c), (e), (f), (h), (i). The complete analysis, which is reported in appendix \ref{appCharge}, shows that the average charge density $\bar{\rho}=(\sum_{n=1}^{L}\rho_n)/L$ corresponding to the internal modes presented in panels (b), (e), (h) is much lower than the corresponding quantity computed for panels (c), (f), (i). This circumstance suggests that zero-bias conductance peak originates from a genuine Majorana mode presenting a weak hybridization with a hole-like non-topological mode, the latter conclusion being supported by the negative sign of $\bar{\rho}$. On the other hand, satellite conductance peaks at higher energy are associated with internal states, presented in panels (c), (f) and (i), with a strong electron-like character (i.e., $\bar{\rho}>0$).\\
As a final comment, we notice that, despite the wave guide spatial configurations corresponding to the (a) and (g) panels are one the mirror reflected of the other with respect to a reflection line coincident with the superconducting electrode, the Hamiltonian of the entire system does not satisfy this symmetry and thus the conductance curves pertaining to these cases are not coincident.
This statement can be easily proven. Reflection transformation of the isolated Kitaev chain Hamiltonian (Equation \ref{HKitaev}) can be implemented by the change of indices $j\rightarrow L+1-j$. The reflection does not modify the structure of the Hamiltonian but transforms the initial pairing term $\Delta$ into $-\Delta$. The transformed Kitaev Hamiltonian is equivalent to the initial one via a $U(1)$ gauge transformation of the fermionic operators ($a_j\rightarrow e^{i\frac{\pi}{2}}b_j$). For the above reasons physical properties of the isolated Kitaev chain are invariant under reflection.
When the KC is connected to external reservoirs the above procedure is still applicable with a different outcome. Indeed, the electrodes Hamiltonians are invariant under reflection transformation, which is only implemented on the Hamiltonian of the Kitaev chain. On the other hand, the N-KC junction position is altered by the transformation (i.e. $n=1 \rightarrow n=121$), while the position of the KC-SC link ($n=61$) remains unaffected since it is a fixed point. Thus the described reflection transformation not only provides a reflected spatial configuration of the initial system but also determines a phase difference between the superconducting pairing terms of the Kitaev chain and the superconducting electrode. Implementing a gauge transformation as the one used for the isolated Kitaev chain does not eliminate the phase gradient (which is now transferred to the N-KC and KC-SC hopping amplitudes), the latter producing observable effects. We have verified that conductance curve of the system described by the transformed Hamiltonian is coincident with the curve shown in Figure \ref{Fig8} (a), obtained for the initial Hamiltonian. Moreover the transformed Hamiltonian does not correspond to the Hamiltonian used in Figure \ref{Fig8} (g), where the mentioned phase gradient is absent. These arguments prove our conclusions and can be equally applicable to the ladder case discussed in Section \ref{sec:junction-ladder} (see Figures \ref{Fig11} (a) and (i)).

\begin{figure}
\centering
\includegraphics[scale=0.55]{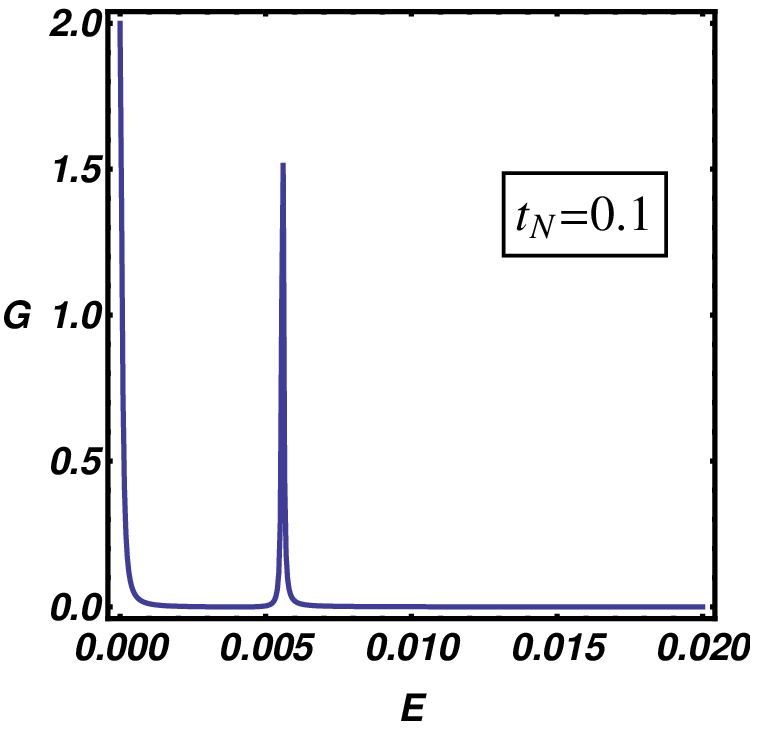}
\includegraphics[scale=0.55]{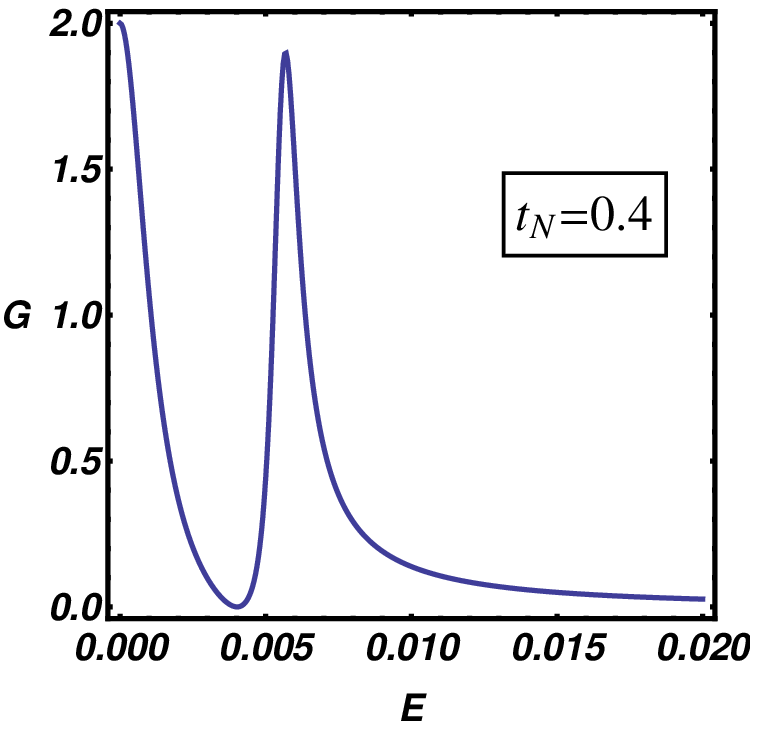}\\
\includegraphics[scale=0.55]{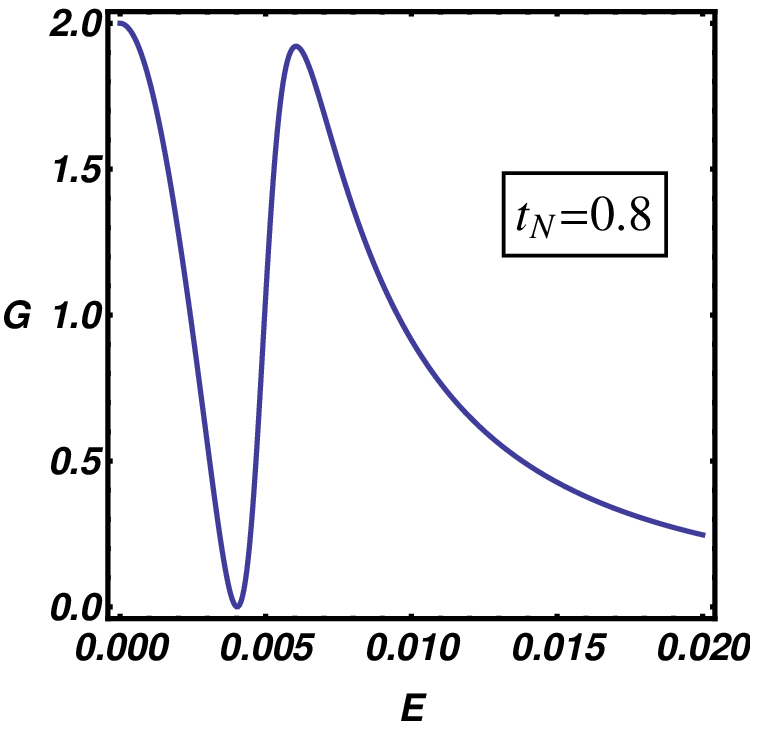}
\includegraphics[scale=0.55]{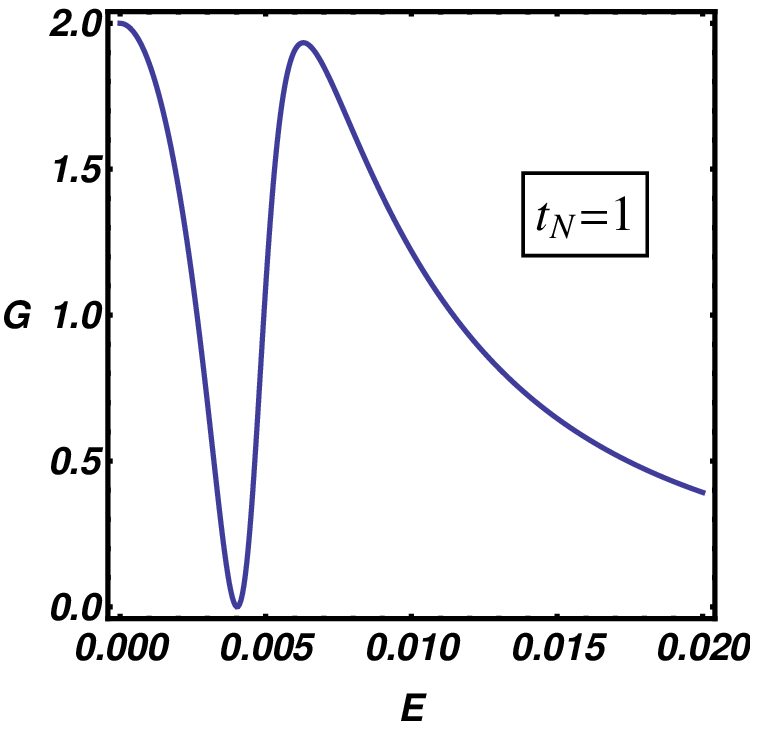}
\caption{N-KC-SC device: Differential conductance  (in units of $\frac{2 e^2}{h}$) as a function of the energy. Different plots are obtained by setting $t_N=0.1, 0.4, 0.8, 1$, while fixing the remaining parameters as: $\Delta=0.02$, $t_S=0.2$, $t=1$, $\mu=0.5$, $\mu_S=\mu_N=0$, $n=2$.}
\label{Fig9}
\end{figure}

Up to now we have shown that the overlap between quantum states of the normal electrode and the internal modes of the Kitaev chain can be changed by varying the normal electrode position along the nanowire. From the experimental viewpoint, an alternative way to modify this coupling consists in changing the N-KC distance, e.g. by modifying the distance of an STM normal tip (normal electrode) from the topological nanowire to be measured. In this way, the transmission at the N-KC interface becomes exponentially suppressed at increasing distance due to the vacuum tunneling phenomenon and the same effect is induced on the overlap between quantum states belonging to the normal electrode and the Kitaev chain. Within the considered transport model the above mentioned effects can be considered by acting on the hopping amplitude $t_N$, which is related to the transparency of the N-KC interface. In order to perform this analysis, in Figure \ref{Fig9} we present the differential conductance curves of the topological device for selected values of $t_N$ going from tunneling ($t_N=0.1$) to the metallic ($t_N=1$) regime. Direct observation of Figure \ref{Fig9} shows that the resonance broadening of the conductance peaks increases when $t_N$ increases, the latter condition being related to a reduced N-KC distance. On the other hand, a smaller hopping amplitude induces a resonance shrinking which reflects a reduced overlap between the tip and system modes, this condition being appropriate to describe an increased N-KC spatial gap.
The analysis of Figure \ref{Fig9} also suggests that the transparent limit of the BTK theory, characterized by a constant subgap conductance plateau sustained by the Andreev reflection, is not accessible within the considered T-stub geometry in which additional scattering events are originated at the waveguide junctions. The BTK transparent limit can be approached using the geometry described in appendix B, which, in general, does not correspond to the experimental conditions of an STM measurement and thus presents a limited relevance in our discussion.\\
To complete our analysis, let us note that the emergence of topological phases in closed systems is commonly discussed in terms of properties of the Hamiltonian spectrum. For the open system considered here, the mentioned classification can be done by studying the scattering matrix properties. Indeed, following the work by Beenakker et al. \cite{Beenakker}, we consider the reflection submatrix $r$ of the scattering matrix $S$:

\begin{equation}
\label{reflection matrix}
r=\left[\begin{matrix}
r_{ee}&r_{eh}\\
r_{he}&r_{hh}
\end{matrix}\right],
\end{equation}

and introduce the topological number $Q=sign[Det(r)]$. We have verified that a numerical estimate of $Q$ gives the same topological phase boundary in the $t_N-\mu$ plane as the one of an isolated Kitaev chain, confirming that one can characterize the non-trivial phase of the system under study via the scattering matrix properties.

\section{Quantum transport through a Normal/Kitaev ladder/Superconductor device}
\label{sec:junction-ladder}
\begin{figure}
\centering
\includegraphics[scale=0.40]{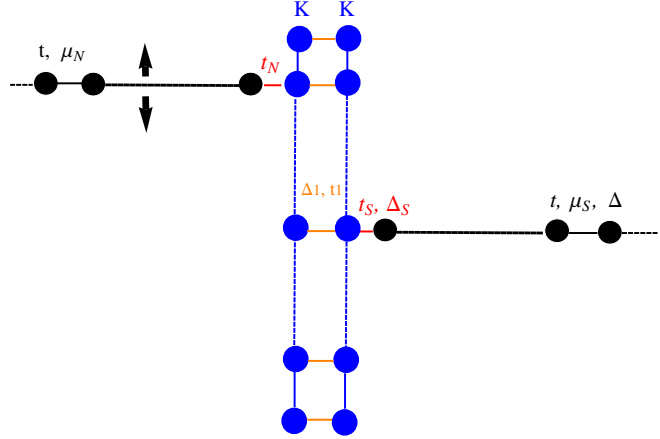}
\caption{N-KL-SC device. A Kitaev ladder (central region) coupled to a movable normal lead and to a superconducting $p$-wave lead.}
\label{Fig10}
\end{figure}

In this Section, we consider the transport properties of the N-KL-SC device depicted in Figure \ref{Fig10}. Proceeding as done for the N-KC-SC device, we have determined the scattering matrix elements by imposing appropriate matching conditions on the incident and transmitted scattering states and we have evaluated the differential conductance in terms of Andreev and normal reflection probability.

\begin{figure}
\centering
\includegraphics[scale=0.22]{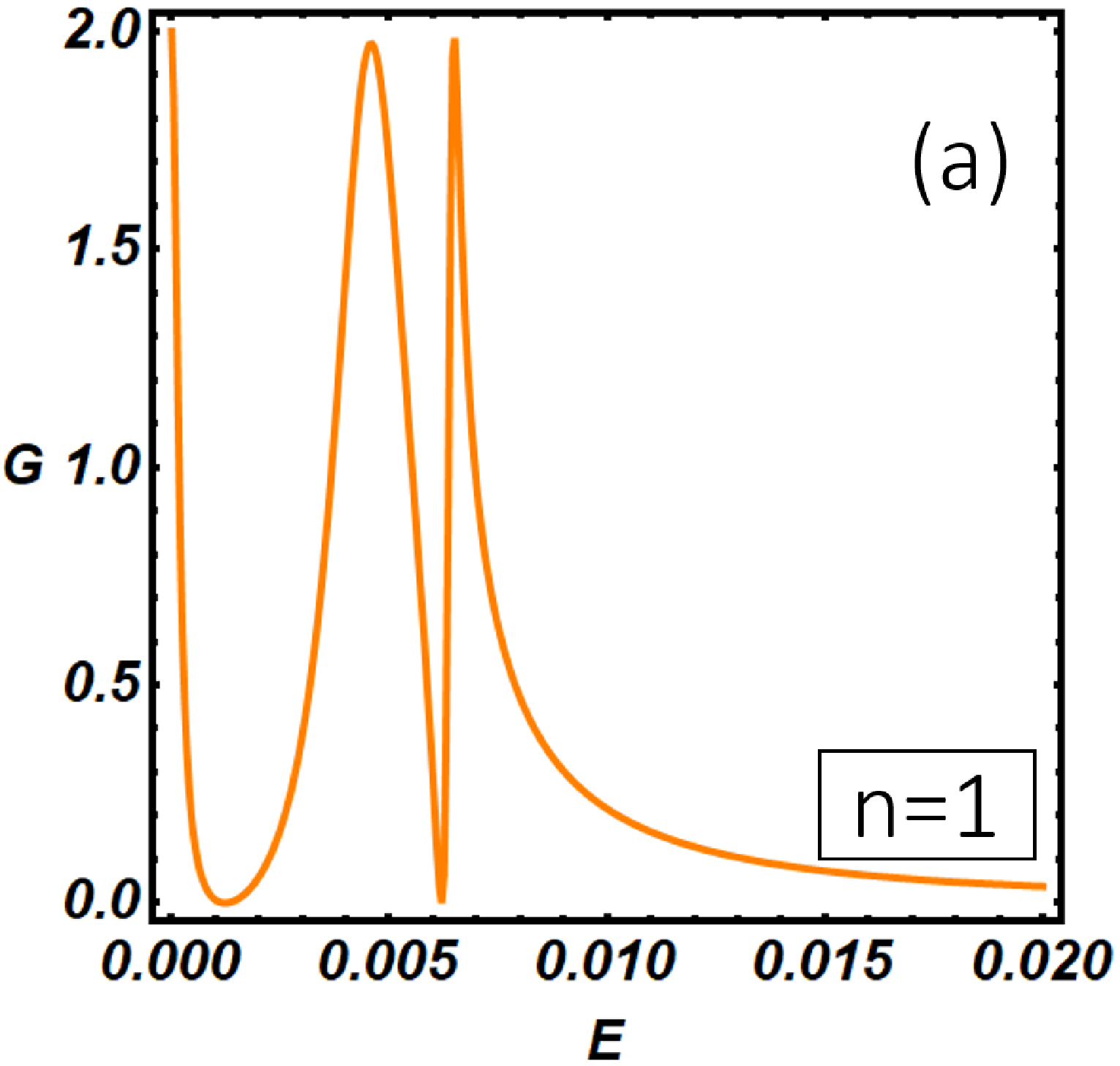}
\includegraphics[scale=0.22]{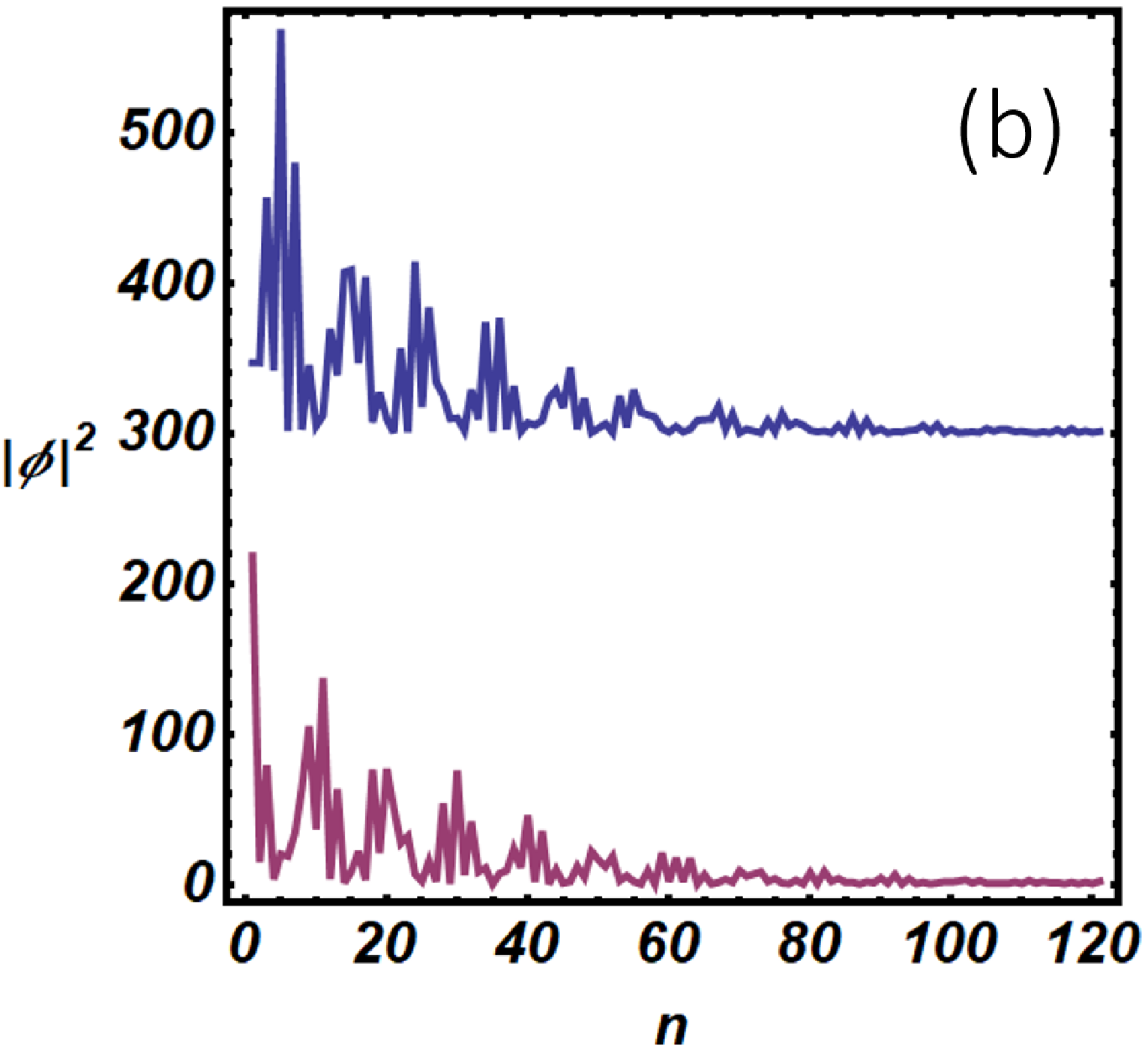}
\includegraphics[scale=0.22]{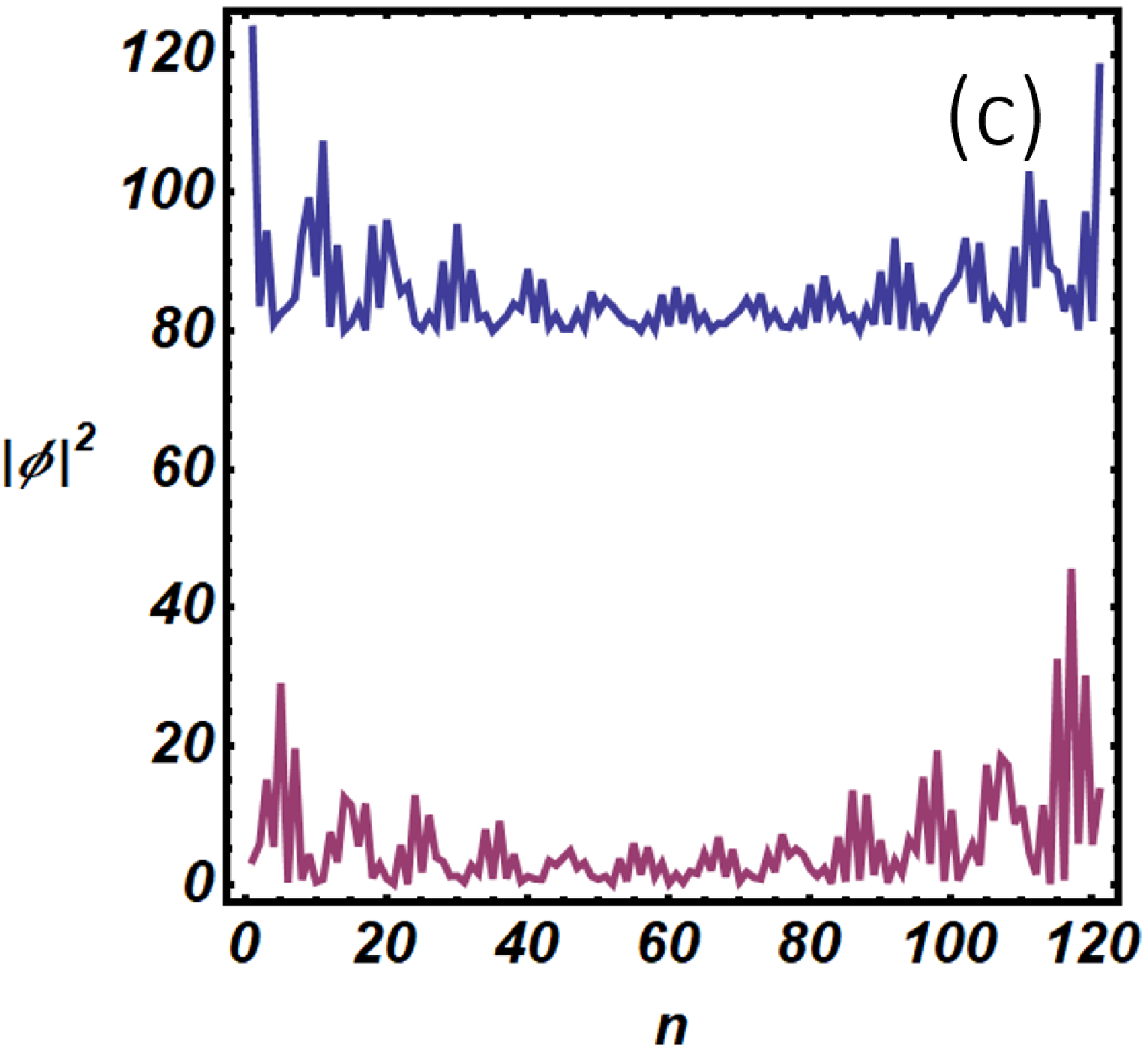}
\includegraphics[scale=0.22]{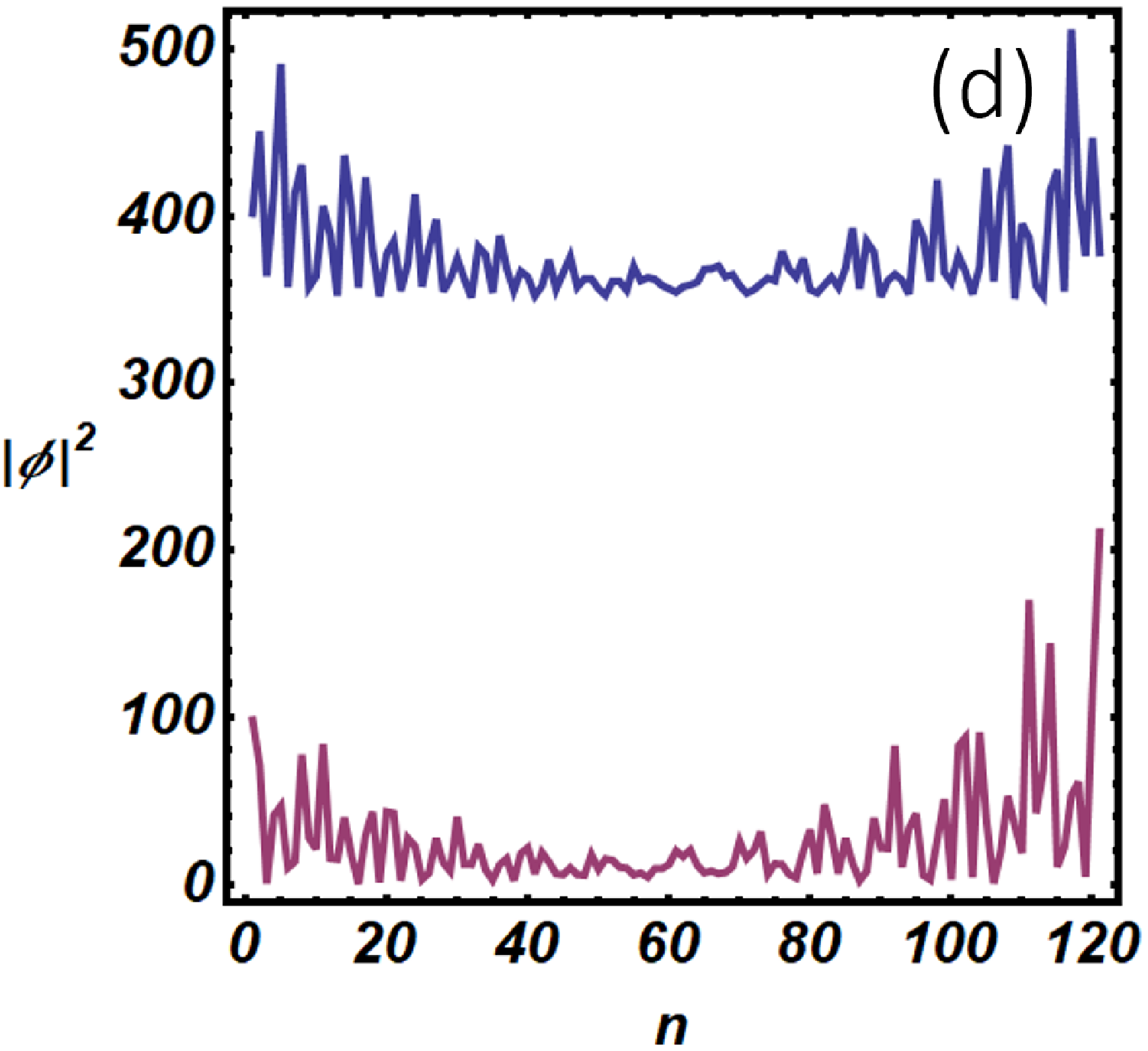}\\
\includegraphics[scale=0.22]{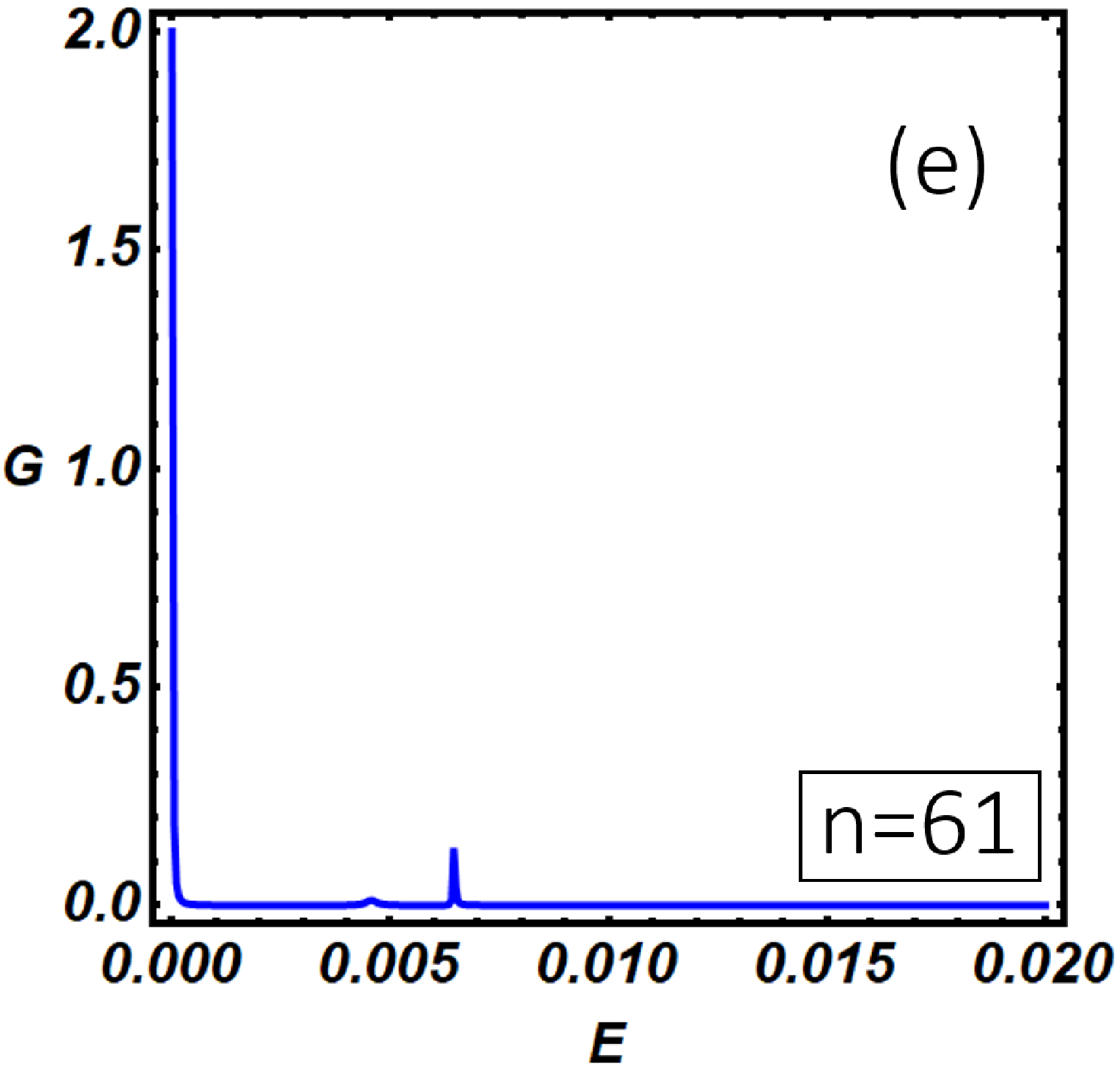}
\includegraphics[scale=0.22]{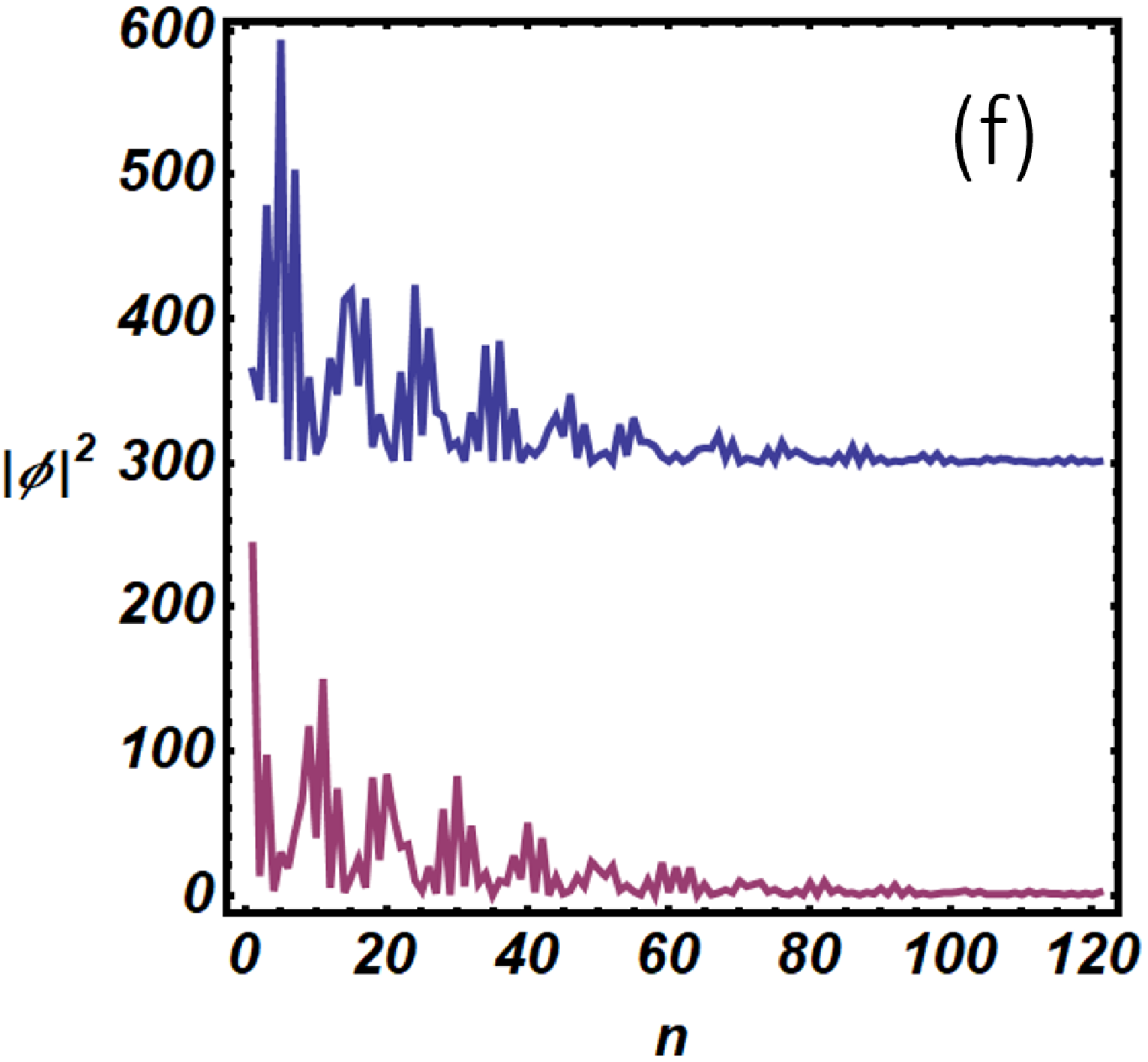}
\includegraphics[scale=0.22]{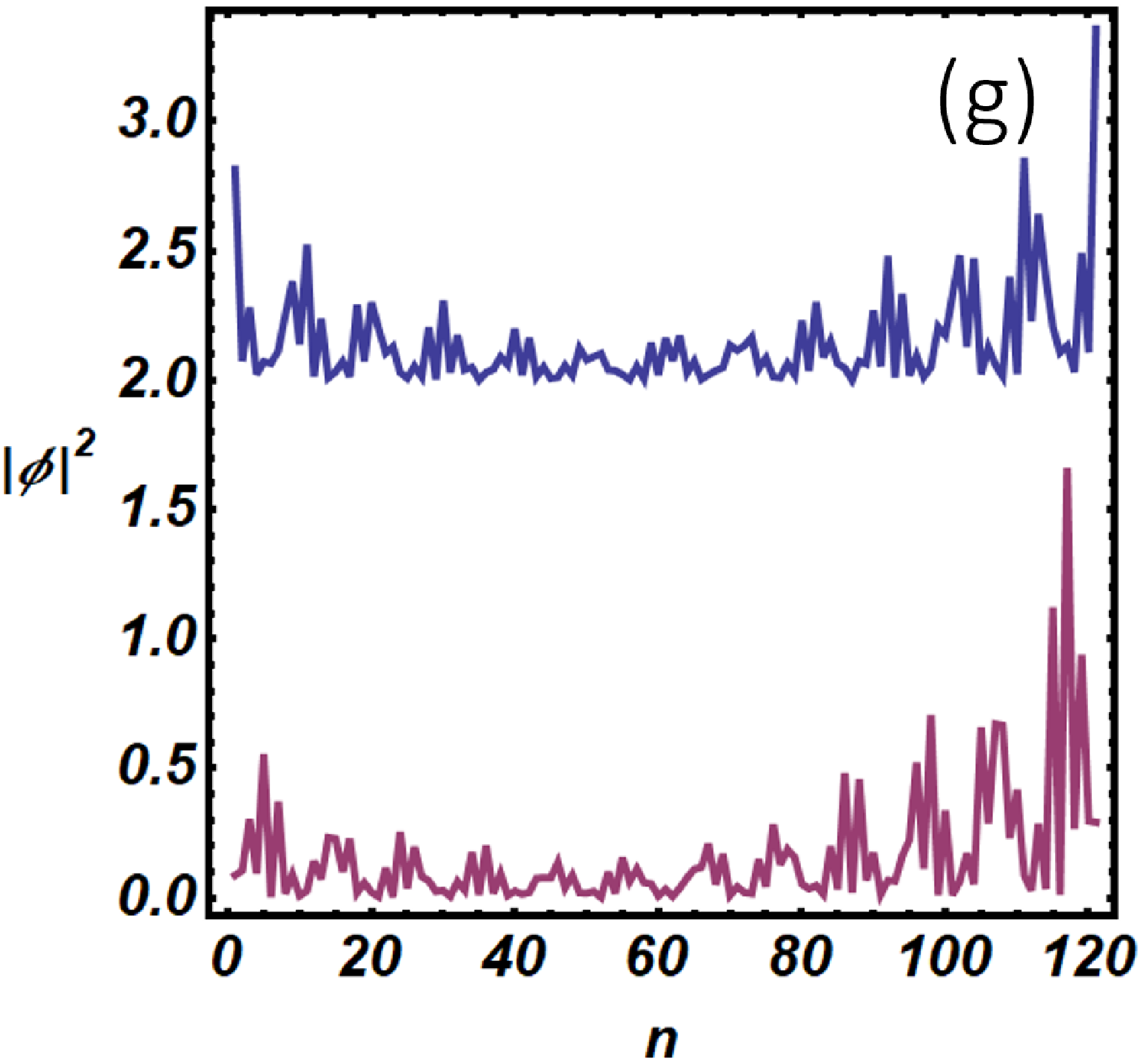}
\includegraphics[scale=0.22]{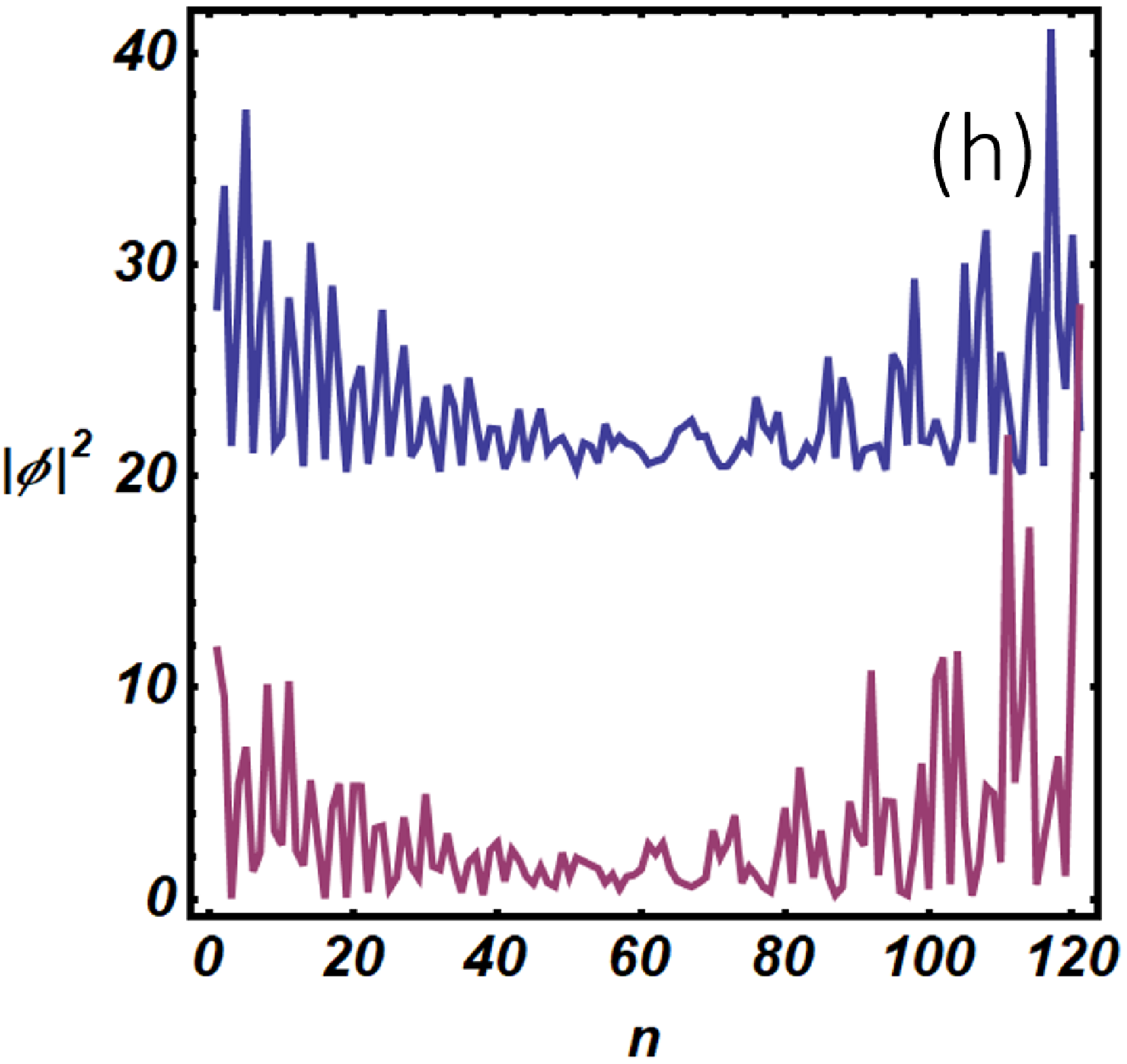}\\
\includegraphics[scale=0.22]{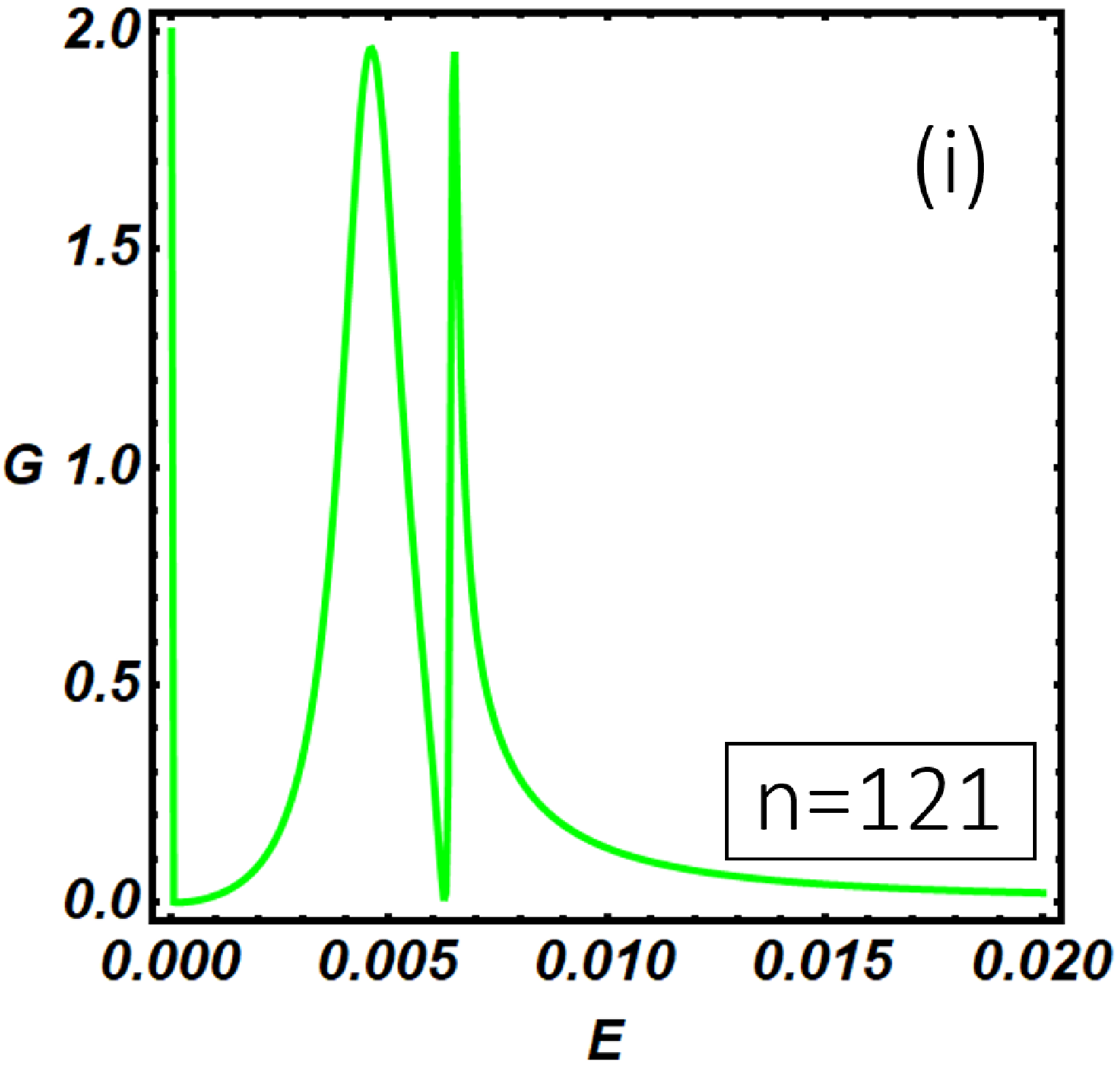}
\includegraphics[scale=0.22]{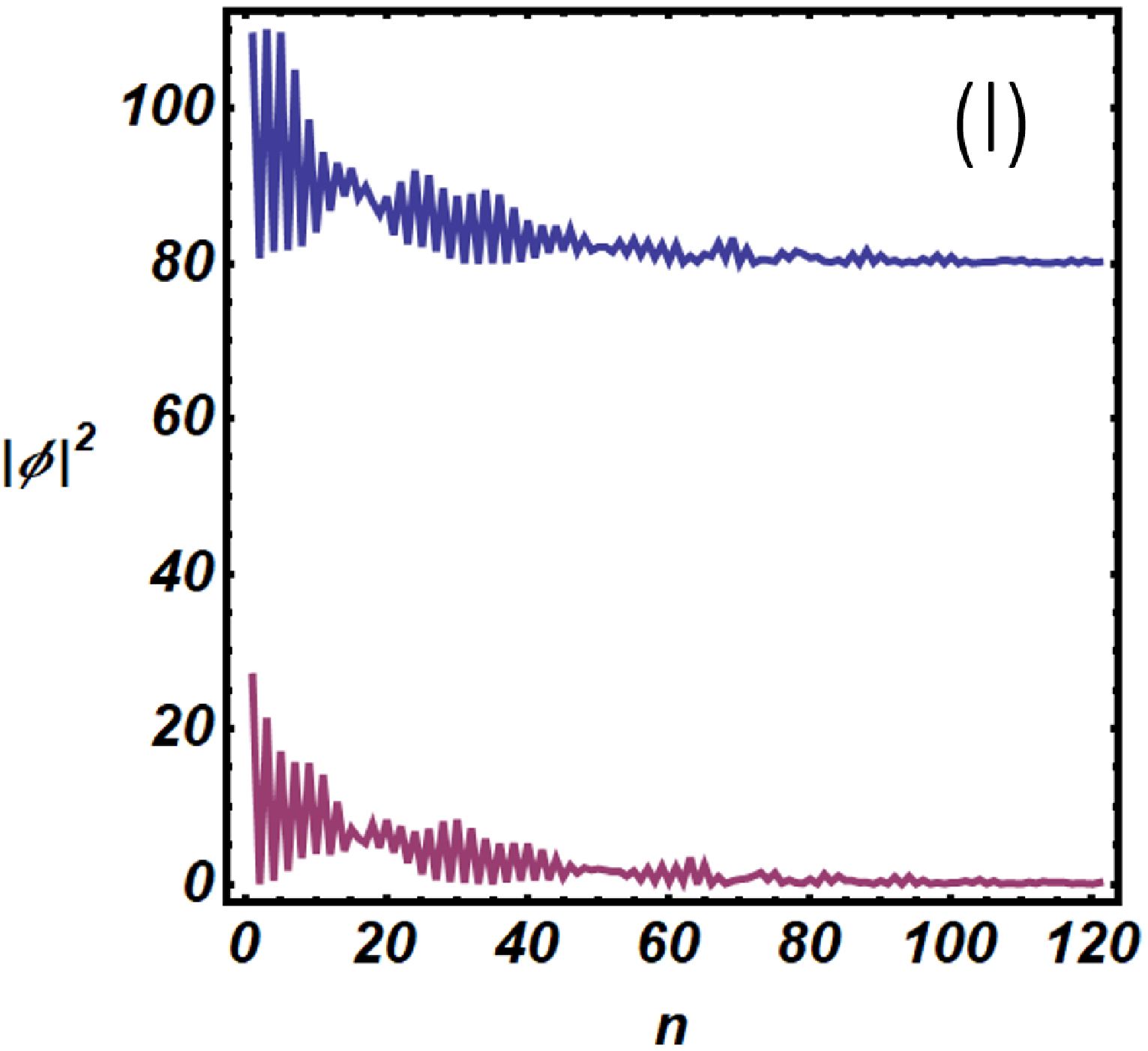}
\includegraphics[scale=0.22]{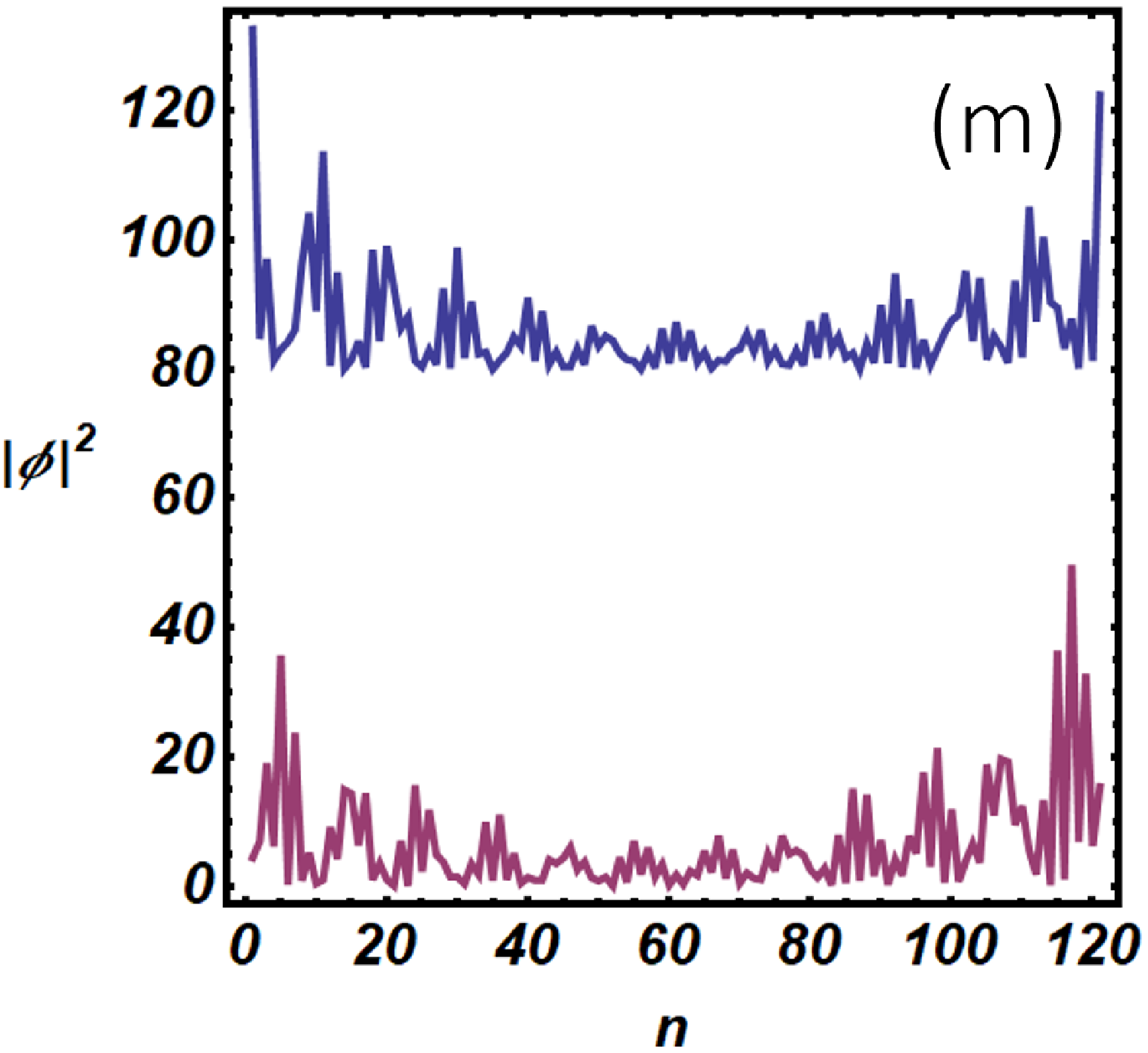}
\includegraphics[scale=0.22]{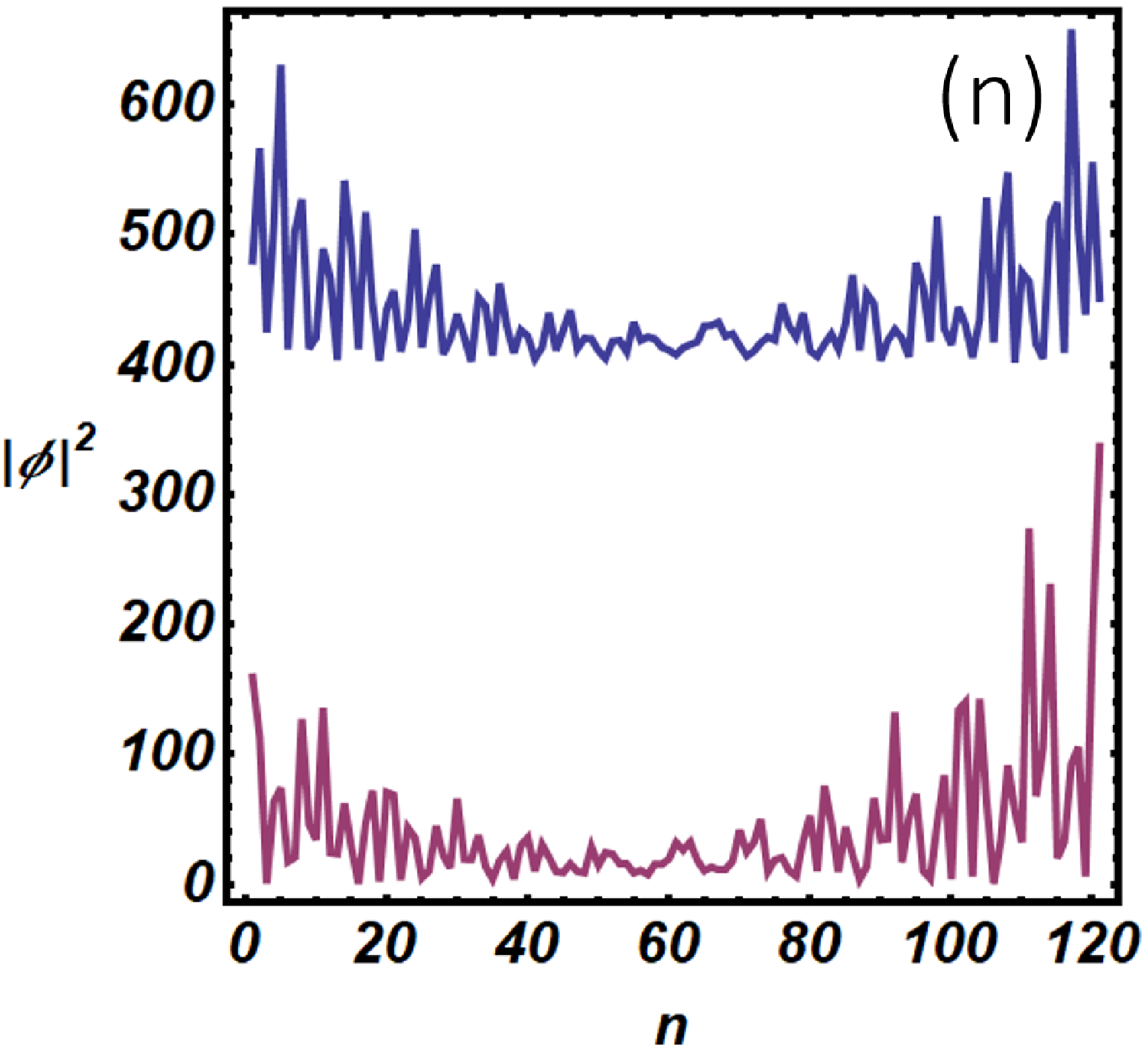}
\caption{N-KL-SC device: zero-temperature differential conductance (in the unit of $\frac{2 e^2}{h}$) as a function of the energy. Different panels, namely (a), (e), (i), are obtained by changing the linking position, given by $n$, between the normal lead and the nearest Kitaev wire of the ladder. The Blue and purple modes in panels (b), (c), (d), (f), (g), (h), (l), (m), (n) represent the modulus squared of the resonant modes on the upper and lower chain of the ladder (shifted by a convenient vertical offset) evaluated at energy values corresponding to the sub-gap conductance peaks. From the left to the right and from the top to the bottom, we have set the following parameters: (b): $n=1$, $E=5\cdot 10^{-5}$. (c): $n=1$, $E=4.5\cdot 10^{-3}$. (d): $n=1$, $E=6.5\cdot 10^{-3}$. (f): $n=61$, $E=5\cdot 10^{-5}$. (g): $n=61$, $E=4.5\cdot 10^{-3}$. (h): $n=61$, $E=6.5\cdot 10^{-3}$. (l): $n=121$, $E=5\cdot 10^{-5}$. (m): $n=121$, $E=4.5\cdot 10^{-3}$. (n): $n=121$, $E=6.5\cdot 10^{-3}$. The remaining model parameters have been fixed as:  $\Delta=0.02$, $t_N=t_S=0.2$, $t=1$, $\mu=0.5$, $\mu_S=\mu_N=0$, $\Delta_1=0.09$, $t_1=0.6$.}
\label{Fig11}
\end{figure}

The results of this analysis are shown in Figure \ref{Fig11} where the conductance curves and the internal modes of the Kitaev ladder are shown. The model parameters used in the computations are those that determine the emergence of a topological phase with two Majorana fermions per edge for the closed system \cite{Maiellaro} discussed in Sec. \ref{sec:ladder} (Figure \ref{Fig6}). The conductance curves (panels (a), (e), (i)) show a multiple peaks structure containing a zero-bias conductance peak accompanied by two satellite peaks at higher energy ($E=4.5 \cdot 10^{-3}$ and $E=6.5 \cdot 10^{-3}$). The zero-bias conductance peak is related to a Majorana mode peaked at the left end of the ladder, while satellite peaks are related with U-shaped internal modes peaked at the ladder edges (see panels (c), (d), (g), (h), (m), (n)). Internal modes pertaining to the satellite peaks can be associated to quantum states coming from hybridization processes of genuine Majorana modes. The hybridization of such states, which is favored by the opening of the system, produces a degradation of the initial topological properties and gives rise to quasi-Majorana states\cite{flensberg}. As discussed before, the degree of hybridization of the ladder internal states can be deduced by measuring the resonance width of the conductance peaks. In close analogy with the discussion reported for the N-KC-SC device, we do observe a clear correspondence between the internal modes wave function at the N-KL interface and the resonance broadening of the conductance peaks. In the next section we describe disorder effects on quasi-Majorana states.

\section{Disorder effects in a Normal/Kitaev ladder/Superconductor device}
\label{disorder}
\begin{figure}
\centering
\includegraphics[scale=0.22]{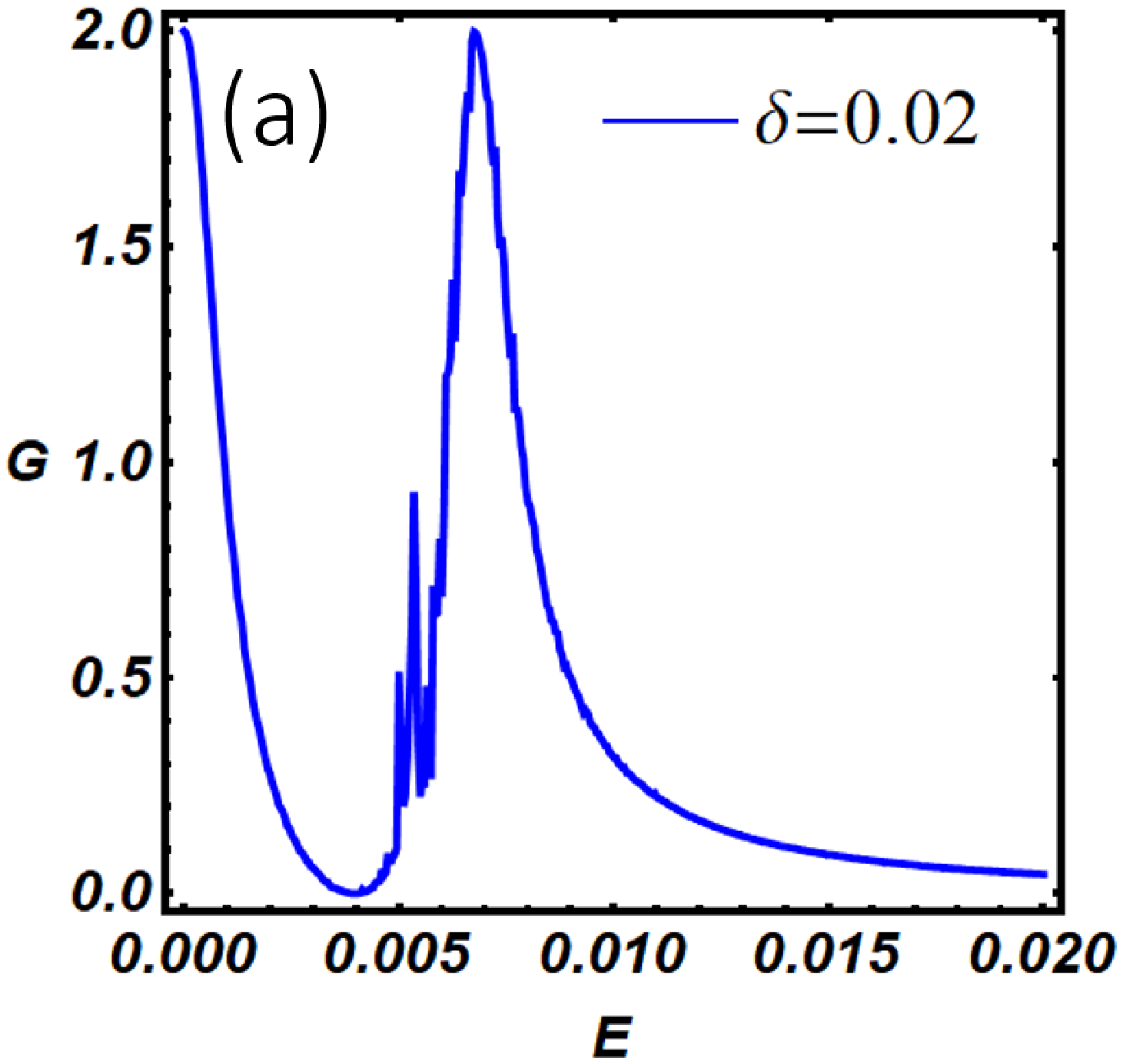}
\includegraphics[scale=0.22]{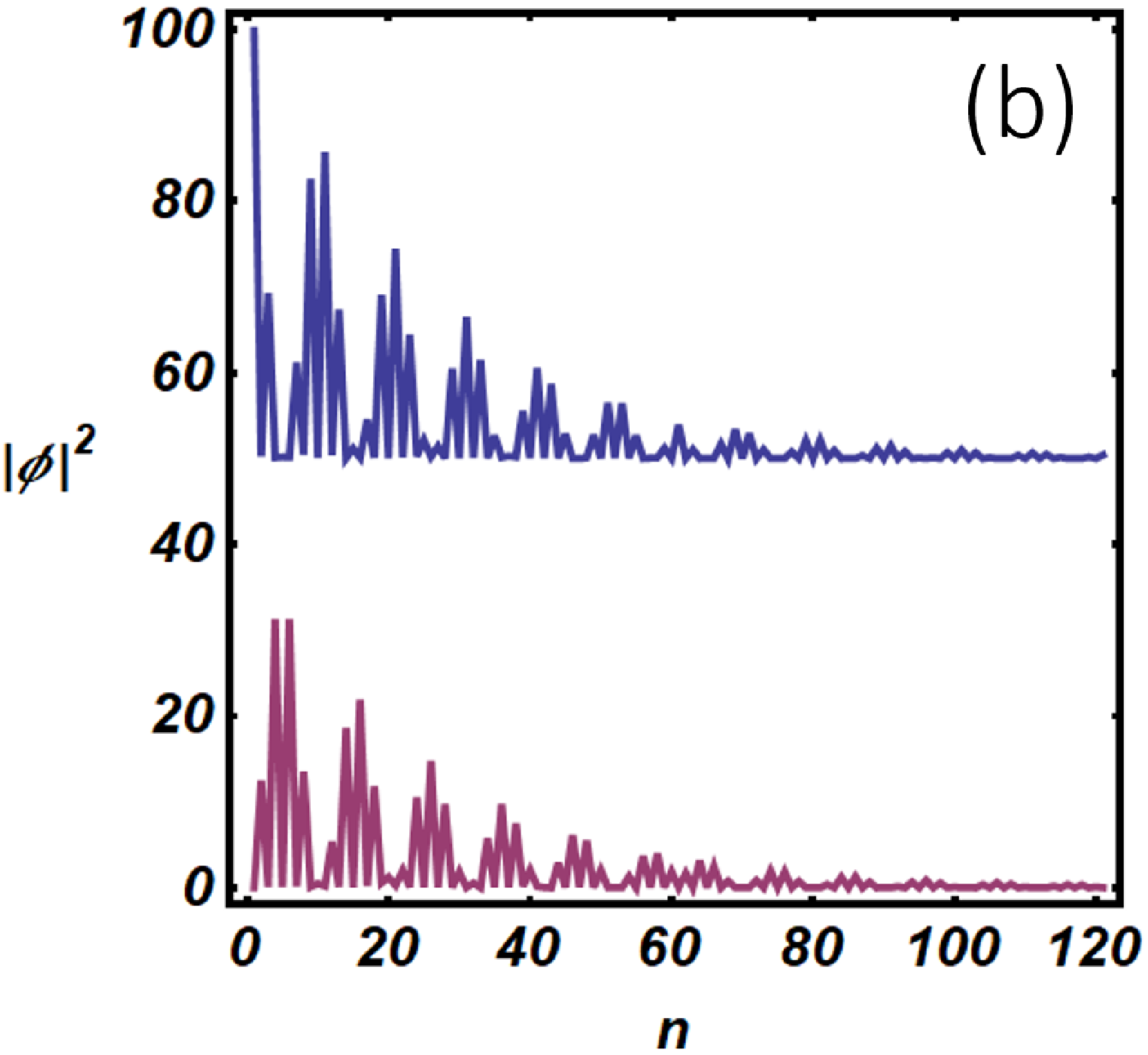}
\includegraphics[scale=0.22]{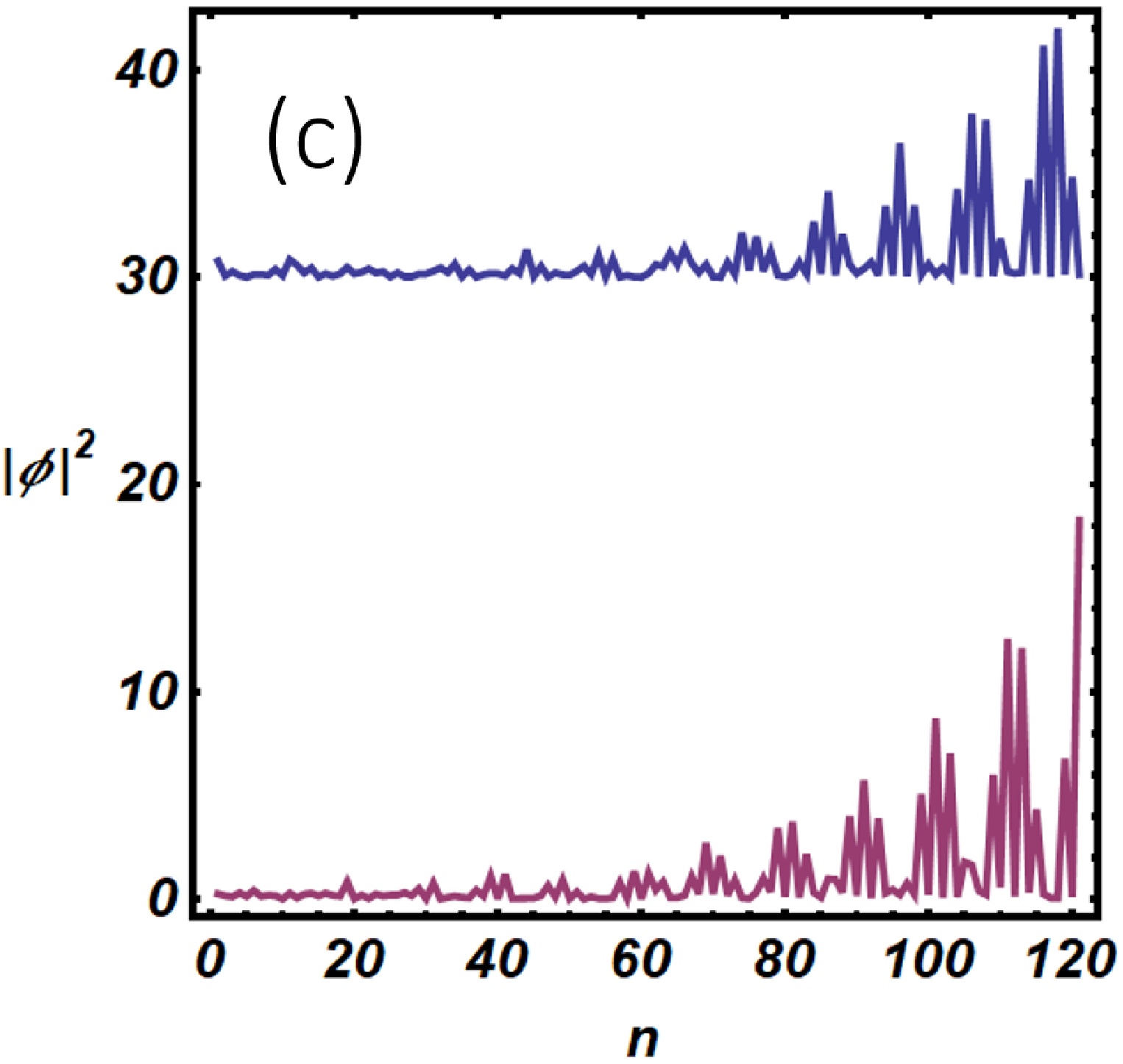}
\includegraphics[scale=0.22]{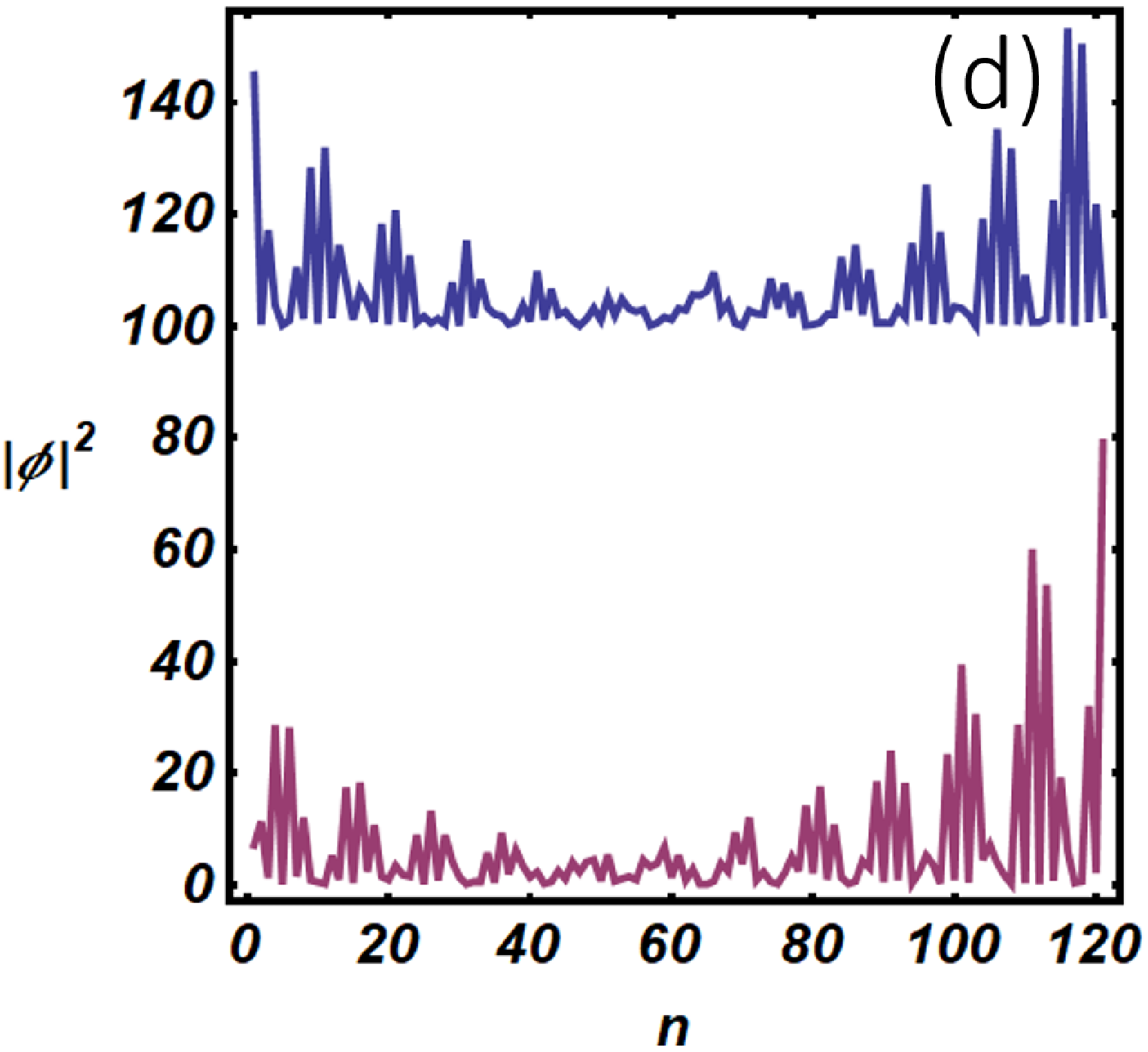}\\
\includegraphics[scale=0.22]{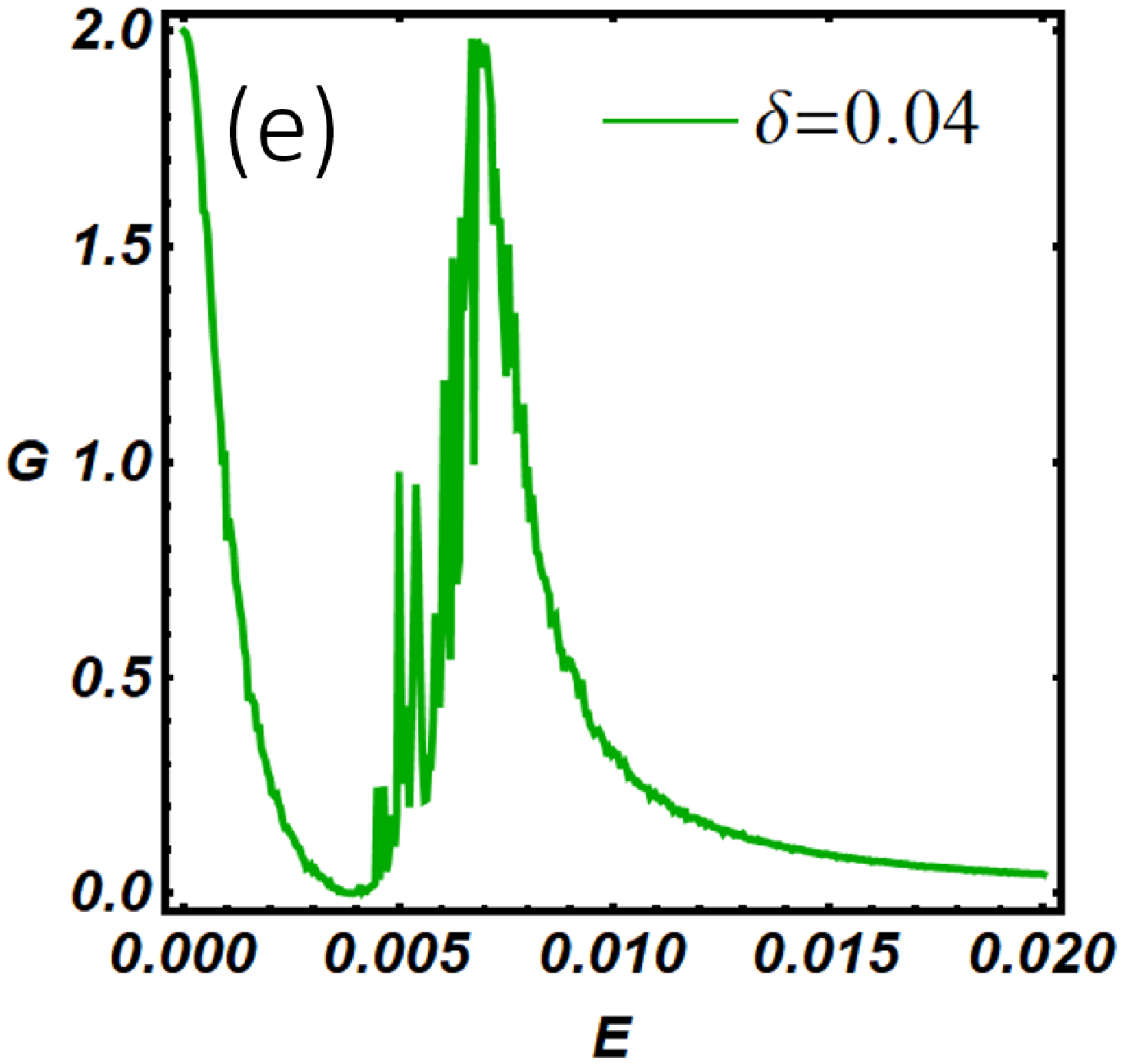}
\includegraphics[scale=0.22]{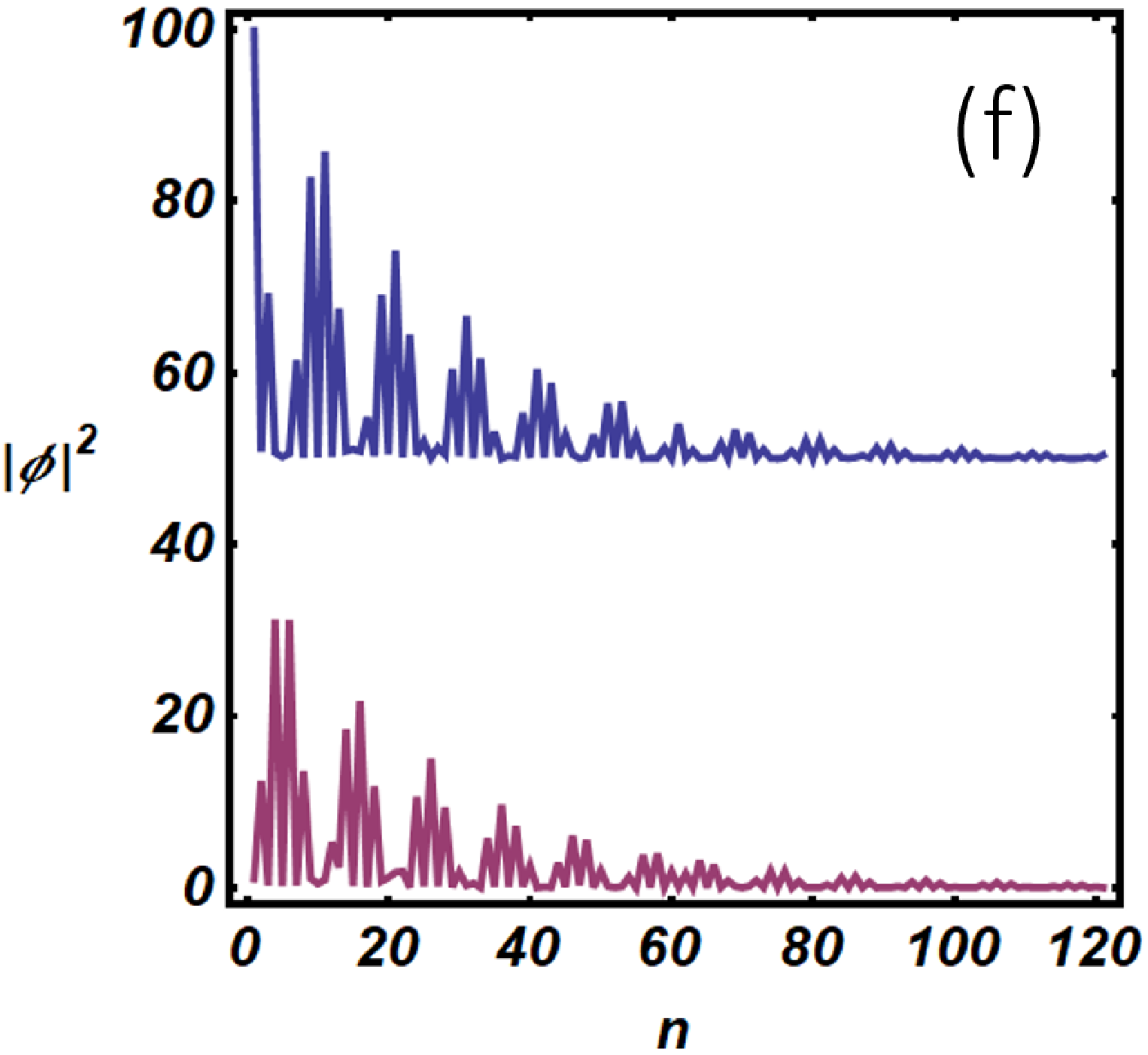}
\includegraphics[scale=0.22]{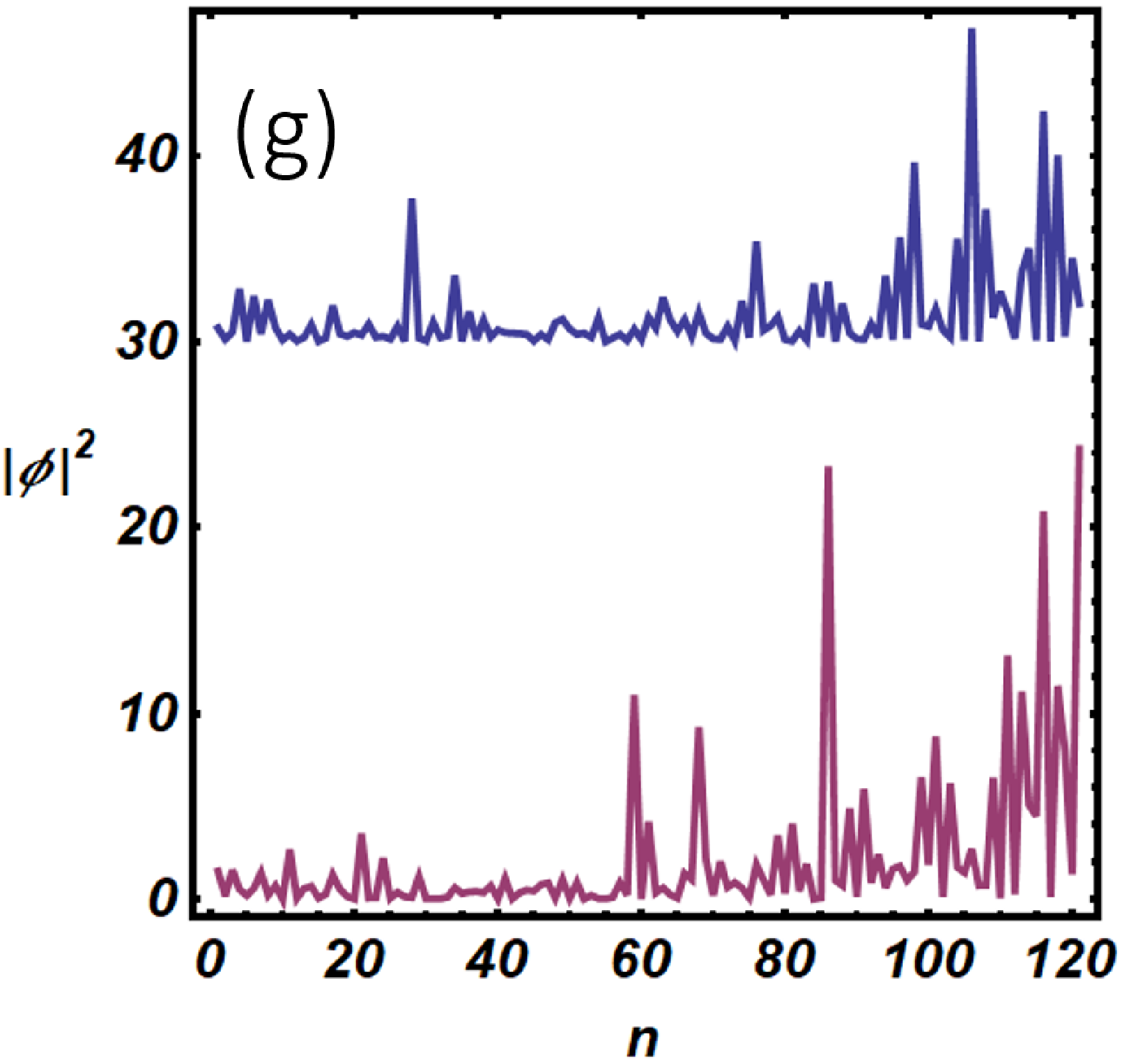}
\includegraphics[scale=0.22]{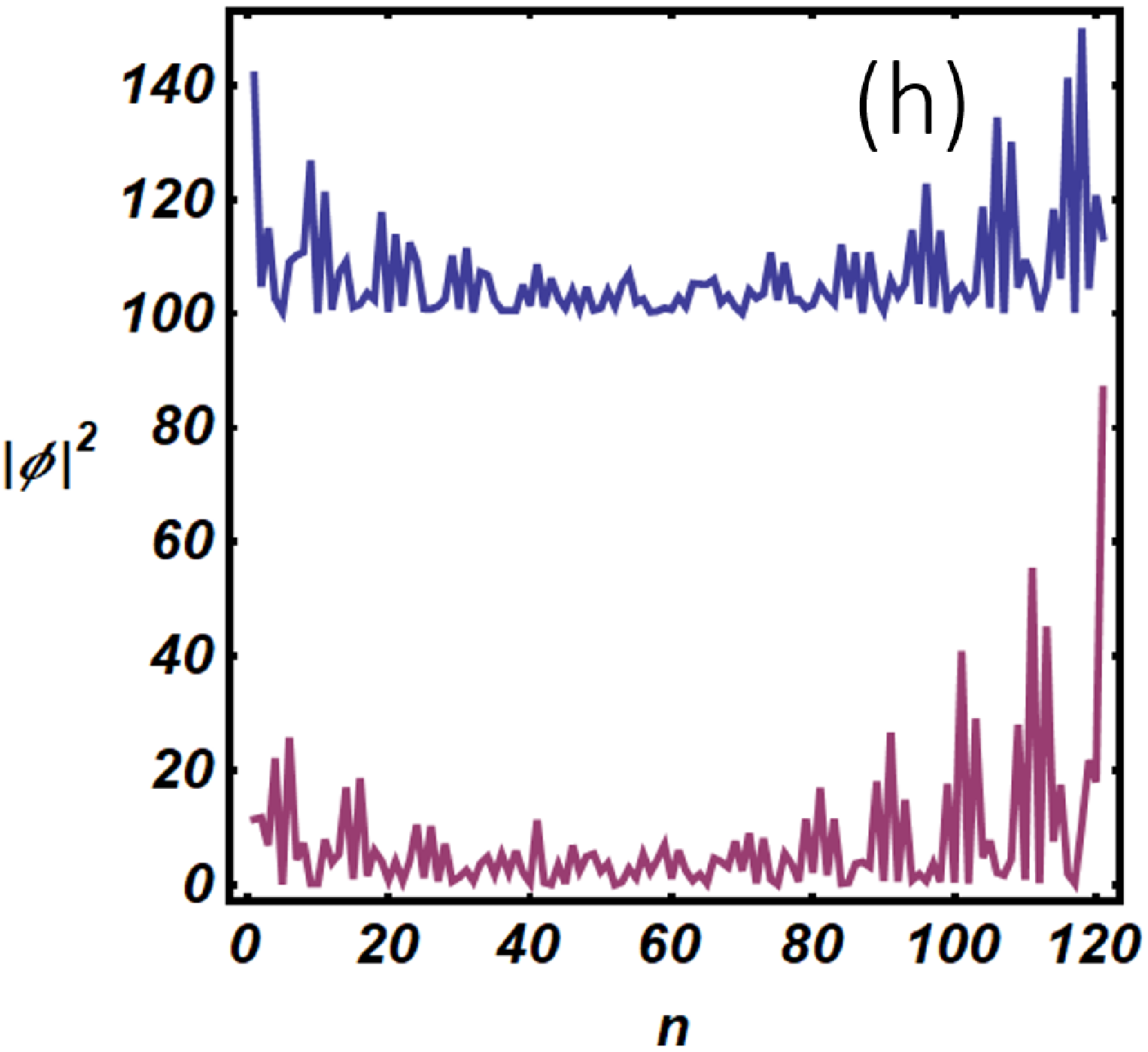}\\
\includegraphics[scale=0.22]{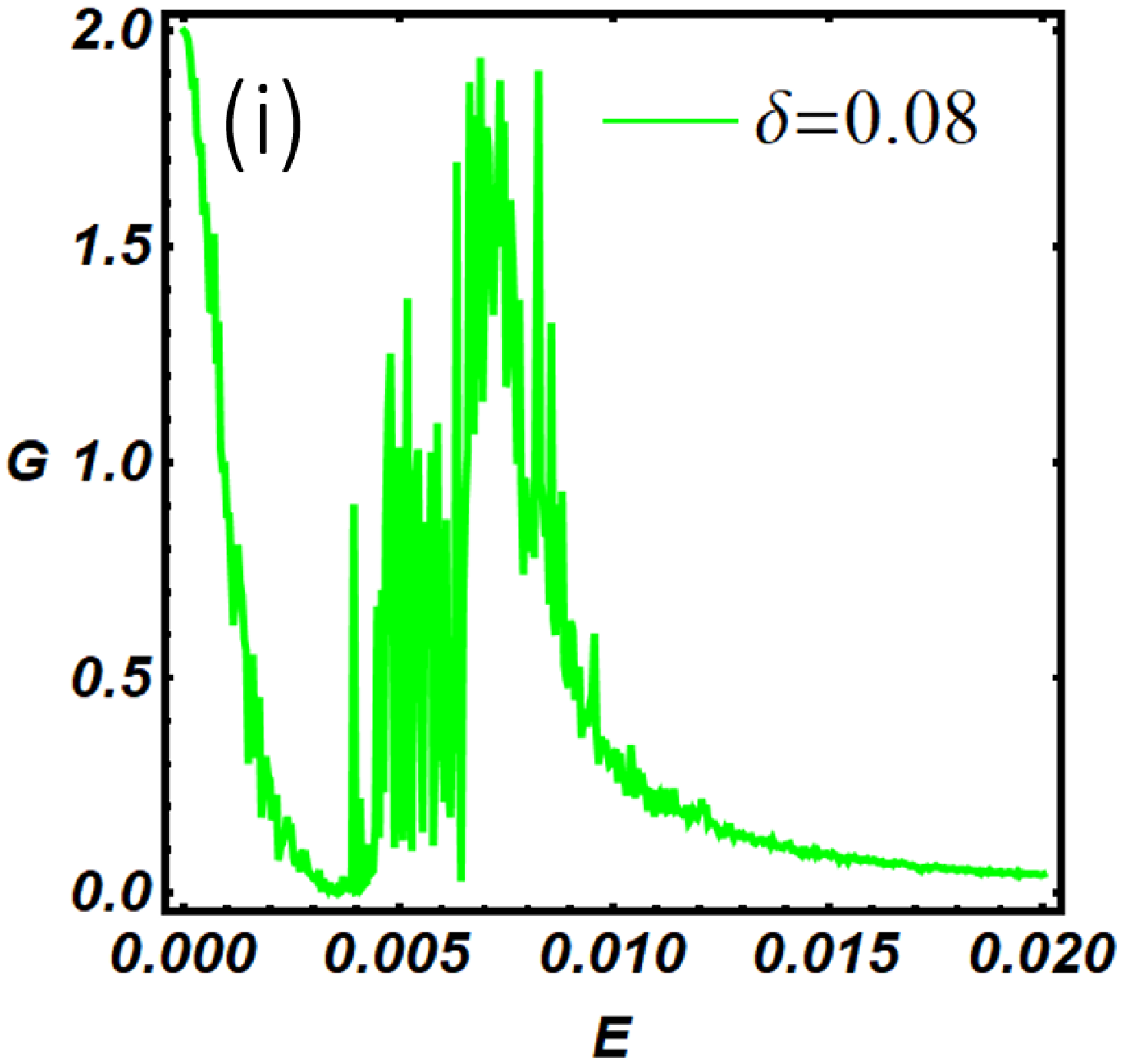}
\includegraphics[scale=0.22]{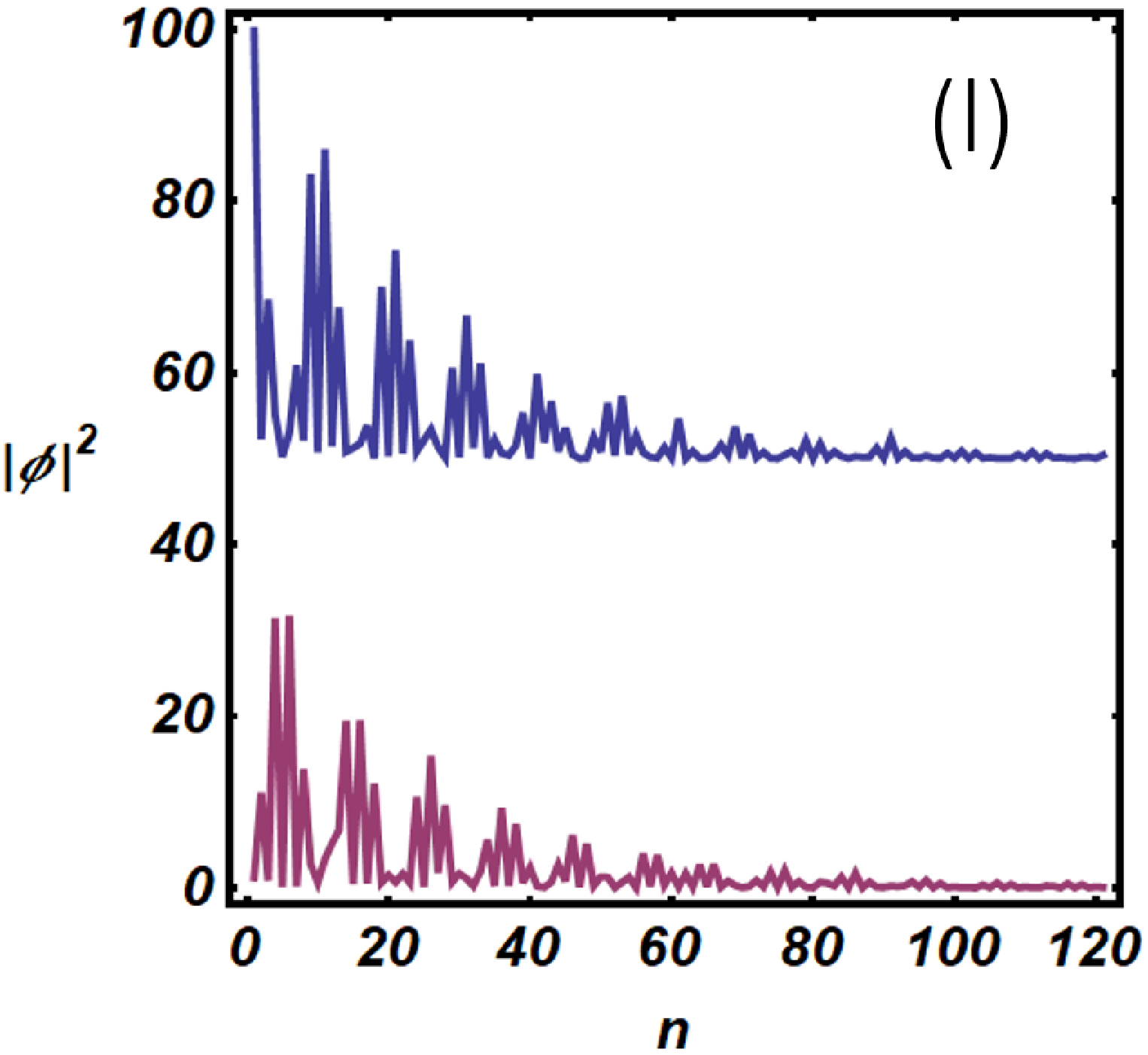}
\includegraphics[scale=0.22]{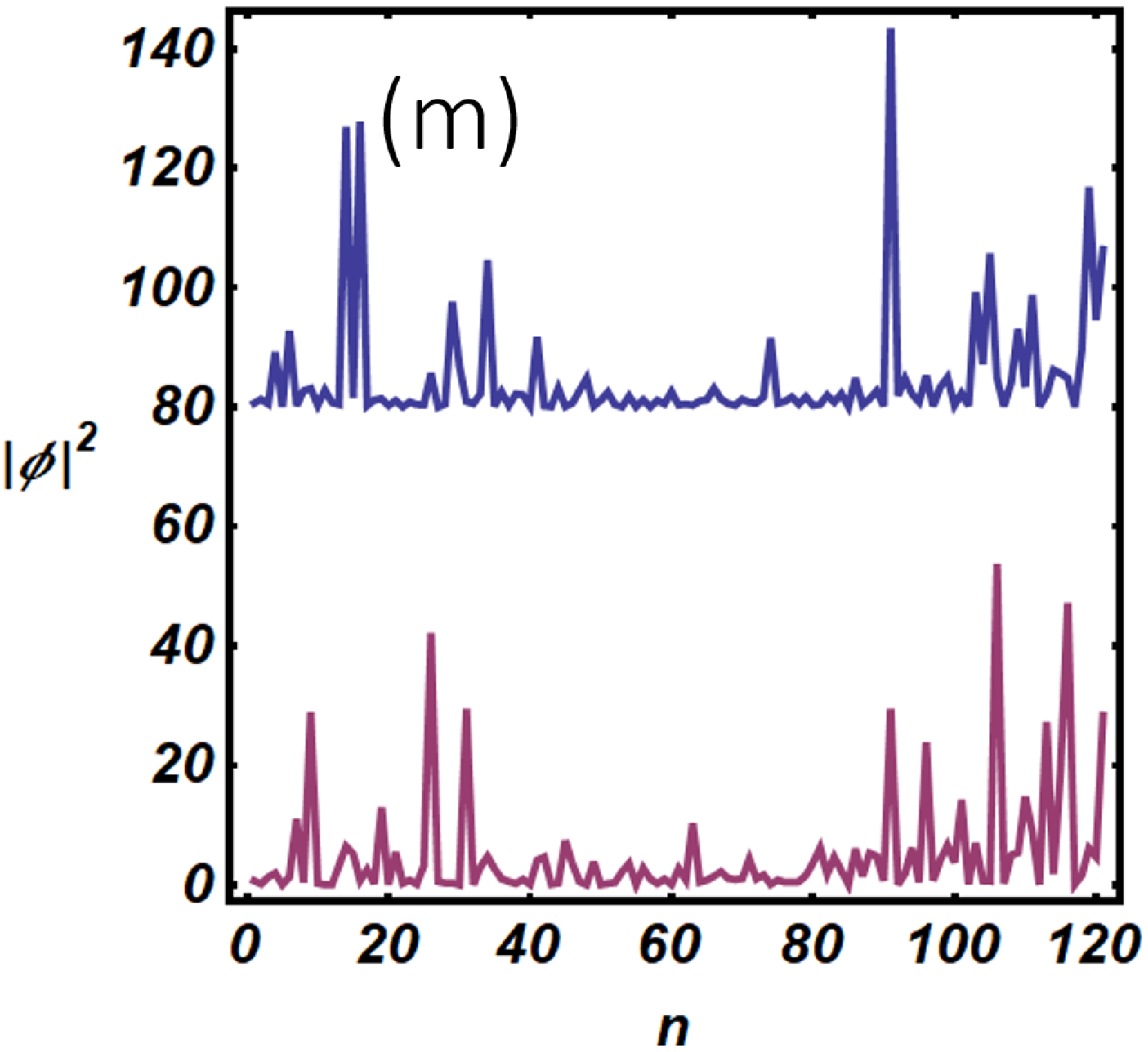}
\includegraphics[scale=0.22]{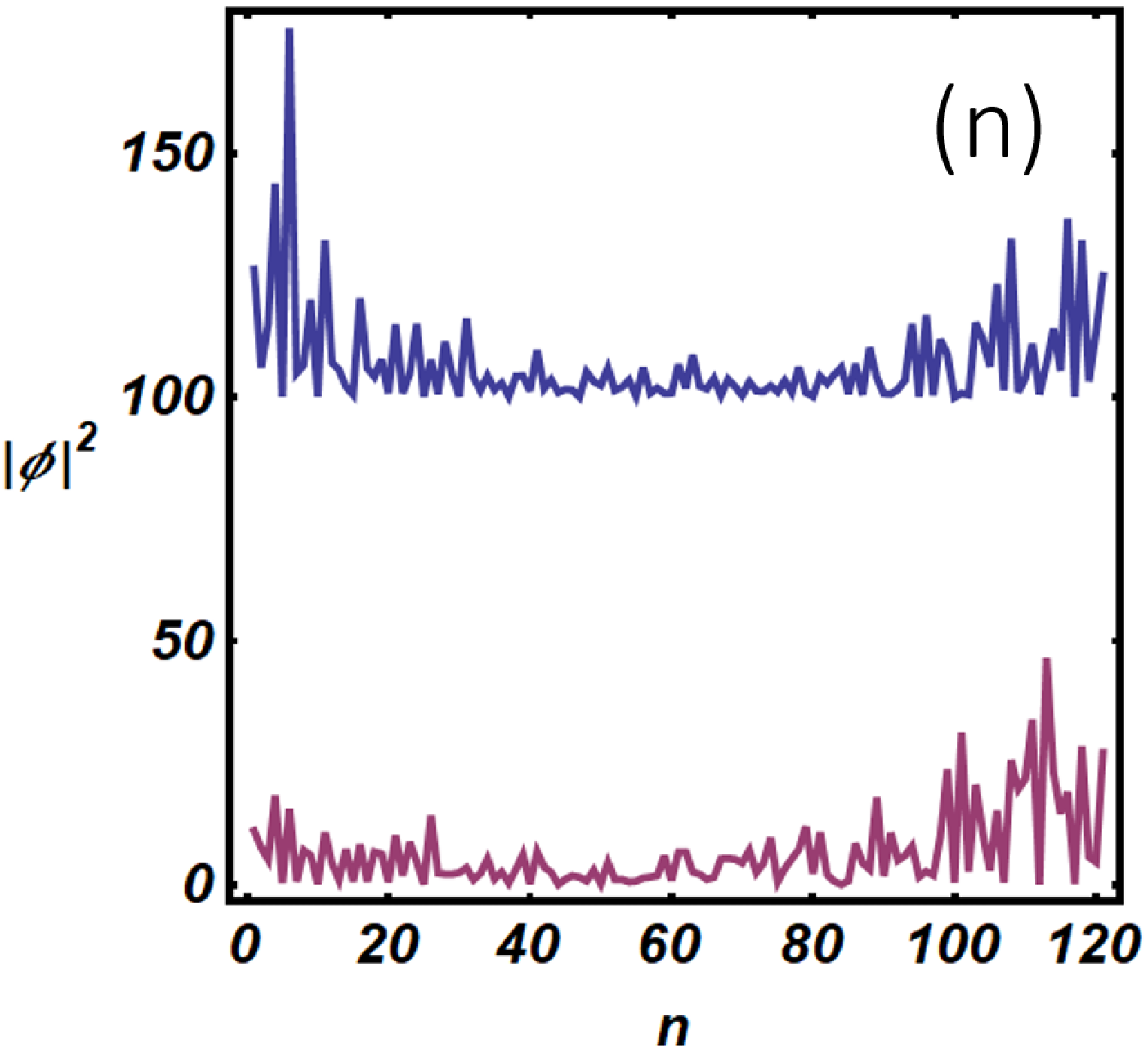}\\
\caption{N-KL-SC device in the presence of disorder. The panels (a), (e), (i) show the zero-temperature differential conductance (in the unit of $\frac{2 e^2}{h}$) as a function of the energy for a normal lead attached to the first site of the first Kitaev wire of the ladder and for the three different values of $\delta$, related to the variance of the random on-site potential. The  panels (b), (c), (d), (f), (g), (h), (l), (m), (n) represent the modulus squared of the resonant modes of the upper and lower chain of the ladder evaluated at energy values corresponding to the sub-gap conductance peaks. Model parameters have been fixed as done in first line of Figure \ref{Fig11}.}
\label{Fig12}
\end{figure}

In order to test the robustness of the topological phase in the N-KL-SC device, we include a random potential in the ladder model which emulates, for instance, the intrinsic effect of impurities and inhomogeneities in semiconducting nanowires. On the other hand, when Majorana quasiparticles are realized using ordered assemblies of magnetic atoms on the surface of conventional superconductors \cite{Li2016,feldman2017}, extrinsic disorder effects can be induced by the random strength of the atom-surface coupling. We model these effects assigning random values of the on-site potential $\mathcal{U}_n$ at any system site; the random realizations have been obtained generating random numbers $\mathcal{U}_n$ with uniform distribution in the interval $(-\delta, \delta)$, with $\delta$ fixed as $\delta=0.02$, $0.04$, $0.08$ (Figure \ref{Fig12}). Accordingly, the statistical average of the on-site potential is zero, while its variance is given by $\overline{\mathcal{U}_n^2}=\delta^2/3$. The remaining model parameters have been fixed as in panel (a) of Figure \ref{Fig11}. When the conductance of a single disordered realization is analyzed (panels (a), (e), (i)), we observe the progressive degradation of the three-peaks structure observed in Figure \ref{Fig11} (a), which has been obtained in absence of disorder. Disorder has a different impact depending on the peak considered. In particular, the zero-bias peak, which corresponds to a robust Majorana mode, remains quantized and disorder only produces a broadening effect on the resonant peak. The above result reflects the fact that disorder introduces a finite mean free path (related to the variance of the random on-site potential) which renormalizes the quasi-particles lifetime and induces a resonance broadening. The the two satellite peaks at quasi-zero energy (see Figure \ref{Fig11} (a)) are fragile against disorder effects. Indeed, in presence of disorder, one peak is completely suppressed, while the one at higher energy appears broadened and no more quantized (at high disorder). The presented phenomenology suggests that quasi-zero energy modes are related to the hybridization of genuine Majorana states. Such states, being distribute along the whole system, are quite fragile to disorder effects. The suppression of the intermediate resonant peak in presence of disorder is easily explained looking at panel (c) of Figure \ref{Fig11} and (c), (g), (m) of Figure \ref{Fig12}. Indeed, the internal mode in Figure \ref{Fig11} (c) (in the absence of disorder) is peaked at the system edges. Accordingly, the latter state is well coupled with the scattering states of the normal electrode. When the disorder is included, the overlap of this state with the normal electrode states goes to zero, producing the disappearance of the conductance peak. This effect originates from the disorder-assisted localization of the ladder resonant mode at a system edge which is not connected with the normal electrode (see panels (c), (g), (m)). The latter effect is not observed for quasi-zero energy mode at higher energy, which starts to be severely degraded only at high disorder ($\delta=0.08$). This is confirmed by the internal modes analysis which has been shown in panels (d), (h), (n). In this case, the analysis evidences that internal modes are peaked at both the system edges and thus are well coupled with the normal electrodes states, the latter being a requisite to observe a finite conductance.\\
The above findings suggest that topological protection of quantum states can be lost as an effect of the opening of the isolated quantum system. Indeed, the connection of a topological system with external reservoirs produces hybridization of the internal modes, which is the prerequisite to manifest fragility against the detrimental effects of disorder.

\section{Conclusions}
\label{sec:conclusions}
We have addressed the problem of the topological phase characterization of open quantum systems by studying the paradigmatic cases of a Kitaev chain and a Kitaev ladder. We have characterized the topological phase diagram for the closed systems and we have found that, differently from the case of a single Kitaev chain, a Kitaev ladder displays topological phases with two Majorana states per edge. Concerning the topological phases with multiple Majorana states, the question of their stability against the opening of the system arises. In order to prove the stability of the topological phases of open systems subject to measuring processes (e.g. tunneling spectroscopy measurements), we have studied the quantum transport through a Kitaev chain and a Kitaev ladder by coupling them to a normal and a superconducting electrode. Using a lattice version of the usual Bogoliubov-de Gennes scattering theory, we have studied the differential conductance of these devices and we have found a correspondence between the conductance peaks and the resonant states confined inside the topological scattering region (i.e. the Kitaev chain or ladder). It has been proven, both for the Kitaev chain and ladder, that zero-energy peaks of the differential conductance correspond to robust Majorana states, while quasi-Majorana quantum states are signaled by almost zero-energy resonant peaks. Quasi-Majorana states originate from the hybridization of Majorana states, the latter being favored by the coupling with the scattering states of the electrodes. By studying the differential conductance of a topological device based on a Kitaev ladder, we have demonstrated that quasi-Majorana states and the associated conductance peaks are fragile against disorder effects. These findings are relevant in studying topological systems with multiple Majorana states, which are supposed to be important in multimode quantum wires or in multi-orbital topological systems.

\vspace{6pt}

%%%%%%%%%%%%%%%%%%%%%%%%%%%%%%%%%%%%%%%%%%
%\authorcontributions{Conceptualization, A.M., F.R. and R.C.; methodology, F.R.; software, A.M.; validation, F.R., C.A.P., V.C. and R.C.; formal analysis, A.M.; writing-original draft preparation, A.M. and R.C.; writing-review and editing, %F.R., C.A.P. and V.C.; supervision, R.C.; funding acquisition, F.R.}

%%%%%%%%%%%%%%%%%%%%%%%%%%%%%%%%%%%%%%%%%%
\acknowledgments{Discussions with Ciro Nappi on the lattice Bogoliubov-de Gennes technique are acknowledged. The authors acknowledge the project Quantox of QuantERA ERA-NET Cofund in Quantum Technologies (Grant Agreement N. 731473) implemented within the European Union's Horizon 2020 Programme.}

%%%%%%%%%%%%%%%%%%%%%%%%%%%%%%%%%%%%%%%%%%
%\conflictsofinterest{The authors declare no conflict of interest.}

\vspace{6pt}

\appendix
\section{The tight binding Bogoliubov-de Gennes equations}
\label{appBdg}
The BdG equations describe the e-like and h-like excitations of a system with superconducting correlations. Hereafter we adopt a lattice BdG approach similar to the one presented in Ref. \cite{bdglattice1} and \cite{bdglattice2} for the s-wave pairing. Thus, in the next subsections, adopting a tight binding formulation specialized to the p-wave superconducting pairing, we derive the scattering states of a normal and a superconducting lead which provide the appropriate boundary conditions for the scattering problem discussed in the main text.

\unskip
\subsection{Tight binding Bogoliubov-de Gennes equations for the normal lead}
\label{appbdgN}
The tight binding BdG equations for a normal lead in the stationary case for a fixed energy $E$ are:

\begin{equation}
\label{BdGNormal}
\begin{cases} -\mu f_j-t(f_{j+1}+f_{j-1})=E f_j \\ \mu g_j+t (g_{j+1}+g_{j-1})=E g_j \end{cases}
\end{equation}

where $f_j$ and $g_j$ are the e-like and h-like components of the Nambu spinor, $\mu$ is the chemical potential of the lead, $t$ is the hopping amplitude for nearest sites and $E$ represents the excitation energy measured from the Fermi level. Requiring translational invariance, we can write the Nambu spinor in the following form:

\begin{equation}
\label{NambuNormal}
\left(
\begin{array}{c}
f_j\\
g_j\\
\end{array}\right)=\left(
\begin{array}{c}
f\\
g\\
\end{array}\right)e^{\pm ikj}.
\end{equation}

Solving the Eq. (\ref{BdGNormal}) using the ansatz in Eq. (\ref{NambuNormal}), we obtain the electron- and hole-like independent solutions, namely $\psi_{e/h}(j)$, with the corresponding energy $E$:

\begin{equation}
\label{Normalsolutions}
\begin{cases} \psi_e(j)=\left(
\begin{array}{c}
1\\
0\\
\end{array}\right)e^{\pm i k j}, & E=-\mu-2 t \cos k, \\ \psi_h(j)=\left(
\begin{array}{c}
0\\
1\\
\end{array}\right)e^{\pm i k j}, &  E=\mu+2 t \cos k \end{cases}.
\end{equation}

Using the dispersion relations for electron and hole solutions of Eq. (\ref{Normalsolutions}), we obtain the wave vectors of main text $k_{e/h}=\arccos\big[(\mu \pm E)/(-2t)\big]$ (Eq. (\ref{wavevectorsNormal})). An electron-like (hole-like) solution describes a particle with group velocity $v_g=\frac{1}{\hbar} \frac{\partial E_k}{\partial k}$ parallel (anti-parallel) to the particle wavevector $k_e$ ($k_h$).

\subsection{Tight binding Bogoliubov-de Gennes equations for the superconducting $p$-wave lead}
\label{appbdgS}
The tight binding BdG equations for a $p$-wave superconducting lead in the stationary case can be written as:

\begin{equation}
\label{BdGSuperconductor}
\begin{cases} -\mu f_j-t(f_{j+1}+f_{j-1})+\Delta (g_{j-1}-g_{j+1})=E f_j \\ \mu g_j+t (g_{j+1}+g_{j-1})+ \Delta (f_{j+1}-f_{j-1})=E g_j \end{cases},
\end{equation}

where the pairing term $\Delta$, responsible for the electron-hole correlation, induces the $p$-wave superconductivity. Due to translational invariance, the system admits translational invariant solutions for the Nambu spinor and thus, reasoning as done before, we can set the following eigenvalues problem:

\begin{equation}
\label{matrixproblem}
\left(\begin{matrix}
-\mu-2 t \cos q&\mp 2 i \Delta \sin q\\
\pm 2 i \Delta \sin q&\mu+2 t \cos q\end{matrix} \right)
\left(
\begin{array}{c}
f\\
g\\
\end{array}\right)
=E\left(
\begin{array}{c}
f\\
g\\
\end{array}\right),
\end{equation}

in which the $\pm$ sign in the off-diagonal blocks of the Hamiltonian is related to the mode propagation direction. The solution of Eq. (\ref{matrixproblem}) gives the excitation energy $E$:

\begin{equation}
\label{energysuperconductor}
E=\sqrt{\xi_q^2+\Delta_q^2}
\end{equation}

and the following relation between the electron and hole component of the spinorial wave function:

\begin{equation}
\label{h-e}
g=\mp \frac{E-\xi_q}{i \Delta_q} f,
\end{equation}

where we have introduced the quantities $\xi_q=-\mu-2 t \cos q$ and $\Delta_q=2 \Delta \sin q$. Using Eq.(\ref{energysuperconductor}) and the explicit expression of $\xi_q$, we obtain the wave vectors of e-like and h-like bogoliubons of the main text:

\begin{equation}
q_{e/h}=\arccos \biggl(\frac{-t \mu \mp \sqrt{(t^2-\Delta^2)(E^2-4 \Delta^2)+\Delta^2 \mu^2}}{2 (t^2-\Delta^2)}\biggr).\nonumber
\end{equation}

Using normalization, we fix the $e$-component of Eq.(\ref{h-e}) and obtain the final expression of the Nambu spinor in the form:

\begin{equation}
\label{spinorIniziale}
\left(
\begin{array}{c}
f\\
g\\
\end{array}\right)=\left(
\begin{array}{c}
\sqrt{\frac{E+\xi_q}{2E}}\\
\pm i\sqrt{\frac{E-\xi_q}{2E}}\\
\end{array}\right).
\end{equation}

Using $\xi_q=\pm \sqrt{E^2-\Delta_q^2}$, derived from Eq. (\ref{energysuperconductor}), Eq.(\ref{spinorIniziale}) can be presented in the form:

\begin{equation}
\label{doppiospinore}
\left(
\begin{array}{c}
f\\
g\\
\end{array}\right)_{\pm}= \left(
\begin{array}{c}
\sqrt{\frac{E \pm \sqrt{E^2-\Delta_q^2}}{2E}}\\
\pm i\sqrt{\frac{E \mp \sqrt{E^2-\Delta_q^2}}{2E}}\\
\end{array}\right).
\end{equation}

Introducing the BCS coherence factors recalled in the main text, we obtain four different solutions for the BdG equations corresponding to e-like and h-like excitations, namely $(u_{e},\pm i v_{e})^t$ and $(v_{h},\pm i u_{h})^t$, characterized by different propagation direction ($\pm$). The above arguments justify the form of the scattering states used in Eq.(\ref{propagatingstate}) of the main text.

\section{Conductance lowering effects in branched quantum waveguides}
\label{appWG}
\begin{figure}
\centering
\includegraphics[scale=0.35]{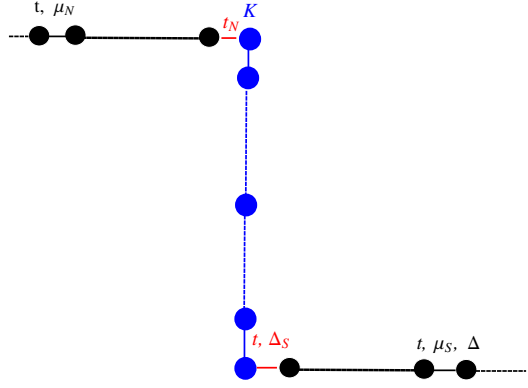}
\caption{In-line configuration of the N-KC-SC device.}
\label{Fig13}
\end{figure}

\begin{figure}
\centering
\includegraphics[scale=0.55]{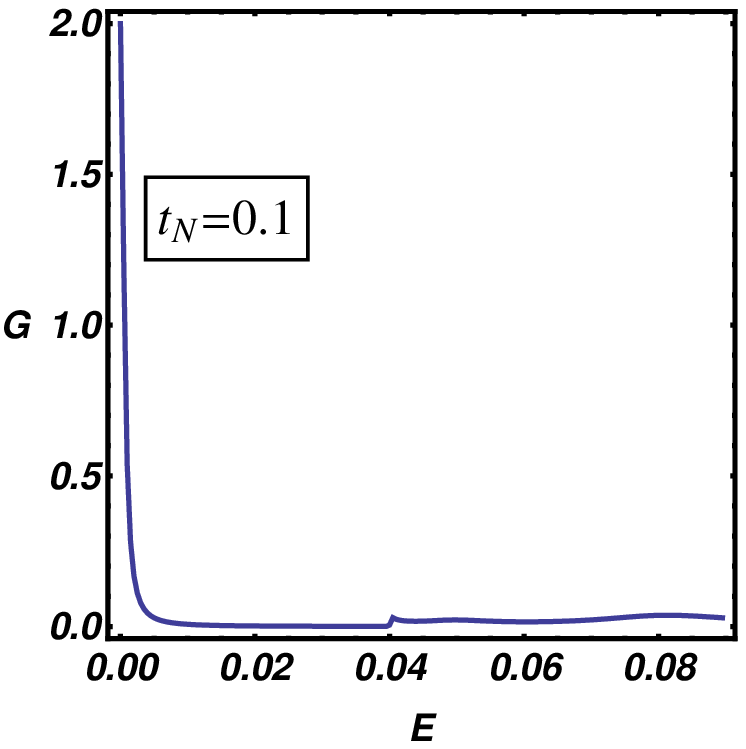}
\includegraphics[scale=0.55]{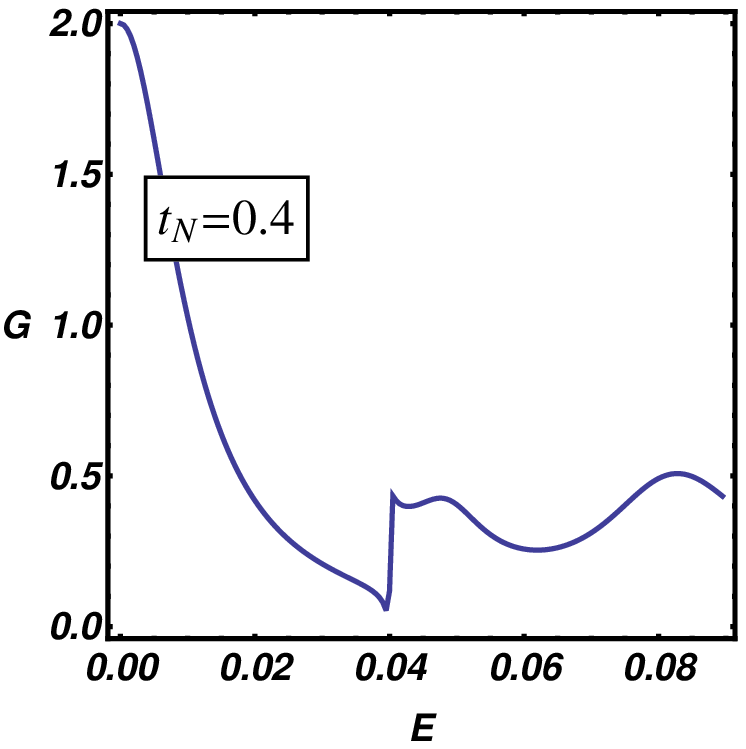}\\
\includegraphics[scale=0.55]{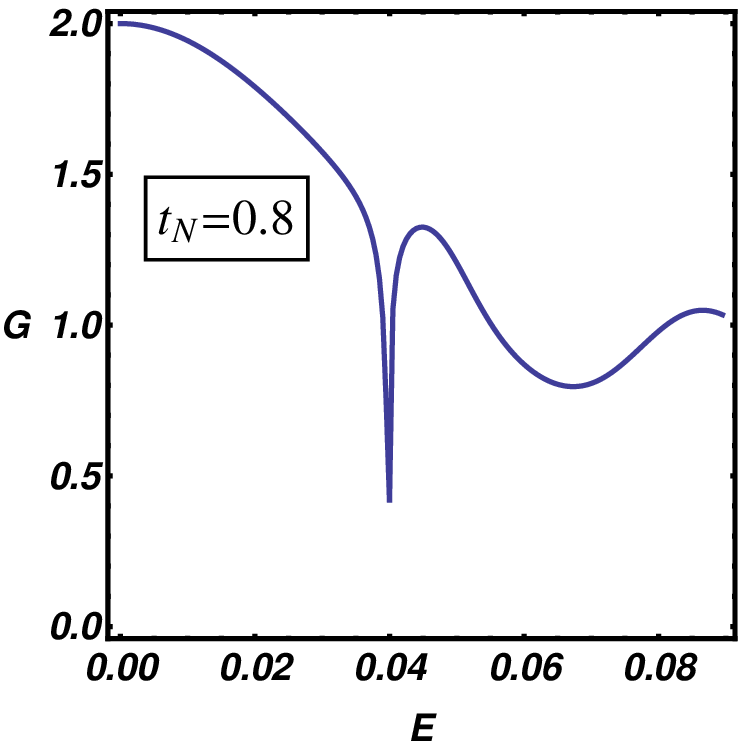}
\includegraphics[scale=0.55]{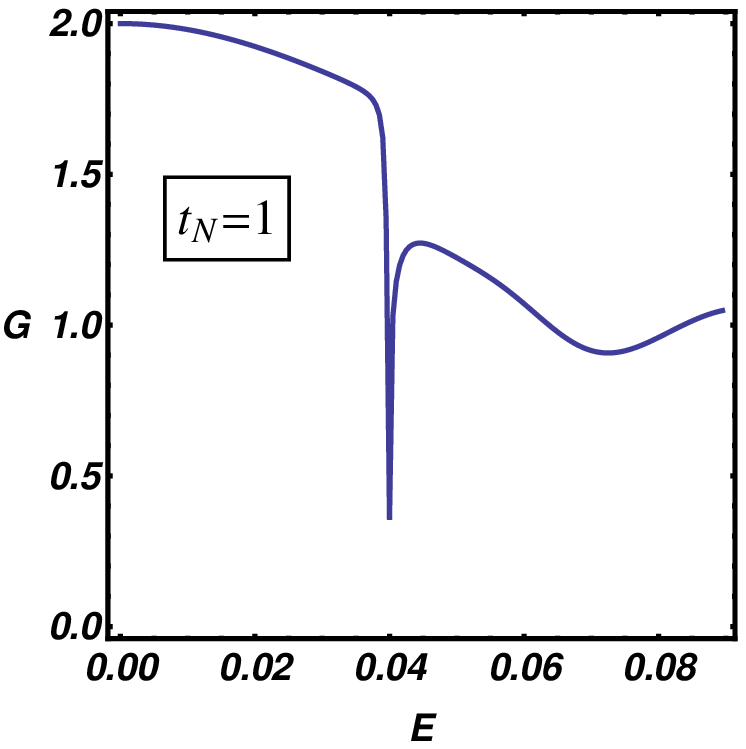}
\caption{N-KC-SC device. Zero-temperature differential conductance of the model depicted in Figure \ref{Fig13}. The model parameters have been fixed as: $\Delta=0.02$, $t_S=t=1$, $\mu=0.5$, $\mu_S=\mu_N=0$, while $t_N=0.1$, $0.4$, $0.8$, $1$ has been used in obtaining the different plots.}
\label{Fig14}
\end{figure}

The conduction properties of the device described in the main text (see Fig. \ref{Fig7}) are affected by the $t_N$ value, describing the interface opacity, and by the network geometry. Indeed, the so-called T-stub configuration in quantum wave guides enhances scattering events and may induce a lowering of the device conductance. According to these considerations, one easily argues that the transparency of the scattering region (i.e. the Kitaev chain) not only depends on the $t_N$ value, which plays the same role of the dimensionless barrier strength $Z$ of the BTK theory \cite{Blonder}, but it is also affected by the wave guide geometry. As a consequence it is plausible that the transparent limit, in which the conductance is maximized, cannot be obtained within the relevant experimental configuration described in the main text.\\
For the sake of completeness, in order to study the transparent limit of the proposed model, we define a new system geometry presenting an in-line configuration which avoids branching points along the current path (see Fig. \ref{Fig13}).\\
We numerically solve the tight binding equations for this model and compute the differential conductance. The results of the analysis are reported in Figure \ref{Fig14}, where analogies with the conductance plots of the BTK theory \cite{Blonder} are present. Differently from the usual BTK conductance curves derived for the s-wave superconducting pairing, a zero-bias peak, reminiscent of the p-wave coupling, is always present in our simulations.


\begin{figure}
\centering
\includegraphics[scale=0.55]{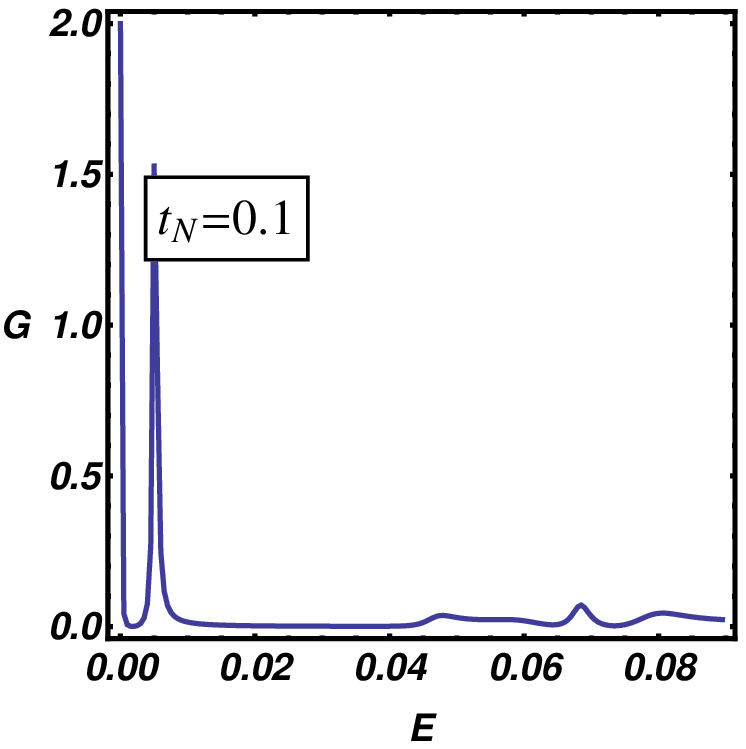}
\includegraphics[scale=0.55]{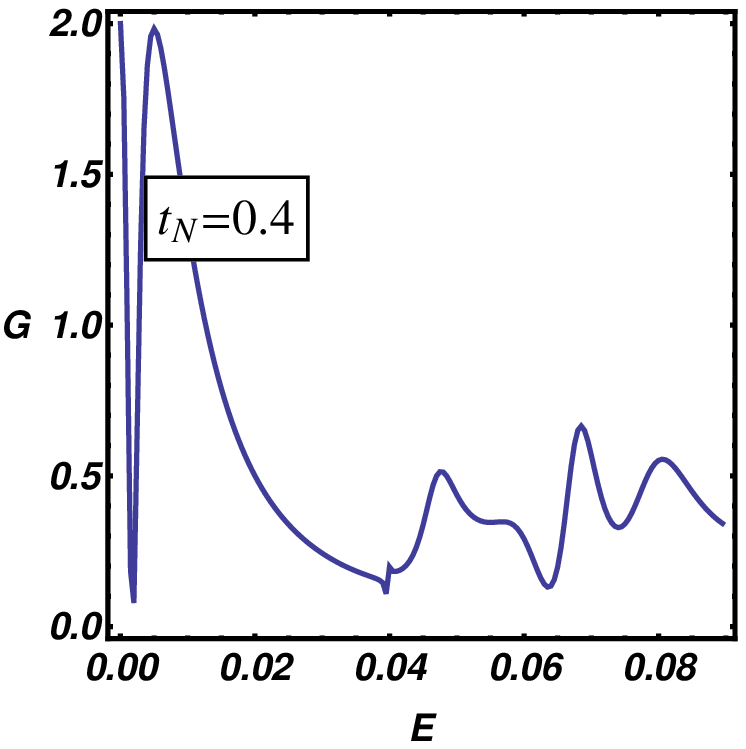}\\
\includegraphics[scale=0.55]{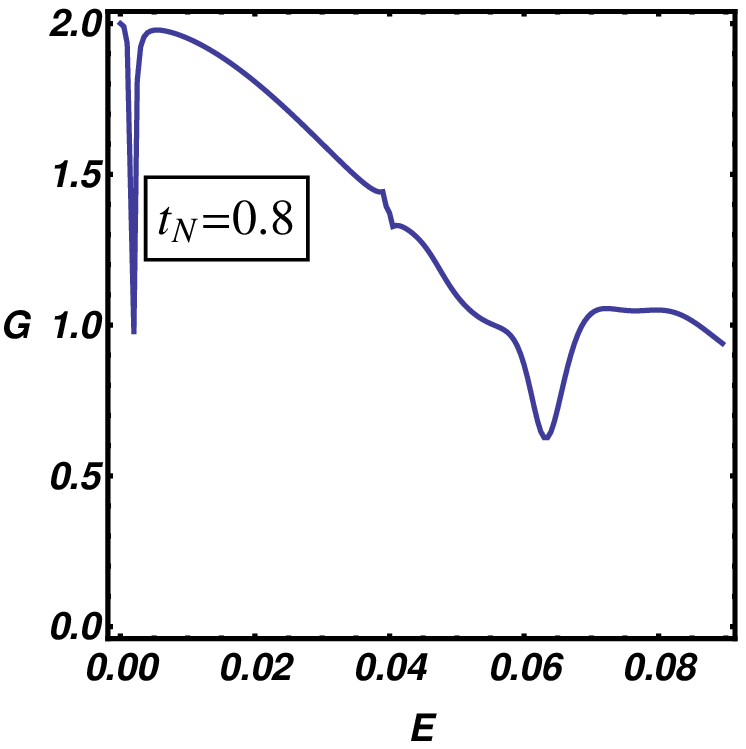}
\includegraphics[scale=0.55]{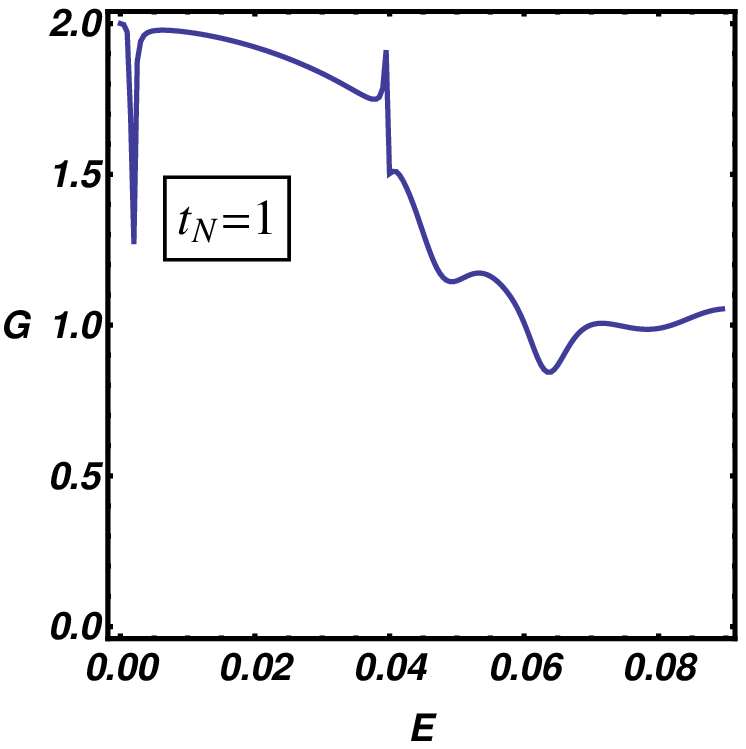}\\
\caption{Transparent limit of the N-KL-SC device. Zero-temperature differential conductance as a function of energy. The model parameters have been fixed as: $\Delta=0.02$, $t=t_S=1$, $\mu=0.5$, $\mu_S=\mu_N=0$, $\Delta_1=0.09$, $t_1=0.6$, while $t_N=0.1$, $0.4$, $0.8$, $1$ has been used in obtaining the different plots.}
\label{Fig15}
\end{figure}

A similar analysis can be performed for the N-KL-SC device (Figure \ref{Fig15}). Compared to the differential conductance of the single Kitaev chain (see Fig. \ref{Fig13}), a peculiar two-peaks structure is present in the ladder differential conductance curve. The two-peaks structure is more pronounced for opaque N-KL interfaces (i.e. for low $t_N$ values), while for higher $t_N$ values the two-peaks structure is substituted by a peculiar conductance minimum originated by the incomplete coalescence of two broadened resonant peaks. Moreover, the zero-bias peak and its satellite at higher energy, constituting the two-peaks structure mentioned before, are related to the presence of a quantum state with Majorana character accompanied by a quasi-Majorana quantum state, as discussed in the main text.

\section{Charge neutrality of quasi-Majorana modes}
\label{appCharge}
\begin{figure}
\centering
\includegraphics[scale=0.41]{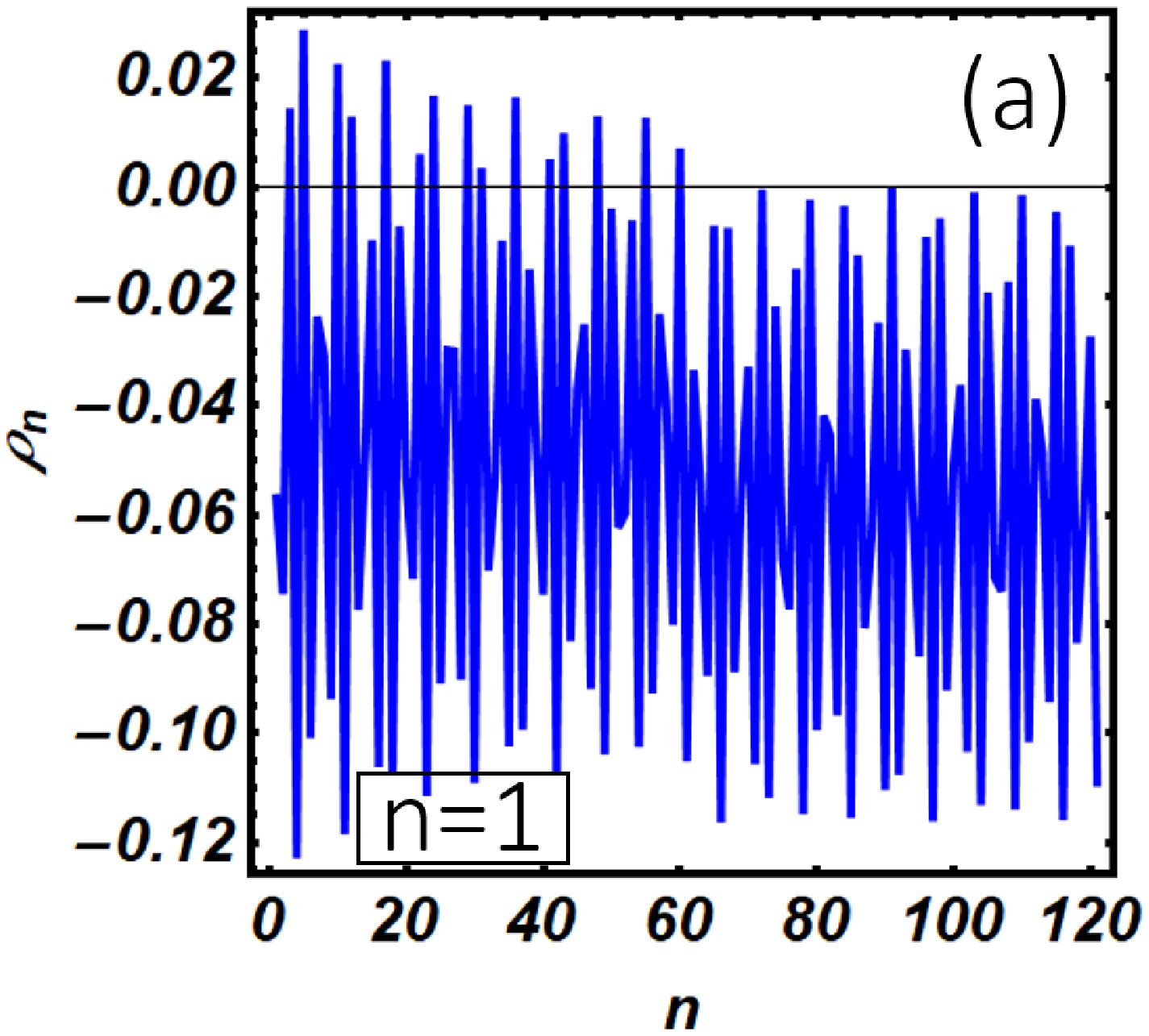}
\includegraphics[scale=0.38]{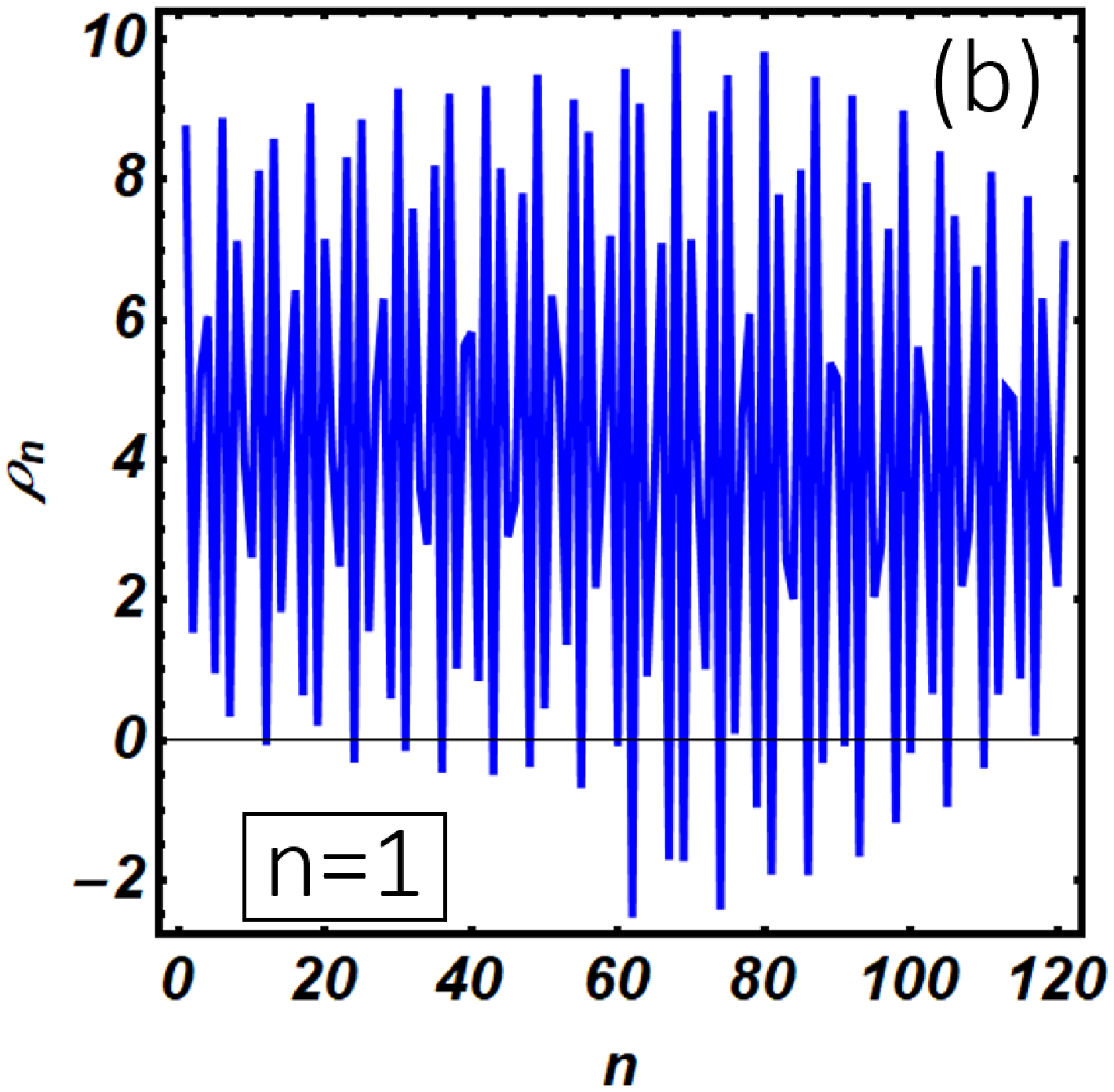}\\
\includegraphics[scale=0.4]{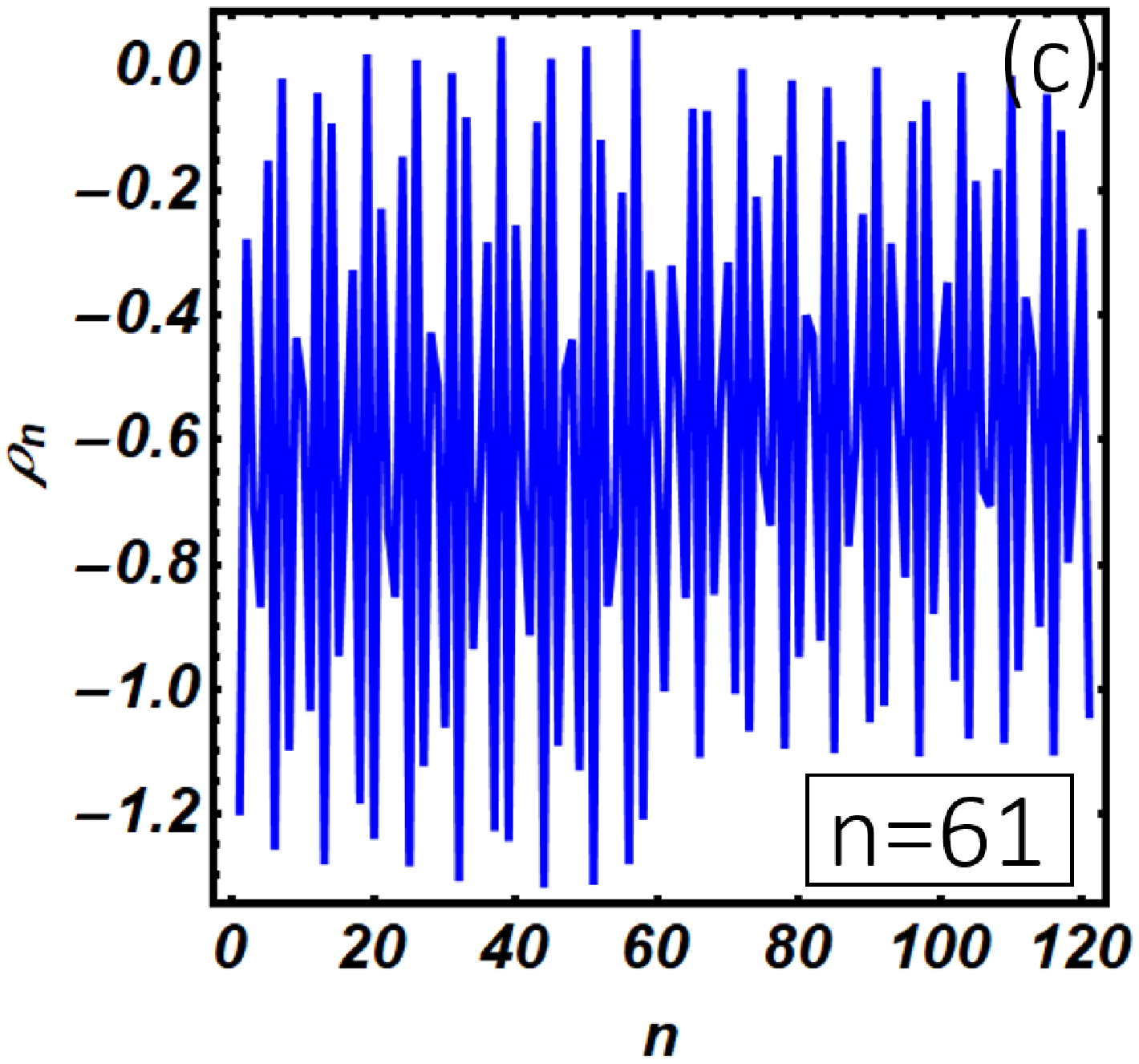}
\includegraphics[scale=0.38]{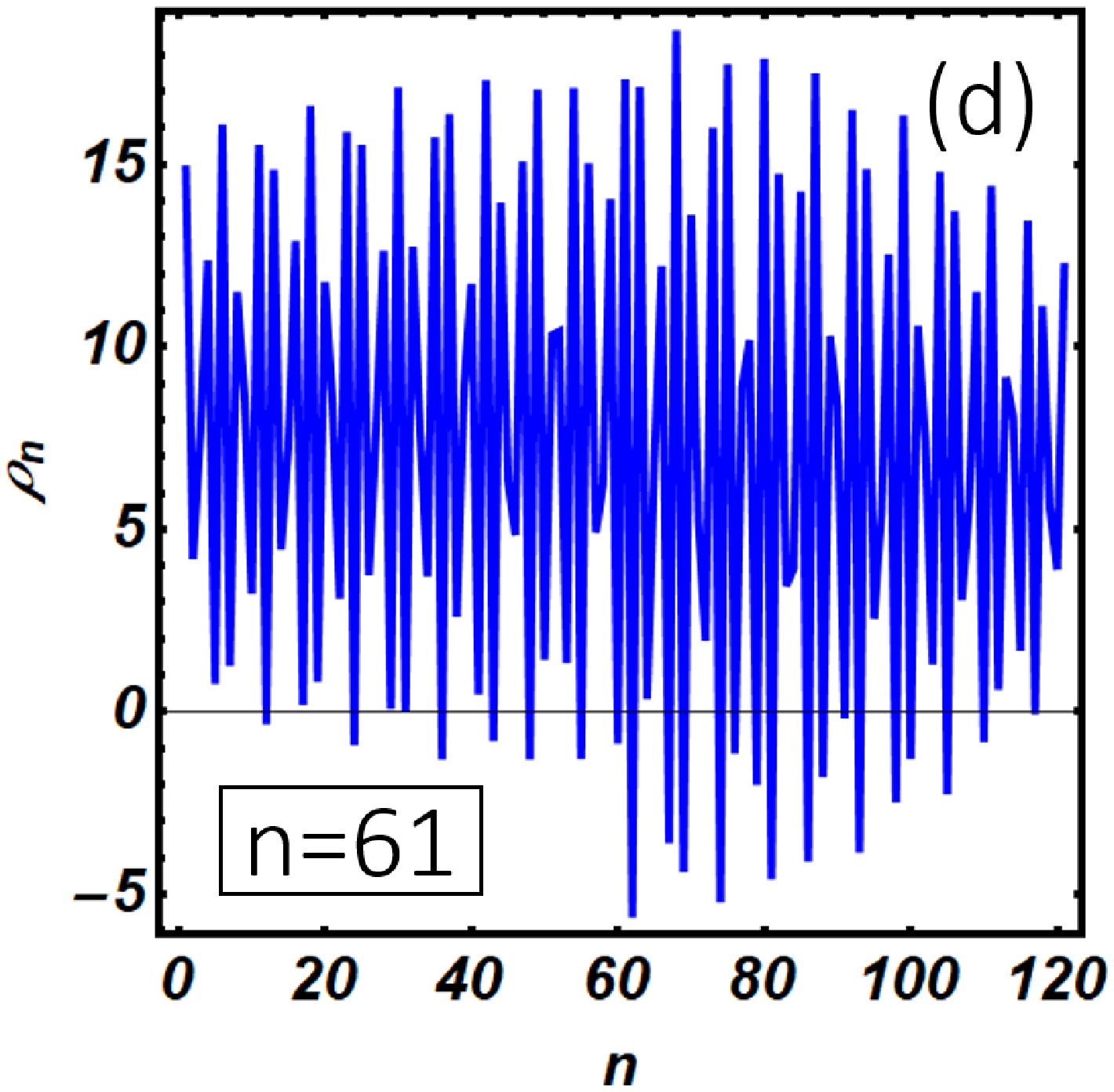}\\
\includegraphics[scale=0.4]{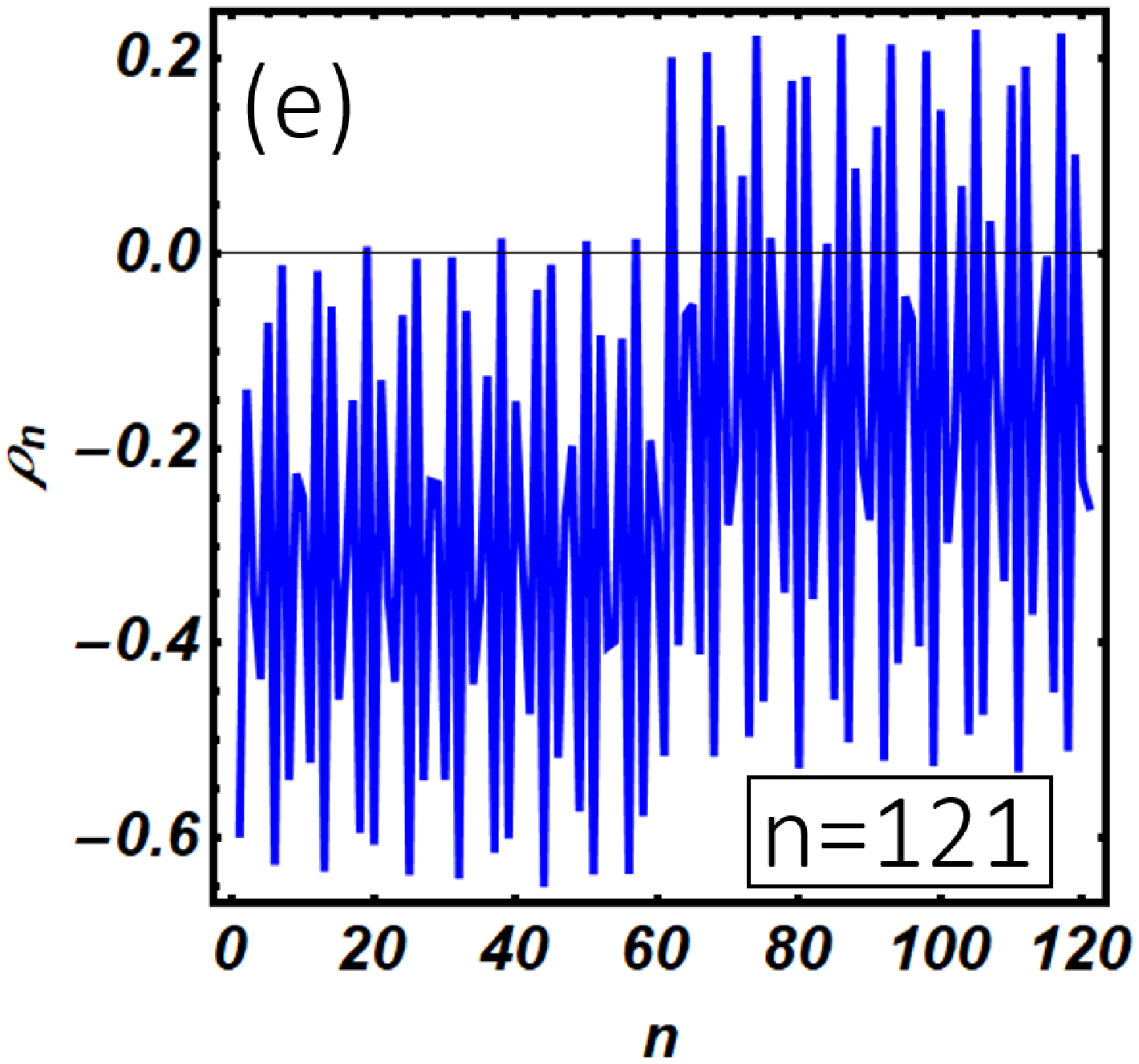}
\includegraphics[scale=0.38]{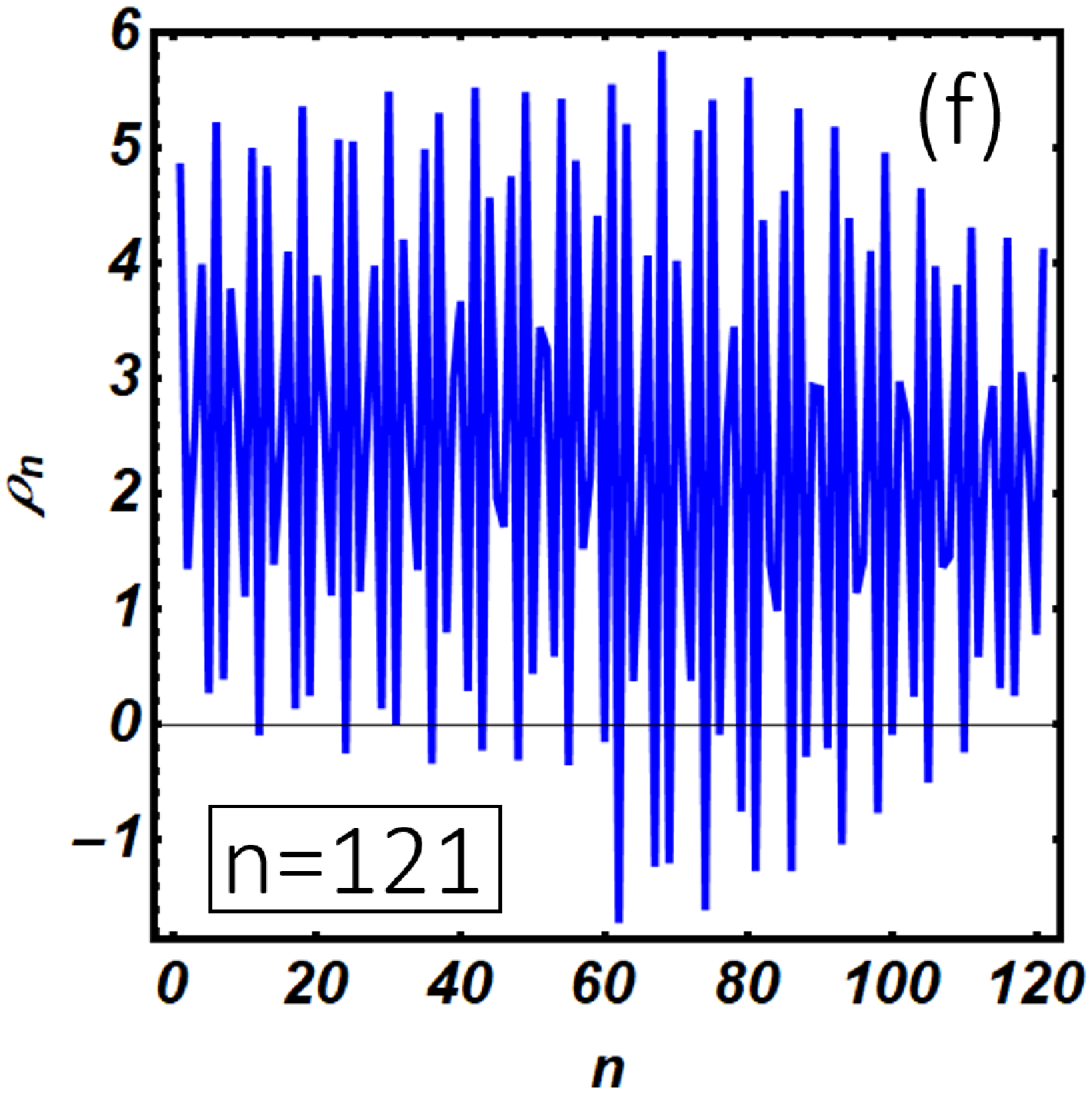}\\
\caption{site-dependent charge density $\rho_n=|f_n|^2-|g_n|^2$ as a function of the position along the Kitaev chain device discussed in Figure \ref{Fig8} of the main text. Curves in panels (a), (c), (e) correspond to the zero-bias conductance peak in Figure  \ref{Fig8} ($E=5 \cdot 10^{-5}$), while the remaining panels correspond to the satellite conductance peak ($E=6 \cdot 10^{-3}$). Panel (a) is computed by setting the model parameters as done in Figure \ref{Fig8} (b). The average charge density is given by $\bar{\rho}\approx -0.05$. Panel (b) corresponds to Figure \ref{Fig8} (c) with $\bar{\rho} \approx 5$. Panel (c) corresponds to Figure \ref{Fig8} (e) with $\bar{\rho} \approx -0.6$. Panel (d) corresponds to Figure \ref{Fig8} (f) with $\bar{\rho} \approx 6$. Panel (e) corresponds to Figure \ref{Fig8} (h) with $\bar{\rho} \approx -0.25$. Panel (f) corresponds to Figure \ref{Fig8} (i) with $\bar{\rho} \approx 2.5$. Strong deviations from the charge neutrality condition evidence an important contamination of topological properties. Hybridized modes with prevalent electron-like (hole-like) character are defined by $\bar{\rho}>0$ ($\bar{\rho}<0$).}
\label{Fig16}
\end{figure}

Majorana modes in closed systems are neutral excitations. Electrical neutrality implies that they cannot transport charge current unless a certain degree of hybridization with other quantum states occurs. These hybridization phenomena are supposed to be relevant when the topological system is connected with external reservoirs and a current is forced to flow through it. Under this circumstance, Majorana modes are expected to be subject to hybridization with the reservoir quantum states in order to warrant the current conservation. The degree of hybridization is expected to be variable with the system geometry, which defines the path followed by the charge current along the device. On the light of these general arguments, in Figure \ref{Fig16} we have investigated the site-dependent charge density $\rho_n=|f_n|^2-|g_n|^2$ (with $f_n$ and $g_n$ the electron-like and hole-like component of the Nambu spinor) corresponding to the internal modes presented in Figure \ref{Fig8} (b), (c), (e), (f), (h), (i). Different $\rho_n$ versus $n$ curves in Figure \ref{Fig16} present a complicate oscillating behavior with the position along the Kitaev chain with average charge density given by $\bar{\rho}=(\sum_{n=1}^{L}\rho_n)/L$.
Our findings confirm that zero-bias peaks in the conductance curves are related to Majorana modes which are weakly hybridized with the electrodes quantum states, the degree of hybridization being related to the deviation of the internal mode charge from the charge neutrality condition. On the other hand, satellites peaks in the differential conductance curves (Figure \ref{Fig8} (a), (d), (g)) are related to internal modes characterized by a relevant deviation from the charge neutrality requirement, the latter condition evidencing a strong contamination of the Majorana character of the internal mode.

\clearpage

%\reftitle{References}

\end{document}